\documentclass{jfm}
\usepackage{graphicx}
\usepackage{epstopdf, epsfig}
\usepackage{amsmath}
\usepackage{amsfonts}
\usepackage{amssymb}
\usepackage{xcolor}

\DeclareMathOperator{\e}{e}
\newcommand{\red}{}
\newcommand{\blue}{}
\newcommand{\green}{}
\newcommand{\Green}{}

\shorttitle{ Turbulence of capillary waves}
	\shortauthor{M. Berhanu et al.}
	
	\title{\red{Turbulence of capillary waves forced by steep gravity waves}}
	
\author{M. Berhanu\aff{1}
  \corresp{\email{michael.berhanu@univ-paris-diderot.fr}}, 
E. Falcon\aff{1},
L. Deike\aff{2,3}}

\affiliation{\aff{1}Mati\`ere et Syst\`emes Complexes (MSC), Universit\'e Paris Diderot, CNRS (UMR 7057), 75013 Paris, France
\aff{2} Department of Mechanical and Aerospace Engineering, Princeton University, USA
\aff{3} Princeton Environmental Institute, Princeton University, USA}

\begin{document}

\maketitle

\begin{abstract}
\red{We study experimentally the dynamics and statistics of capillary waves forced by random steep gravity waves mechanically generated in laboratory. Capillary waves are produced here by gravity waves from nonlinear wave interactions. Using a spatio-temporal measurement of the free-surface, we characterize statistically the random regimes of capillary waves in the spatial and temporal Fourier spaces. For a significant wave steepness ($0.2-0.3$), power-law spectra are observed both in space and time, defining a turbulent regime of capillary waves transferring energy from large scale to small scale. Analysis of temporal fluctuations of spatial spectrum demonstrates that the capillary power-law spectra result from the temporal averaging over intermittent and strong nonlinear events transferring energy to small scale in a fast time scale, when capillary wave trains are generated in a way similar to the parasitic capillary wave generation mechanism. The frequency and wavenumber power-law exponents of wave spectrum are found to be in agreement with those of the weakly nonlinear Wave Turbulence Theory. However, the energy flux is not constant through the scales and the wave spectrum scaling with this flux is not in good agreement with Wave Turbulence Theory. These results suggest that theoretical developments beyond the classic Wave Turbulence Theory are necessary to describe the dynamics and statistics of capillary waves in natural environment. In particular, in presence of broad scale viscous dissipation and strong nonlinearity, the role of non-local and non-resonant interactions could be reconsidered.}
\end{abstract}

\section{Introduction}

\red{Disordered patterns of waves are easily seen on a choppy sea, a consequence of both the wave dynamics and the wind forcing. Due to the large number of degrees of freedom and to the exchanges between these scales permitted by nonlinear effects, the dynamics becomes complex and unpredictable. The relevant approach is therefore a statistical analysis of the free surface, considering random propagation of dispersive surface waves interacting nonlinearly in presence of wind forcing and dissipation. At large scale the main restoring force is gravity, whereas at scale below $1\,$cm, surface tension is dominant, and the waves are said capillaries or capillary waves. Although the energy carried by capillary waves is significantly smaller compared to gravity waves, the study of capillary wave dynamics is important to describe the exchanges between the sea and the atmosphere. They also increase tremendously the water surface roughness, necessary for the radar scattering monitoring of sea waves \citep{Hwang2013}, and contribute to the overall dissipation of gravity waves \citep{Zhang2002,Tsai2010,Caulliez2013,MelvilleJFM2015,DeikeJFM2015cap}. Therefore a statistical  description of random waves at the capillary scales is important in environmental fluid dynamics and oceanography.}

\red{For small wave amplitudes at the air-water interface, in absence of current and vorticity, wave propagation obeys the linear dispersion relation:
 \begin{equation}
\omega^2=\left[g\,k + (\gamma / \rho) \, k^3 \right]\tanh (k\,h_0)\, ,
\label{RDL}
\end{equation} 
where $\omega=2\pi\,f$ is the angular frequency, $k=\frac{2\pi}{\lambda}$ the wavenumber modulus, $g=9.81$\,m/s$^2$ the gravity acceleration, $\gamma$ the surface tension at the air-water interface, $\rho$ the water density and $h_0$ the depth of the fluid layer. In the following we consider the deep-water limit, where $ k\,h_0 \gg 1$. At higher wave amplitude, nonlinear effects must be taken in account, leading to modifications of the dispersion relation \citep{Whitham,Crapper} and to wave interactions between different scales. A dimensionless nonlinear parameter is introduced to quantify the relevance of nonlinear effects: the wave steepness $a\, k$, where $a$ is the wave amplitude and $k$ the typical wavenumber. It represents the typical slope of the deformed free-surface.}

\red{Three-Wave interactions for frequencies close to the gravity-capillary crossover are prone to produce capillaries from gravity waves \citep{Hammack1993,Aubourg2015} in weakly nonlinear regime. For a wave-field having frequency wave components $f_1$ and $f_2$, the quadratic nonlinearity of the equations describing surface wave propagation induces a product of these components. An excitation is obtained at the sum frequency $f_1 + f_2$. For large system, or for long observation time, only interactions which are resonant have a net contribution to energy transfers. The resonance condition for waves following the dispersion relation $\omega (k)$ with $k=||\mathbf{k}||$ consists in the following relations:}
 \red{\begin{eqnarray}
\mathbf{k_1} \pm \mathbf{k_2}  \pm \mathbf{k_3} &=&0 \label{Resonancek}\\ 
\omega ({k_1}) \pm \omega ({k_2})   \pm \omega ({k_3})&=&0
\label{Resonance}
\end{eqnarray}}
\red{Non-resonant interactions  (\textit{i.e.} not validating the above conditions) product oscillating contributions with a period (spatial or temporal) decreasing with the mismatch to the resonant condition. After averaging over this period, energy transfer by non-resonant interactions vanish. These interactions are thus most of the time neglected. However, experimentally, especially in finite domains, the resonances cannot be exact.} \blue{Therefore, we define quasi-resonant interactions in the wave-field dynamics, as interactions with small differences to the resonant conditions \citep{Aubourg2015,PanYueJFM2017}. The tolerance is usually justified by a nonlinear broadening of the dispersion relation. For quasi-resonant interactions, the mismatch to the resonance is sufficiently small, that the average on a finite size domain leads to a net energy transfer.}

\red{At a higher level of nonlinearity, \textit{i.e.} higher wave steepness, other mechanisms of generation of capillary waves must be considered. Parasitic capillary wave generation is the most studied mechanism. It describes appearance of capillaries on the crest of a steep gravity propagating wave \citep{Longuet-Higgins1963,Longuet-Higgins1995, FedorovJFM1998,FedorovPOF1998}, by a fast mechanism transferring energy directly from large scale to small scale \citep{MelvilleJFM2015}. A such nonlinear interaction between a long wave and a short wave is said non local in wavenumber space. Parasitic capillary waves are typically observed for gravity waves with a wavelength between $5$ and $40$\,cm and a steepness larger or equal to $0.05$ \citep{Zhang2002,FedorovJFM1998,FedorovPOF1998}. Similar capillary wave patterns occur also on the front of  gravity waves at higher steepness, before spilling breaking events \citep{Duncan1999,Duncan2001,DeikeJFM2015cap,MelvilleJFM2015} and also on the crest of steep standing gravity waves \citep{Schultz1998}. Nevertheless parasitic capillary wave generation mechanism is described to date theoretically only for unidirectional, propagative and monochromatic long gravity wave.} 

\red{The Wave Turbulence Theory \citep{Zakharovbook,Nazarenkobook,Newell2011,NazarenkoRev} provides an analytical description in the Fourier space of dynamics of random dispersive waves interacting by resonant wave interactions in a weakly nonlinear regime. It predicts the capillary waves elevation spectra as a power law, $S_\eta (f) \propto f^{-17/6}$ in time and $S_\eta (k) \propto k^{-15/4}$ in space (after integration on all the orientations) \citep{Zakharov1967,Pushkarev2000}. The hypotheses used in the theoretical derivation are drastic, in particular scale separation between forcing and dissipation, i.e. negligible dissipation in the inertial range, the infinite homogeneous system and the randomness of the wave field. Moreover, analytic results are provided only for pure gravity waves or pure capillary waves. Therefore the applicability of wave turbulent concepts in experimental and natural systems remains questionable. Beyond this theory, in order to describe statistically spectra of gravity-capillary wave field, few authors consider the dynamics of wave action spectrum modeled with a kinetic equation including Three-Wave interactions numerically \citep{Dulov2009,Kosnik2010} and analytically \citep{Stiassnie1996}.}
\red{Finally to study statistically gravity-capillary waves at higher nonlinearity an interesting approach in Fourier space consists in enabling a given mismatch to resonant conditions in wave interactions \citep{WatsonJFM1993,WatsonJFM1996,WatsonJFM1999}. Parasitic waves can be numerically reproduced by considering non-resonant unidirectional Three-Wave interactions  \citep{WatsonJFM1996,WatsonJFM1999}. Therefore, at sufficiently high steepness, energy transfer to the small scales could occur through non-resonant Three-Wave interactions under the form of parasitic capillary wave generation.}

\red{In this article, we study experimentally random fields of capillary waves forced by random steep gravity waves mechanically generated in a finite sized tank. Small scale properties of a choppy sea surface is in this way mimicked, by replacing the chaotic forcing by gravity waves by a stochastic one. Using a resolved spatial temporal measurement of the free-surface, we characterize statistical and dynamical properties of capillary wave in real and Fourier space.}

 \red{This study follows a previous work, using the same experimental device \citep{Berhanu2013}. At high enough level of excitation by a spatio-temporal analysis, we characterize the capillary wave turbulence, forced by gravity waves. We reported an agreement of the exponents of temporal and spatial power-law spectra with the predictions of Wave Turbulence Theory. Nevertheless in these conditions, we observed that several hypotheses of the theory are not met. The nonlinearity is not weak, the dissipation in not negligible for capillary waves and the wave field is not isotropic even at small scale \citep{Berhanu2013}.}
 
 \red{Here we perform a deeper and more complete analysis for wave steepness varying from $0.15$ to $0.34$. In the real space, the wave field displays capillary wave trains on the crests of large gravity waves. First, we demonstrate by a spatial Fourier analysis, that the capillary wave phases are uniformly distributed, justifying the stochasticity of the wave field at small scale. Then we compute space-time Fourier spectra of wave elevation. We obtain experimentally continuous dispersion relations displaying a significant nonlinear broadening. To describe energy transfer from gravity waves to small scale capillary waves, the decays of spectral energy as a function of the frequency and of the wavenumber are analyzed. We observe power-law spectra, that we interpret as a signature of turbulence of capillary waves. The spectral exponents are close to the predictions of Wave Turbulence Theory. Then by studying higher order correlations in Fourier space, we detect substantial presence of Three-Wave interactions. We note the broad width of the dispersion relation may permit interactions far from the resonant conditions. However by studying the temporal fluctuations of the spatial spectrum, we show that the energy transfers occur through strong nonlinear events generating capillary wave trains, similarly to the parasitic capillary wave mechanism. A departure from the Gaussian distribution of wave amplitude is reported, corresponding to the presence of intermittent large scale coherent structures, having a broad spectral contribution. This strong capillary wave turbulence differs thus from the weak capillary wave turbulence described by the Wave Turbulence Theory, although the exponents of the power-law of temporal and spatial spectra are the same in both cases. A quantitative comparison of our results with predictions of the Wave Turbulence Theory is then performed. Using the computation of the dissipated power, the energy flux is estimated and is found non-conserved through the scales due to the broad band dissipation.} \Green{The scaling of the wave spectrum with the mean energy flux is tested and is found approximate and depending on the wave amplitude.} \red{Thus, for turbulent capillary waves, broad scale dissipation and strong nonlinearity should be taken into account in theoretical analyses. Moreover, due to the significant viscous dissipation, we show that finite size effects are negligible for capillary waves.}  

This paper is organized as follows. First we present in \S 2, the methods of this work: the experimental device, the \red{forcing} protocol and a brief characterization of the velocity field. Then in \S 3, we analyze the turbulent wave-field in stationary regime in Fourier space. In \S 4, the capillary wave turbulent regimes are studied as a function of time to characterize the fluctuations. In \S 5, we \red{compare} our results with the predictions of Wave Turbulence Theory. \red{Then we discuss the applicability of the Wave Turbulence Theory to capillary waves in experiments. Finally, in \S 6, before giving the conclusion, we discuss the role of viscous dissipation and gravity-capillary crossover on turbulent regimes of capillary waves.} Moreover, a first appendix demonstrates that due to viscous dissipation, finite size effects are negligible for capillary waves.

\section{Methods}

\subsection{Experimental Setup}    
  
\begin{figure}
 \begin{center}
\includegraphics[width=8.cm]{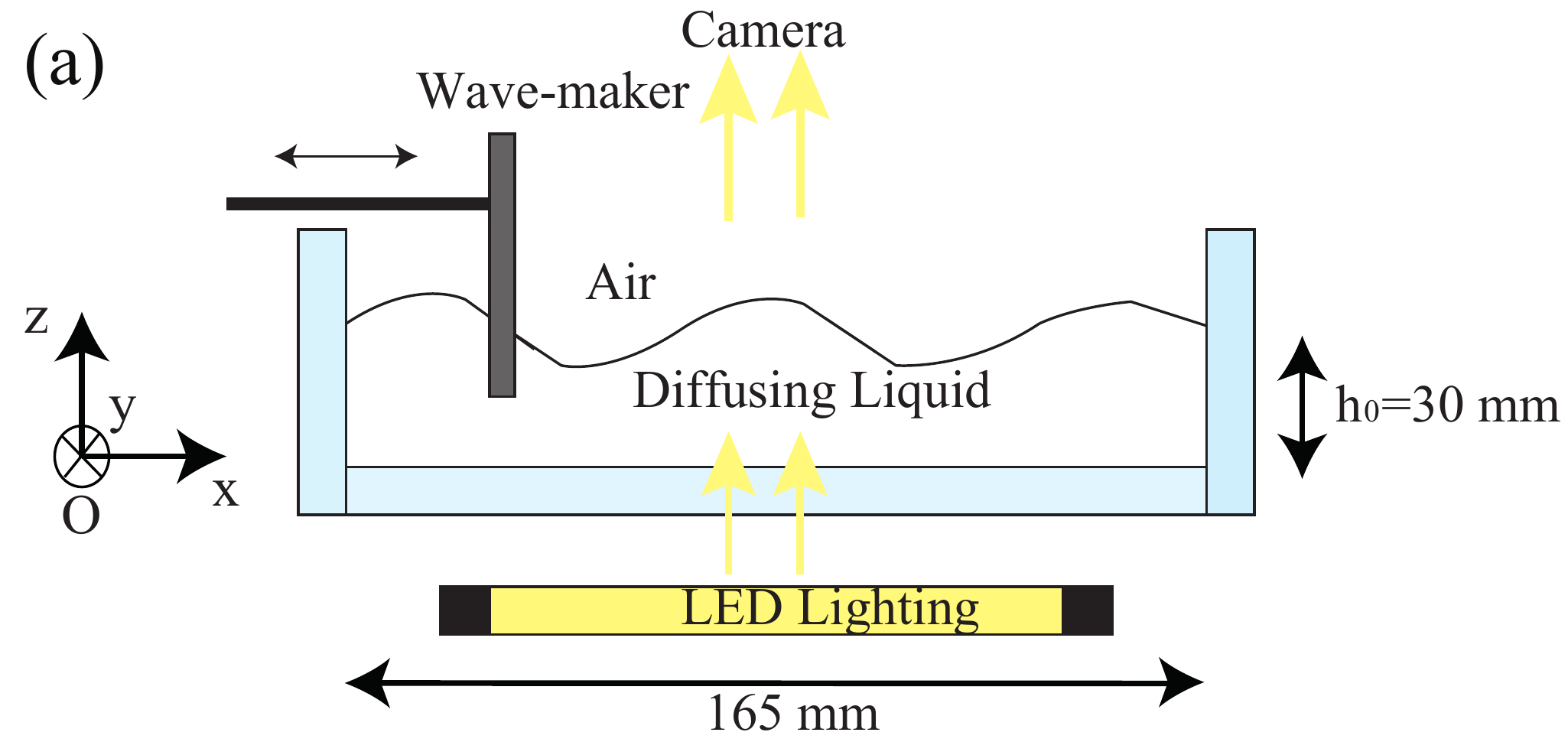}
\includegraphics[width=5cm]{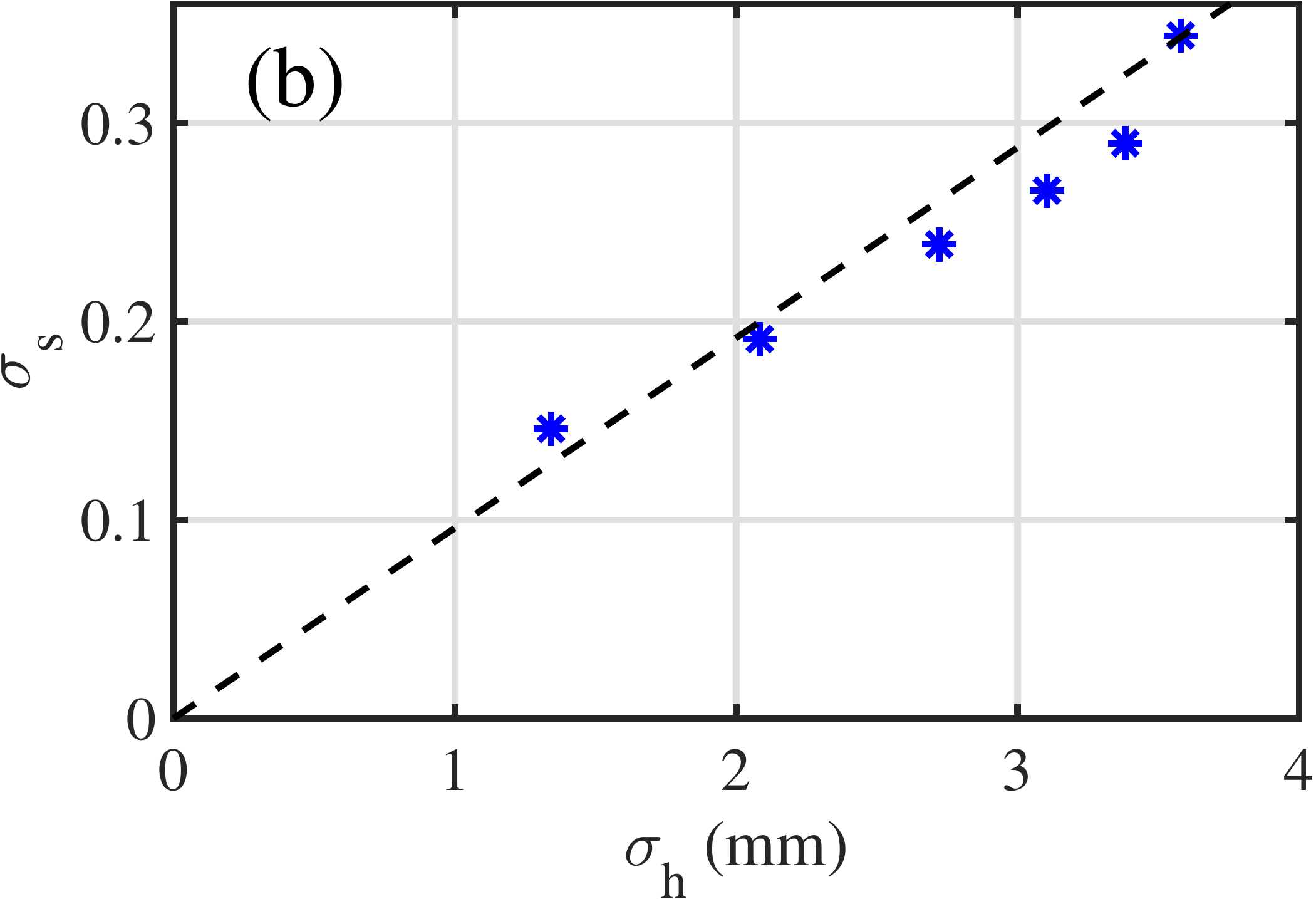}
 \caption{(color online) (a) Experimental setup. Surface waves are produced by the horizontal motion of a rectangular paddle and the free-surface is measured in space and time using \textit{Diffusing Light Photography} (DLP). (b) Typical wave steepness $\sigma_s$ versus the typical wave amplitude $\sigma_h$. The dashed line depicts the usual estimation of steepness by $\sigma_h \,k$, with $k$ the wavenumber obtained with the linear dispersion relation for $f=5\,$Hz (central forcing frequency). }
    \label{schemab}
       \end{center}
 \end{figure}  
The experimental setup and measurement techniques are similar to those described in \citet{Berhanu2013} and the experimental device displayed in Fig.~\ref{schemab} is thus identical. We recall here only the essential points. The free-surface elevation is directly measured in space and time using the \textit{Diffusing Light Photography} (DLP) \citep{Putterman1996,Putterman1997,Xia2012} technique. A Plexiglass tank ($165\times 165$\,mm$^2$) is filled with a diffusing liquid ($1$\,L of distilled water with $6$\,mL of Intralipid 20\%) up to a height  $h_0=30\,$mm. Intralipid 20~\% (Fresenius Kabi \texttrademark)  is a commercial lipidic emulsion of microspheres, whose aqueous solutions are used as model diffusing media with characterized optical properties \citep{Staveren}. Due to the low dilution, the fluid viscosity and density are close from the pure water values ($\nu=10^{-6}$\,m/s$^{2}$ and $\rho=1000\,$kg/m$^{3}$, respectively). \green{The surface tension has been measured statically with a Du No\"uy ring tensiometer, providing a value of $53.6$ mN/m. However, a dynamic value of surface tension $\gamma=60$ mN/m is obtained from the spatio-temporal measurements \citep{Berhanu2013}.  The dynamical value appear nevertheless more relevant to analyse wave propagation \citep{Hammack1993}}. Surface waves are produced by the horizontal motion of a rectangular paddle ($130$\,mm in width and $13 $\,mm in immersed depth) driven by an electromagnetic shaker (LDS V406). A LED device Phlox ($100 \times 100$\,mm$^2$) ensures a homogeneous lighting below the transparent tank. A 16-bits camera (PCO EDGE), one meter above, is focused on the liquid free-surface and records with $1024 \times 1120$ pixels and a $200\,$Hz frame rate on an observation area $\mathcal{S}$ of $89 \times 96$\,mm$^2$.

 \subsection{Random forcing of gravity waves}
 \label{randomforcing}
 \begin{figure}
 \begin{center}
 \includegraphics[width=12cm]{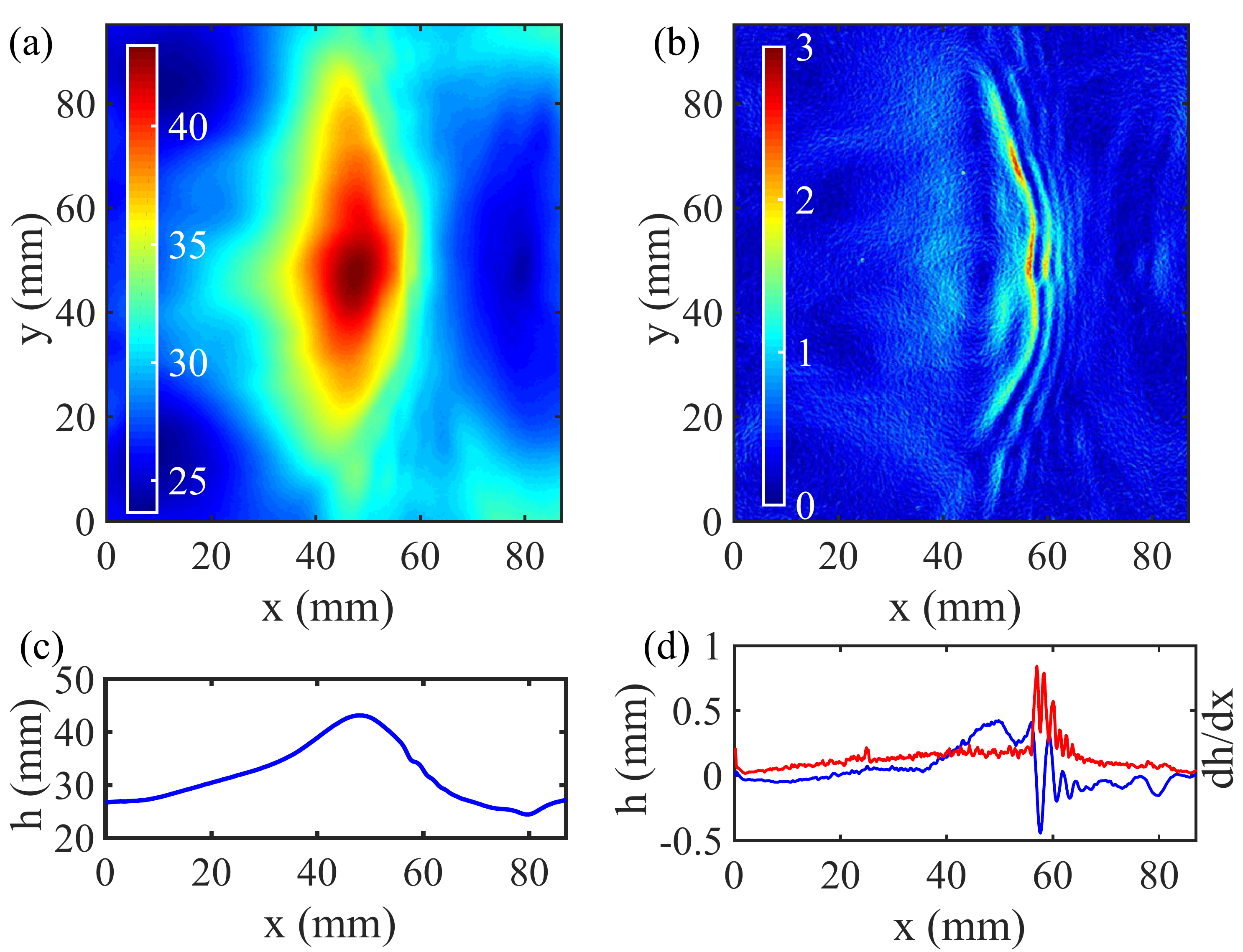} 
        \caption{(color online). (a) Snapshot of the wave field $h(x,y)$ at $t=1.51\,$s. $\sigma_h=3.6$\,mm and $\sigma_s=0.34$. Colorscale is in mm. Wave maker is parallel to the $y$-axis and located at $x=-12$\,mm. (b) Snapshot of the spatial gradient of wave elevation $||\mathbf{\nabla}h(x,y)||$ at the same instant. Colorscale is dimensionless. A train of capillary waves is visible on the forward front of the large carrier wave. (c) Wave profile $h(x,y=48\text{mm})$. (d) Blue, wave profile $h(x,y=48\text{mm})$ after a high-pass spatial filtering keeping the capillaries (cut-off $k=113$\,m$^{-1}$). Red (light gray), slope profile of $||\mathbf{\nabla}h(x,y=48\text{mm})||$.}
    \label{Snapshot}
       \end{center}
 \end{figure}    
 
The waves are forced at large scale through the random motion of the paddle. The electrical signal sent to the shaker is obtained by band pass filtering an initial white noise between $4$ and $6$ Hz. The resulting excitation is random in amplitude and phase and has an autocorrelation time of order $1$\,s. It generates waves belonging to the gravity wave range, the corresponding wavenumbers given by the linear dispersion relation being respectively $k=65.3$\,m$^{-1}$ ($\lambda=96.2$\,mm) and $k=131$\,m$^{-1}$ ($\lambda=47.9$\,mm). By enhancing the initial mixing of waves, this type of random forcing is known to produce \red{power-law spectra interpreted as gravity-capillary wave turbulence} in laboratory experiments \citep{Falcon2007,Herbert2010,Cobelli2011,Deike2012,Deike2014}. \red{At small scale, surface tension dominates gravity as the restoring force for $k$ larger than the inverse of capillary length $l_c=\sqrt{\gamma / (\rho\,g)} \approx 2.5$\,mm. Capillary waves are thus observed for $k > k_c = 404$\,m$^{-1}$ ($\lambda_c=15.5$\,mm).}

The paddle is parallel to the $y$ axis, close to one boundary, and the resulting wave field on the center of the tank is thus mainly directed along the $x$ axis and results from the waves emitted by the paddle and also by the waves reflecting on the opposite boundary. \red{Due to the reflections on the walls, the wavelengths which are a divisor of the container length ($165\,$mm), are the eigenmodes of the square tank, corresponding to possible standing waves. Using the linear dispersion relation (\ref{RDL}), the eight first eigenmodes frequencies are: $2.79$, $4.38$, $5.53$, $6.57$, $7.60$, $8.65$, $9.75$ and $10.90$\, Hz.} For each measurement, $8192$ images are recorded corresponding to a duration of $41$\,s. Using the DLP method, the deformation of the free-surface is reconstructed for each image, providing the spatio-temporal wave-field $h(x,y,t)$. The intensity of transmitted light measured on the camera decreases indeed as a function of the local depth $h$ \citep{Putterman1996}. After calibration, the free surface deformation is obtained with a horizontal spatial resolution of $0.5\,$mm, a vertical sensitivity of less than hundred microns even for high steepness of the surface and a temporal resolution given by the acquisition frequency of the camera ($200\,$Hz).
To characterize and compare the measurements, we define the following parameters, the typical wave amplitude \begin{equation} \sigma_h\equiv \left\langle \sqrt{\frac{1}{\mathcal{S}} \int_{\mathcal{S}} h^2(x,y,t) \mathrm{d}x \mathrm{d}y -\left( \frac{1}{\mathcal{S}} \int_{\mathcal{S}} h(x,y,t) \mathrm{d}x \mathrm{d}y \right)^2} \right\rangle \end{equation} \\ and the typical wave steepness  \begin{equation}\sigma_s \equiv \left\langle \sqrt{ \frac{1}{\mathcal{S}} \int_{\mathcal{S}} {||\mathbf{\nabla}h(x,y,t)||}^2 \mathrm{d}x \mathrm{d}y -\left( \frac{1}{\mathcal{S}} \int_{\mathcal{S}} {||\mathbf{\nabla}h||} \mathrm{d}x \mathrm{d}y \right)^2} \right\rangle, \end{equation} where $\left\langle \cdot \right\rangle $ denotes a temporal averaging and $\int_{\mathcal{S}}$ a spatial integration on the surface $\mathcal{S}$. This computation of steepness for a random field of waves, extends the classical definition $\sigma_h\,k$ for a monochromatic wave and evaluates the amplitude of nonlinear effects in wave propagation. The set of experiments described here, corresponds to $\sigma_h = $1.3, 2.1, 2.7, 3.1, 3.4 and 3.6 mm, and $\sigma_s=0.15$, 0.19, 0.24, 0.27, 0.29, and 0.34. \red{The  global steepness of the wave-field is above $0.1$ and the nonlinearity level is thus not small.} The maximal forcing is set to avoid breaking events in the observation area, for which the free-surface becomes multivalued. Whereas waves with a such steepness are intrinsically unstable, due to the limited size of the container, they are experimentally observed in the field of view of the camera.  In Fig.~\ref{schemab} (b), $\sigma_s$ is plotted as a function of $\sigma_h$. The typical steepness is well described by the relation $\sigma_h \,k$, with $k$ the wavenumber obtained with the linear dispersion relation for $f=5\,$Hz, which is the central frequency of the forcing range.\\

\begin{figure}
 \begin{center}
 \includegraphics[width=12cm]{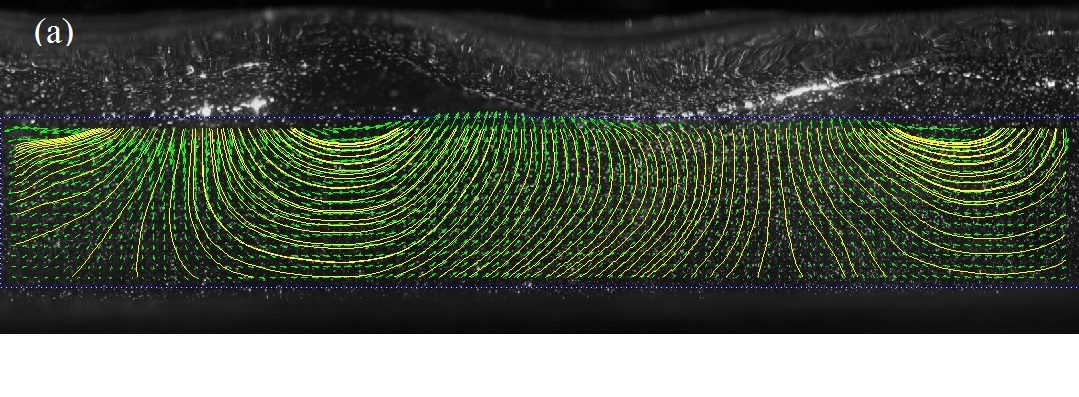} 
 \includegraphics[width=7.1cm]{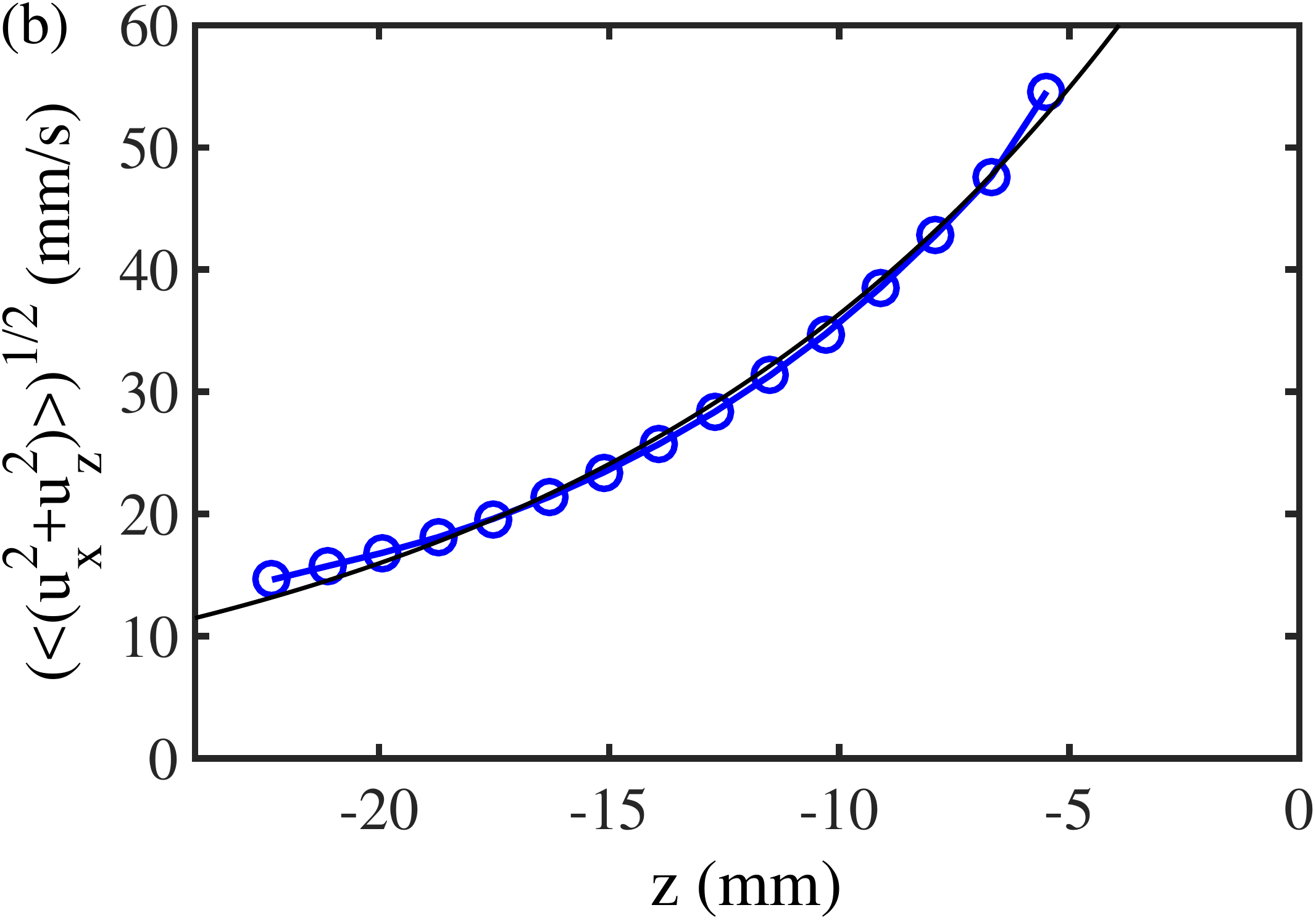} 
   \includegraphics[width=12.5cm]{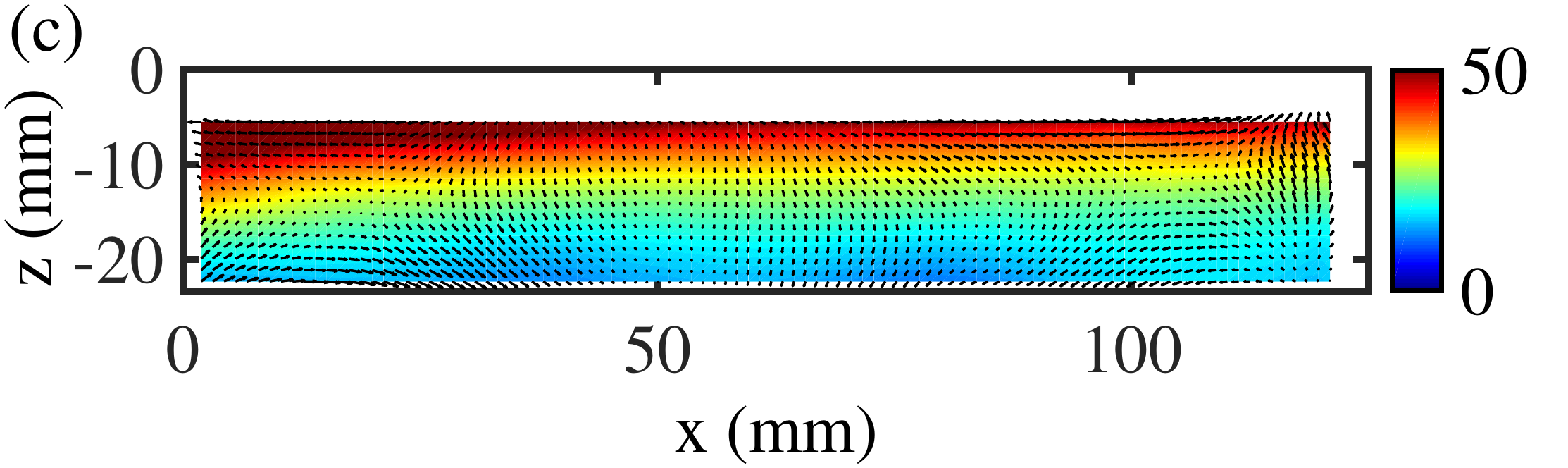}
    \includegraphics[width=12.5cm]{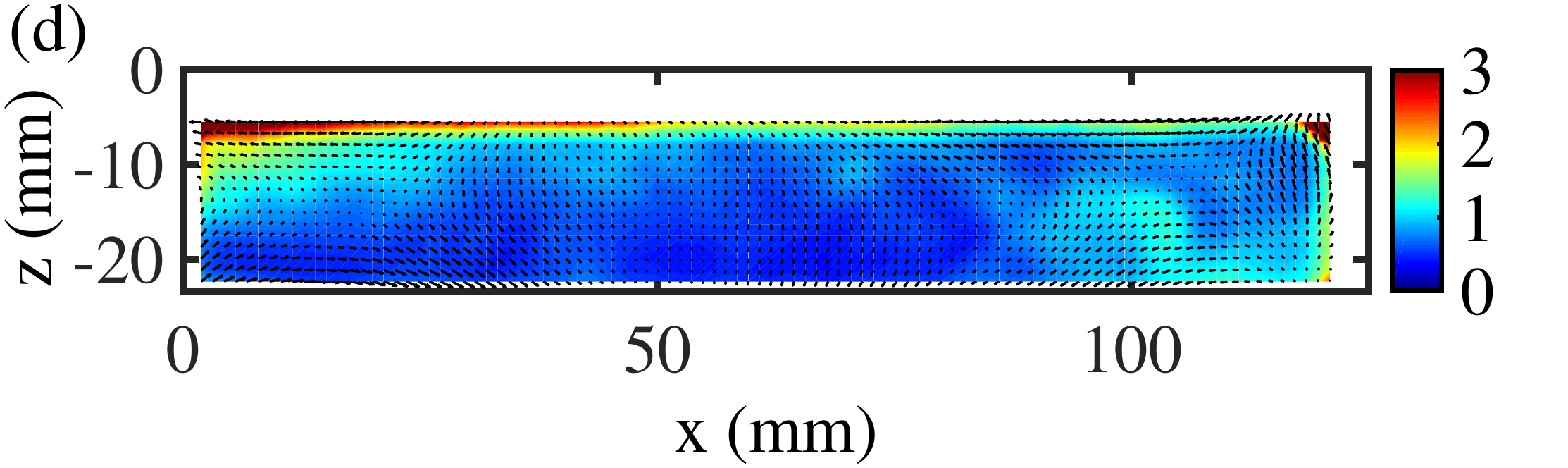} 
             \caption{(color online) (a) Superimposition of an image used for PIV measurement, with the velocity field found in water (green arrows) and the corresponding streamlines (yellow) for \red{$\sigma_h=2.4\,$mm}. The width of the image is $121$\,mm. (b) Averaged vertical profile of  r.m.s. velocity (blue circle). The black line displays an exponential fit $\e^{- k |z|}$ with \red{$k=82.4$}\,m$^{-1}$. $z=0$ corresponds to the free-surface position at rest. (c) Temporally averaged velocity field (arrows) and r.m.s. velocity in colorscale (mm/s). The paddle producing the wave is located on the \red{left} side.  (d) Temporally averaged velocity field (arrows) and temporally averaged r.m.s. vorticity field in colorscale ($s^{-1}$). A significant vorticity is present in a top layer. }
    \label{PIV}
       \end{center}
 \end{figure}

 An example of the free-surface reconstruction is depicted in Fig.~\ref{Snapshot} (a) for the highest forcing amplitude  $\sigma_h =3.6\,$mm at $t=1.51$\,s. In this example, a large gravity wave appears in the center of the observation area, propagating from left to right along the $x$ axis. \red{A 3D plot movie ({Movie1}) corresponding to the same measurement is available in supplementary.} By plotting the map of the modulus of the free-surface gradient $||\mathbf{\nabla}h(x,y,t)||$ in Fig.~\ref{Snapshot} (b), to enhance the small scale variations, a train of capillary waves on the forward face become visible. The corresponding wave profiles along the $x$-axis taken for $y=48\,$mm confirms this observation, for the wave height in Fig.~\ref{Snapshot} (c),  for the high pass filtered wave height in Fig.~\ref{Snapshot} (d) (blue) and the gradient profile $||\mathbf{\nabla}h||$ in Fig.~\ref{Snapshot} (d) (red). This description corresponds to the standard picture of generation of parasitic capillary waves by gravity waves \citep{Zhang1994,FedorovPOF1998,Perlin2000}. The capillary wave train of higher group velocity remains visible on a duration of order $0.1\,$s, before exiting the observation area or disappearing when the trough of the gravity wave is reached. Due to the important reflections along the $x$ axis, strong generation of parasitic capillary waves are also observed when two gravity waves are crossing in the area of observation or when transiently a standing wave is excited in the middle of the square tank. In these cases, parasitic wave generation is not theoretically described in the literature which considers only propagative wave, but a similar experimental observation was reported for standing waves of high amplitude \citep{Schultz1998}. \red{ During a $41\,$s measurement, typically two hundred events of parasitic capillary wave generation occur in our experiment.} Due to the randomness of the forcing and to the multiple reflections, it becomes hazardous to make a direct comparison between theoretical models of parasitic wave generation \citep{Longuet-Higgins1963,Longuet-Higgins1995,FedorovJFM1998} and the statistical analysis of the turbulent random wave field obtained with this forcing protocol.

 \subsection{Velocity field characterization}
 \label{PIVtext}
In similar conditions but not simultaneously, 2D-velocity field has been measured by Particle Image Velocimetry (PIV) in the vertical plane $O_{xz}$. $50$ micron particles were used to seed the water. A vertical laser sheet produced by a 2 W continuous laser lights the plane corresponding to the middle of the tank. PIV fields are computed through the open access software PIVlab \citep{PIVlab}. Due to the moving free-surface, PIV would require to use a dynamical mask, to avoid to compute the correlation outside the water, which implies a simultaneous measurement of the free surface in the vertical plane as it was performed in experiments with a smaller free surface displacement  \citep{Jamin2015}. Due to the absence of dynamic mask implementation, PIV is performed on a window located $5$\,mm below the mean-free surface. $3195$ images are recorded with a fast camera during $12.78$\,s with a frame rate of $250\,$Hz. 2D velocity field from PIV obtained for a forcing level corresponding to \red{$\sigma_h= 2.4\,$mm} are depicted in Fig.~\ref{PIV} and provides a qualitative complement to the 2D free-surface deformation measurement with DLP.

The velocity field $\textbf{u}(x,z)$ appears to follow the structure given by the large gravity waves. \red{An instantaneous snapshot is displayed in Fig.~\ref{PIV} (a)}.  After temporal averaging, the presence of mean velocity field appears of \red{one centimeter per second, depicted by black arrows in Fig.~\ref{PIV} (c) and (d)}. This observation may be associated with the transient excitation of a standing wave mode during the measurement or nonlinear streaming induced by the wave-maker. \red{In the top layer the averaged horizontal velocity  is found $\langle u_x \rangle =0.26$\,mm/s. In contrast the average root-mean-square (r.m.s.) velocity $\langle \sqrt{u_x^2+u_z^2} \rangle$ is larger and is depicted in colorscale in Fig.~\ref{PIV} (c)}. The r.m.s. velocity decreases with the distance to the free surface. The r.m.s velocity follows an exponential law whose wave number \red{$k=82.4$\,m$^{-1}$} belongs to the forcing range, showing that the velocity field is mainly induced by gravity wave propagation. 

Then the r.m.s. vorticity along $y$ axis, $ \Omega = {\nabla}  \times \textbf{u}$ appears of significant amplitude (about 2 $s^{-1}$) in the top layer, close to the free surface. This observation is unexpected because surface wave propagation is theoretically described only for potential flows. Nevertheless the large scale picture shows that the random gravity wave excitation is not associated with bulk hydrodynamics turbulence, as expected. The top layer with high vorticity, despite not directly in contact with the free-surface, could be related to the vorticity generation by parasitic capillary waves as it was reported \citep{FedorovJFM1998,Lin2001}. The order of magnitude is coherent with the estimation $\Omega \approx 2 (\sigma_h\,k)^2\,\omega $ \citep{Longuet-Higgins1992,DeikeJFM2015cap}, \red{by taking $k \approx 100$\,m$^{-1}$ as the typical forcing wavenumber and $\omega \approx 2\pi\,5\,$s$^{-1}$ the typical forcing pulsation}. The horizontal motion of the paddle generating the waves could also inject vorticity in the bulk. Finally the average 2D kinetic energy per volume and mass unit $E_{c h}=1/2 \, \langle\textbf{u}^2\rangle$ of the flow is found to be $E_{c h}\approx 18\cdot10^{-4}$m$^2 $.s$^{-2}$, whereas the viscous dissipated power per volume and mass unit $D_\Omega= \nu \, \langle \Omega^2  \rangle \approx 3.3\cdot10^{-4}$m$^2 $.s$^{-3}$. The energy dissipated during a wave period is thus smaller than the kinetic energy $E_{c h}$. \red{By fitting the decay of the vorticity with $z$ by a decreasing exponential, a typical scale $l_\Omega = 1.59\,$ mm is found . The potential flow assumption should be thus valid for waves whose wavelengths are larger than $l_\Omega$ or $1/\lambda <630\,$m$^{-1}$. We assume in the following}, that the hydrodynamic flow is mostly the potential flow induced by wave propagation.
\section{\red{Analysis of stationary regimes in Fourier space}} 
\label{Cascade}
\subsection{\red{Random statistics of wave-field}} 
  
\begin{figure}
 \begin{center}
\includegraphics[width=6.3cm]{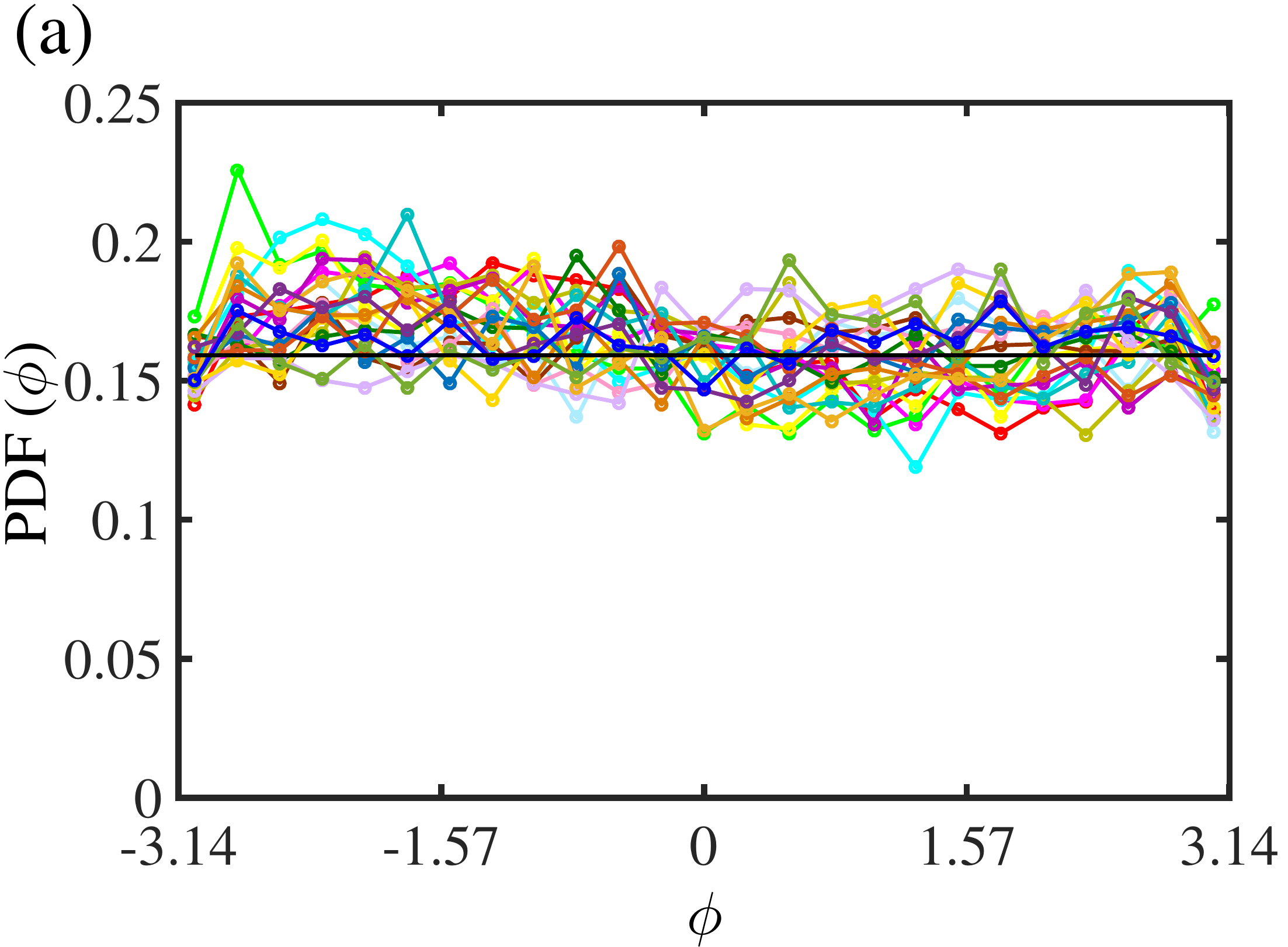}
\includegraphics[width=6.3cm]{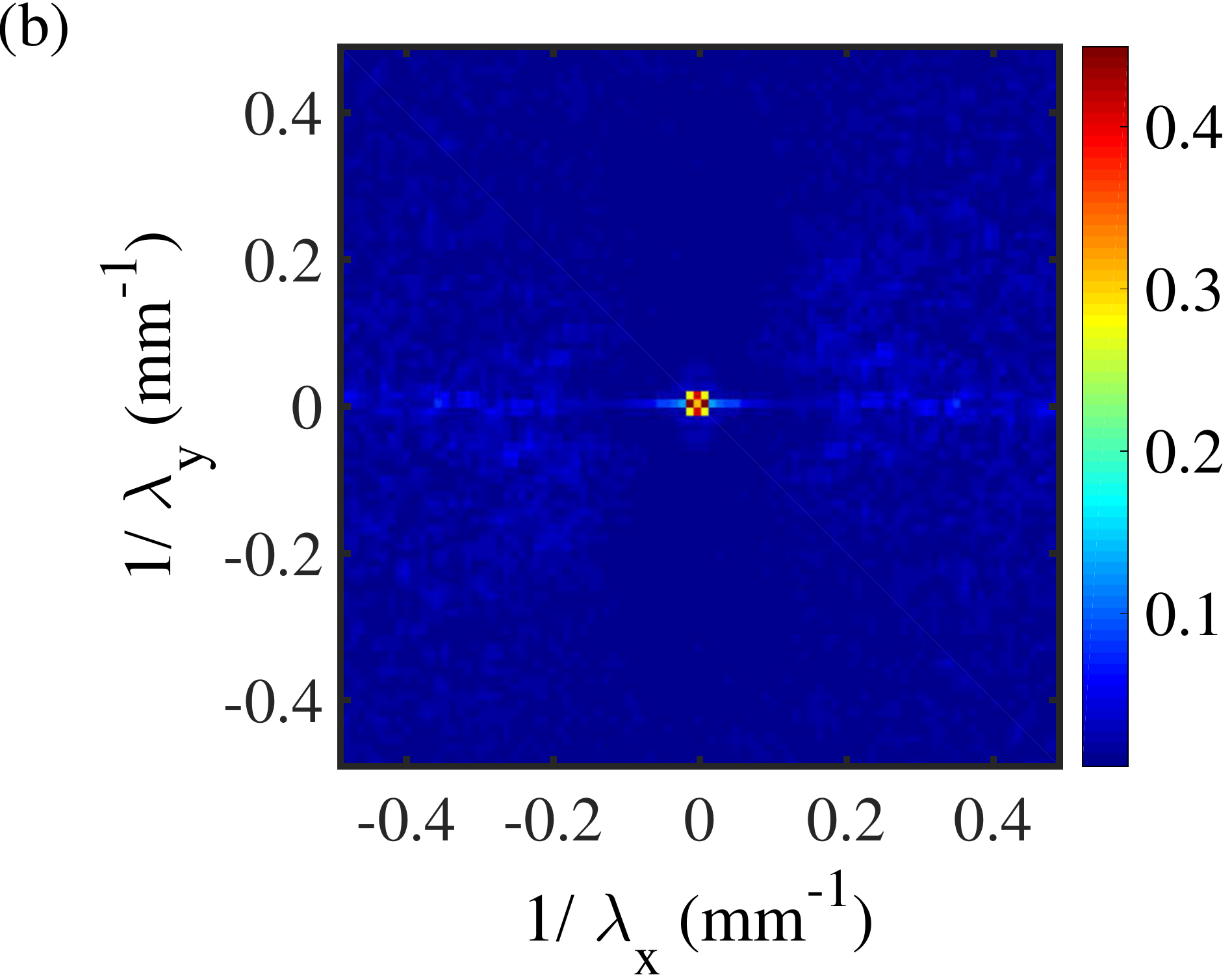} 
\hspace{.4cm}
        \caption{(color online). \red{(a)  PDF (probability density function) of the temporal fluctuations of the phase $\phi$ of $\tilde{\eta} (\mathbf{k},t)$, for $21$ equally spaced values of $\mathbf{k}=2\pi\,(\lambda_x^{-1},\lambda_y^{-1})$ varying from $(41.6,45.5)$ to $(874,956)$\,m$^{-1}$. The black line is the uniform distribution $1/(2\pi)$. The case displayed corresponds to $\sigma_h=3.6$\,mm. (b) Colormap of the absolute  distance to the uniform distribution $\langle|\mathrm{PDF}(\phi) -1/(2\pi)|\rangle $ as a function of $\lambda_{x}^{-1} $ and $\lambda_{y}^{-1} $. Except at large scale, for $\mathbf{k}$ close to $(0,0)$, phases of free surface deformation are uniformly distributed, as the distance to the uniform distribution is close to zero.}}
    \label{phase}
       \end{center}
 \end{figure}

\red{First, we aim to verify that there is no phase correlations in the wave-field and that waves propagate independently. To measure the phase of the waves, we perform a spatial Fourier analysis of the field of free-surface deformation $\eta=h(x,y,t)-\langle h \rangle$, as it was done for bending waves in a plate in the study of elastic wave turbulence \citep{Mordant2010}: }
 \red{\begin{equation} \tilde{\eta} (\mathbf{k},t)=\dfrac{1}{{2\pi}}\, \int_{0}^{L_y} \int_{0}^{L_x} \, \eta (x,y,t)\,\e^{-\mathrm{i} (k_x\,x+k_y\,y)} \,\mathrm{d}x\,\mathrm{dy}\,\end{equation}}
\red{The decomposition in a 2D Fourier space define wave modes $\tilde{\eta} (\mathbf{k},t)$ as a complex function evolving in time for a given $\mathbf{k}$, whose phase evolutions $\phi_{\mathbf{k}}(t)$ are obtained by taking the argument of $\tilde{\eta} (\mathbf{k},t)$. For the highest amplitude measurement, the probability density function (P.D.F.) of $\phi$ are given in Fig.~\ref{phase} (a) for a large number (21) of equally spaced values of $\mathbf{k}=2\pi\,(\lambda_x^{-1},\lambda_y^{-1})$ varying from $(41.6,45.5)$ to $(874,956)$m$^{-1}$. The phase distributions fluctuate around the value $1/(2 \pi)$, which corresponds to the uniform distribution. For the duration of experiments ($41\,s$), phases appear randomly distributed. To generalize this observation to all wave modes, the temporal average of the absolute distance to the uniform distribution is plotted in color-scale in Fig.~\ref{phase} (b). Except in the center at large scale, the random distribution of phases holds. Still at large scale, we note a relatively higher level along the axes, which could be caused by boundary effects.}

  \begin{figure}
 \begin{center}
\includegraphics[width=6.3cm]{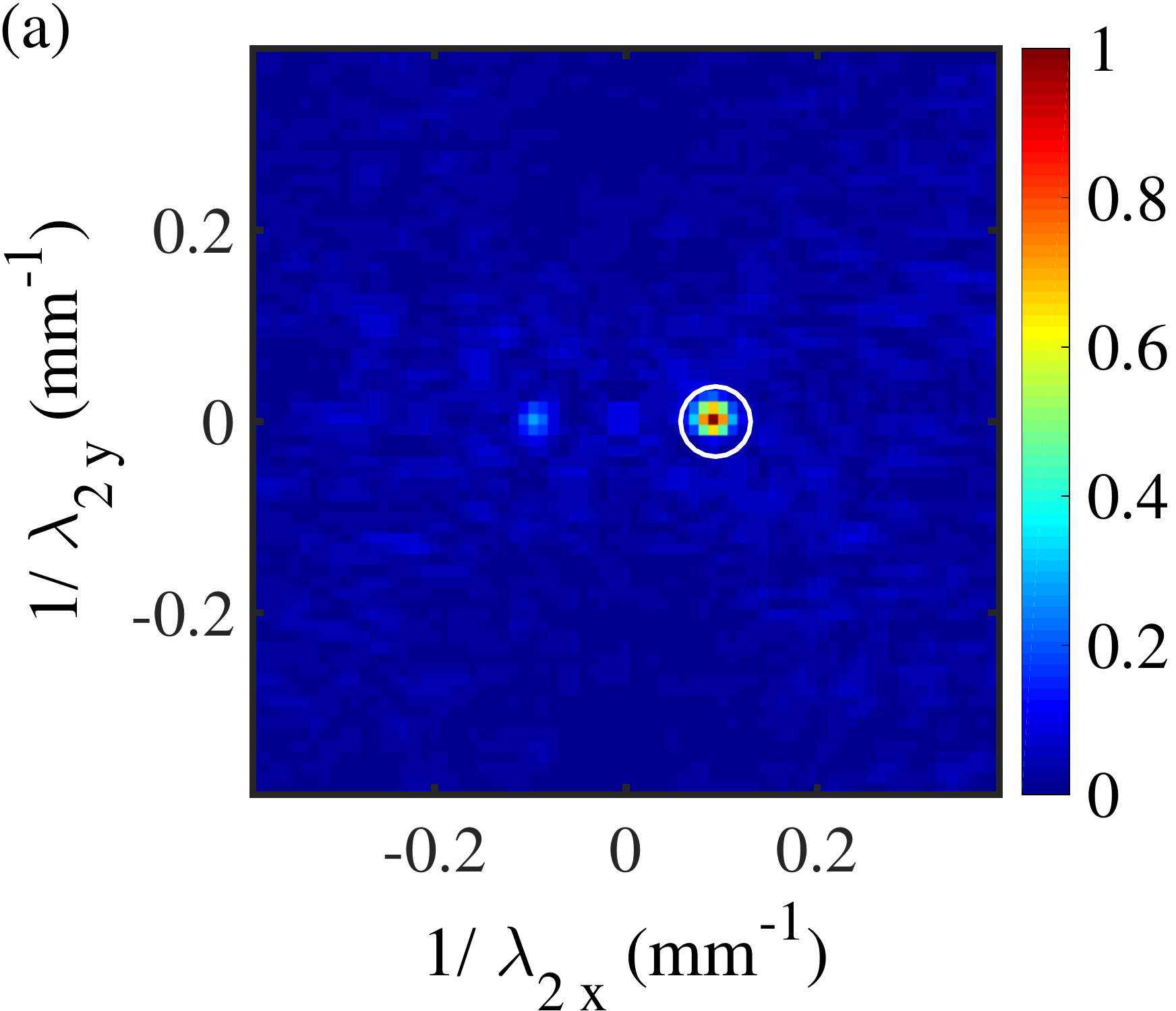}  \hfill
\includegraphics[width=6.3cm]{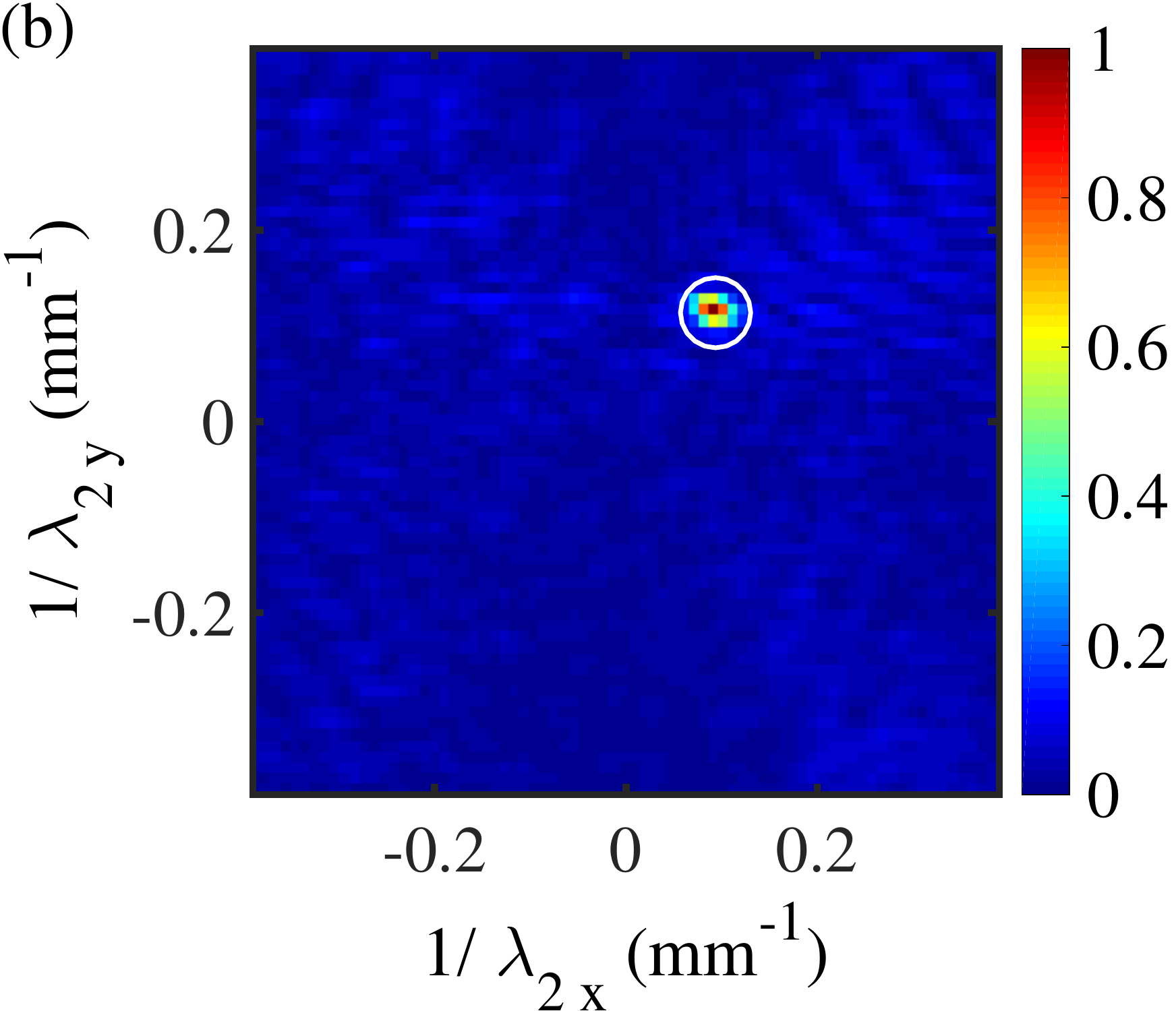} 
\hspace{.4cm}
        \caption{(color online). \red{(a) Magnitude of normalized correlation (see (\ref{correlFourier})) between spatial Fourier modes  $\tilde{\eta} (\mathbf{k_1},t)$ and $\tilde{\eta} (\mathbf{k_2},t)$, for $\mathbf{k_1}/(2\pi)=(0.094,0)$\,mm$^{-1}$ and variable $\lambda_2^{-1}= k_2/(2\pi)$. $\sigma_h=3.6$\,mm. The white circle depicts the position of $\mathbf{k_1}$. (b) Same for $\mathbf{k_1}/(2\pi)=(0.094,0.1024)$\,mm$^{-1}$. For $\mathbf{k_1} \neq \mathbf{k_2}$, the correlation is essentially zero at the level of convergence. The finite size of the correlation peak, could be due to the resolution of spatial Fourier transform due to the size of the images.}} 
    \label{phase2}
       \end{center}
 \end{figure}
 
   \red{The independence of wave modes at leading order is a substantial consequence of a random distribution of wave phases. The statistics of the wave-field in Fourier space can therefore be described using only the power spectrum. This condition is verified if the correlations of wave modes writes: 
   \green{\begin{equation} \langle\tilde{\eta} (\mathbf{k_1},t)\, \tilde{\eta} (\mathbf{k_2},t) \rangle \propto \delta (\mathbf{k_1}-\mathbf{k_2}), \end{equation}}
where $\delta$ is the Dirac delta function. This property is tested for two examples for the measurement at highest amplitude and shown in Fig.~\ref{phase2}, by computing the normalized correlation between Fourier modes:
\begin{equation}
\dfrac{ \vert \langle\tilde{\eta} (\mathbf{k_1},t)\, \tilde{\eta} (\mathbf{k_2},t)^\star \rangle \vert}{\sqrt{\langle\vert \tilde{\eta} (\mathbf{k_1},t)\vert \rangle \langle\vert \tilde{\eta} (\mathbf{k_2},t)\vert \rangle }}
\label{correlFourier}
 \end{equation}}
\red{For $\mathbf{k_1} \neq \mathbf{k_2}$, the correlation is essentially zero at the level of convergence. The finite size of the correlation peak, could be due to the resolution of spatial Fourier transform due to the size of the images.}

\red{At the leading order, phases of wave modes appear uncorrelated, especially for the scales of capillary wave propagation, \textit{i.e.} $\lambda^{-1} > 0.064\,$mm$^{-1}$ ($k\geq 100$\,m$^{-1}$), which justifies a description using mainly power spectra. At the third order, nonlinear wave interactions are investigated using phase correlations in \S \ref{bispectrum}. Note that the absence of phase correlation related to finite-size tank can appear as a surprising result. Despite the random excitation, multiple reflections on the wall of the tank could build phase correlations in the spatial Fourier space. But, due to viscous dissipation, we show in Appendix A, that waves with $k\geq 275$\,m$^{-1}$ ($f \geq 10$ Hz) are damped before experiencing reflections on the walls. The significant dissipation may thus explain the absence of wave correlations and of quantization of wave-numbers in a finite-size tank.}

 \subsection{Experimental dispersion relation}
\label{RDpart}
The analysis at small scale of a random wave-field is performed in stationary regime in the Fourier space  $(\omega,\textbf{k})$. The space and time power spectrum of wave elevation $ S_\eta (\omega,\textbf{k})$ is computed from the set of free-surface images $h(x,y,t)$ ($\eta=h-\langle h \rangle$), by performing successively a two-dimensional Fourier transform in space, and a Fourier transform in time, \red{and then by taking the square modulus}. 
 \red{\begin{equation} S_\eta (\omega,\textbf{k})=\dfrac{1}{L_x\,L_y\,T}\,\big{\vert}   \tilde{\eta}(k_x,k_y,\omega)\,  \big{\vert}^2 \end{equation}
 \begin{equation}\mathrm{with}\quad \tilde{\eta}(k_x,k_y,\omega)\, =\int_0^T \int_0^{L_y} \int_0^{L_x} \, \eta (x,y,t)\,\e^{-\mathrm{i} (k_x\,x+k_y\,y+\omega\, t)}  \,\mathrm{d}x\,\mathrm{dy}\,\mathrm{d}t\,\end{equation}}
\red{The result is then converted in radial coordinates $(k=||\textbf{k}||,\theta)$ and integrated over the different directions of $\mathbf{k}$ to obtain the one-dimensional spectrum.
\begin{equation}  S_\eta (\omega,k)=\int_0^{2\pi}S_\eta (\omega,k,\theta)\,k\,\mathrm{d}\theta \end{equation}}

 \begin{figure}
 \begin{center}
\includegraphics[width=13.5cm]{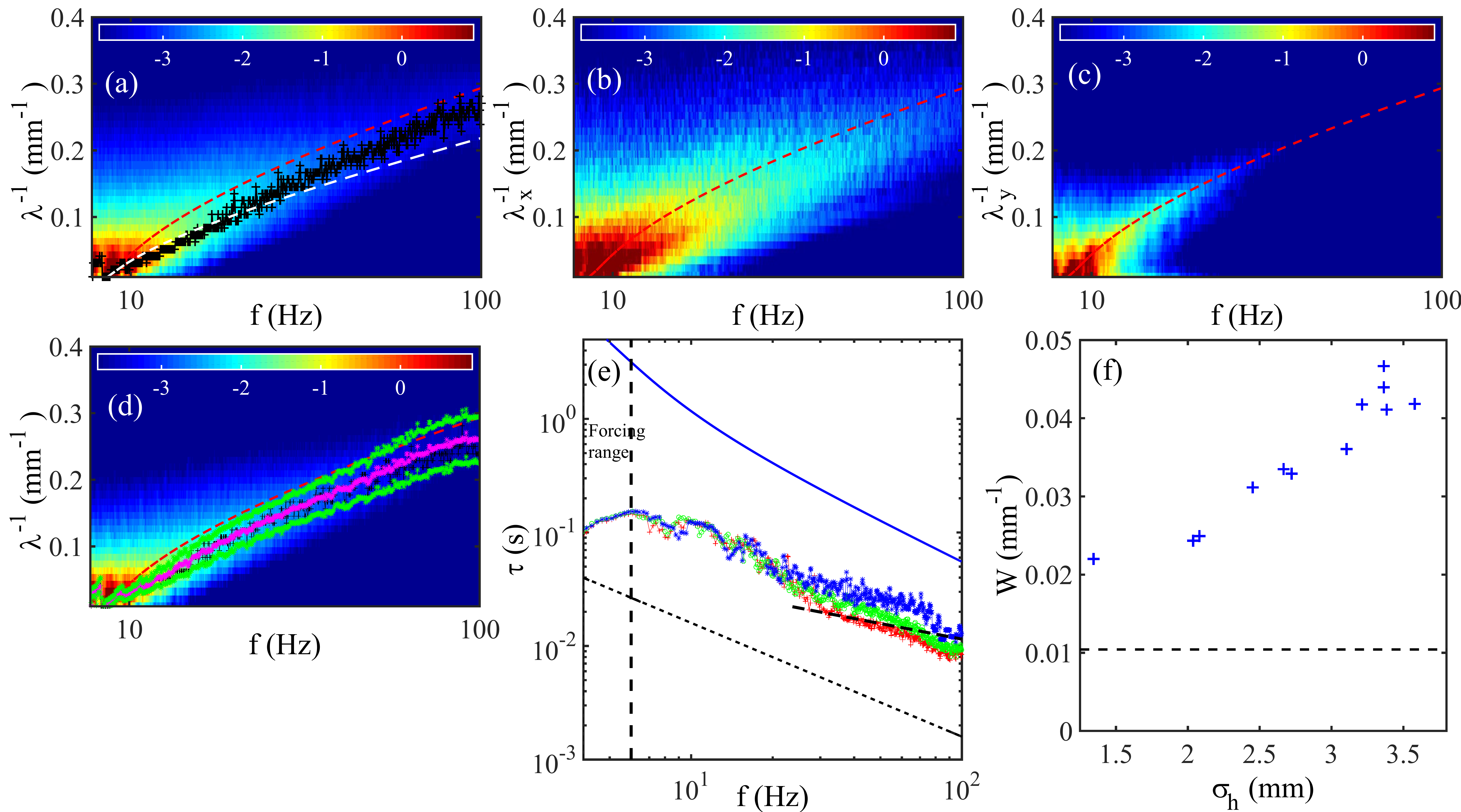} 
\caption{(color online) (a) Spatio-temporal spectrum of wave elevation $S_\eta (\omega,k)$ for $\sigma_h =3.6\,$mm. ($+$) Experimental dispersion relation extracted from the maxima of $S_\eta (\omega,k)$. Red (dark gray) dashed line: linear theoretical dispersion relation (\ref{RD}). White dashed line: nonlinear dispersion relation with $a=\sigma_h$ (\ref{RDNL}). Colorscale corresponds to $\log_{10}(S_\eta (\omega,k))$ (b) Spatio-temporal spectrum of wave elevation $S_\eta (\omega,k_x)$ in the forcing direction with the same parameters. (c) Spatio-temporal spectrum of wave elevation $S_\eta (\omega,k_y)$ in the direction perpendicular to the forcing with the same parameters. (d) Spatio-temporal spectrum of wave elevation $S_\eta (\omega,k)$ for $\sigma_h =3.4\,$mm. The ($*$) center (magenta) and approximate upper and lower limits (green) of the experimental dispersion relation. (e) Typical correlation times $\tau_c$ extracted from the width of the dispersion relation, for $\sigma_h =3.6\,$mm (red $+$),  for $\sigma_h =3.4\,$mm (green $\times$) and for  $\sigma_h =2.1\,$mm (blue $*$). Dashed black line, linear time $1/\omega$. Blue line, linear dissipative time. Black dashed thick line, power law $f^{-1/2} $. (f) Dependency of dispersion relation width $W(f^*)$ with the wave amplitude for $f^*=30\,$Hz. Dashed horizontal line, $\delta \lambda^{-1} = 1/ L$ resolution due to the finite size of the image.}
    \label{RD}
       \end{center}
 \end{figure}  

The spectrum $ S_\eta (\omega,k)$ displayed in Fig~\ref{RD} (a), reveals the spreading of wave energy in time and space. For sufficiently large wave amplitude, the spectrum appears continuous, without privileged scales, which constitutes one criterion of \red{turbulent regimes.} By finding the maxima of $ S_\eta (\omega,k)$ for each $\omega$ value, the experimental dispersion relation is found. \red{Wave propagation is indeed characterized by a concentration of signal energy on a curve in  the $(\omega,k)$ space.} In this case (with $\sigma_h=3.6\,$mm), we observe a small but significant departure from the linear dispersion relation (\ref{RDL}), with an increase of frequency at a given wavelength. The other measurements at lower amplitude, show that this shift increases with $\sigma_h$ (not displayed here), arguing for a nonlinear mechanism. A similar nonlinear shift was also reported independently in similar conditions~\citep{Aubourg2016}. Moreover this dispersion relation shift is anisotropic. $ S_\eta (\omega,k_x)$ is plotted in Fig.~\ref{RD} (b) and $ S_\eta (\omega,k_y)$ in Fig.~\ref{RD} (c). The deviation is only visible in the forcing direction $x$, for which the spectrum is more intense. \red{Typically for $1/\lambda=0.1$\,mm$^{-1}$, the frequency is increased by $10$\,Hz or the inverse of the wavelength is reduced by $0.03$\,mm$^{-1}$.} 

\red{A complete explanation is still missing, but this observation might be explained by the following mechanisms. First, a Doppler shift due to the presence of a mean horizontal current $\mathbf{u}$, which modifies the linear dispersion relation
 \begin{equation}
(\omega-\mathbf{u}\cdot \mathbf{k})^2=\left(g k+\frac{\gamma}{\rho}k^3 \right)\tanh (k h_0)
\label{RDDOP}
\end{equation}}
 \red{The mean horizontal velocity measured using PIV for $\sigma_h=2.45\,$mm in \S. 2.3 was found $\langle u_x \rangle = 0.26$\,mm/s. The corresponding frequency shift for $1/\lambda=0.1$\,mm$^{-1}$, which would be of order $0.03$\,Hz and would not be detectable.} Second, the shift could be explained by the nonlinear corrections explicitly computed for monochromatic waves, in the gravity range as Stokes waves \citep{Whitham} and in the capillary range as Crapper correction \citep{Crapper}. Considering these two corrections, the dispersion relation reads
\begin{equation}
\omega^2=\left(g k\left[1+(a k)^2\right]+\frac{\gamma}{\rho}k^3\left[1+\left(\frac{a k}{4}\right)^2\right]^{-1/4} \right)\tanh (k h_0).
\label{RDNL}
\end{equation}
By taking the amplitude $a$ equal to $\sigma_h$, the magnitude of the shift, is reproduced. But this estimation does not take in account the strong decrease of wave amplitude as $k$ increases. The nonlinear correction appears too small, to justify the shift at moderate $k$ and does not explain the anisotropy. \red{Another similar mechanism could be related to the current induced by the Stokes drift due to large gravity waves. A simplified estimation of the horizontal velocity due to the Stokes drift reads $u_s=\omega\,k\,{\sigma_h}^2 \approx 39$ mm/s, with $f=5$\,Hz, $k=96$\,m$^{-1}$ and $\sigma_h=3.6$\,mm. The corresponding frequency shift at $1/\lambda=0.1$mm$^{-1}$ would be $u_s / \lambda \approx 3.9$\,Hz, which is the correct order of magnitude. Finally a frequency up-shift, was also noticed in a simplified numerical model studying capillary waves excited by gravity waves \citep{WatsonJFM1999}. The wave-field was decomposed in the spatial Fourier space in modes interacting through Three-Wave interactions. The shift of magnitude consistent with our observation, is interpreted as a modulation by longer gravity waves (a "drag") of short capillary waves, which are not bound waves. The capillary waves simulated with this method are qualitatively close to parasitic capillary waves. Presence of capillary wave trains may indeed explain the observed shift in experiments. We note also, that the model of Fedorov and Melville \citep{FedorovJFM1998} predicts an increase of the phase velocity of parasitic capillary wave of class 2 (a pressure maximum is associated to the crest) which can reach as much as $20$\, \%, as a function of gravity wave steepness. A nonlinear increase of phase velocity is indeed equivalent to a frequency up-shift of dispersion relation. Several elements indicate thus, that the observed shift is likely caused by the modulation of parasitic capillary waves by the long and steep gravity wave.} 

\red{Another important feature of $S_\eta (\omega,k)$ is the width of the experimental dispersion relation. In comparison with elastic waves \citep{CobelliMordant2009,Mordant2010,DeikeJFM2013} and gravity wave experiments \citep{Herbert2010,Cobelli2011,Aubourg2016}, the width of the dispersion relation appears broad. In order to quantify this observation, we apply in $k$ space for each frequency value $f^*$ a Gaussian fit:
\begin{equation} S_\eta (f^*,k) = A \exp \left[  \left( \dfrac{-(k-k_{DR}(f^*))}{2 \pi\, W(f^*)} \right)^2\right]  \end{equation}
We extract the width of the dispersion relation $W(f^*)$. In Fig.~\ref{RD} (d), above the spectrum $S_\eta (\omega,k)$ for $\sigma_h=3.4\,$mm, the fit is displayed with magenta stars for each frequency the central $k_{DR}(f^*)/(2\pi)$ corresponding to the dispersion relation and also $k_{DR}(f^*)/(2\pi)\pm W(f^*)/2$ (green stars). We observe in Fig.~\ref{RD} (f), that $W(f^*)$ increases with the wave amplitude, showing that the nonlinear effects contribute to the wave-broadening. The figure is displayed for $f^*=30\,$Hz, but a similar behavior is observed for other frequencies in the capillary range. The horizontal dashed line shows the spatial resolution $\delta (1/\lambda) = 1/(2\,L)$ with $L=96\,$mm the size of the window of observation.}
 
\red{The substantial width of the dispersion relation can be interpreted in several ways. In presence of parasitic capillary wave generation, a significant broadening of the dispersion relation for such steepness level could be caused by the modulation by Doppler effect of the frequencies of capillary by the orbital velocity field \citep{FedorovPOF1998,FedorovJFM1998,WatsonJFM1999}. However by considering a r.m.s. velocity of $50$\,mm/s, estimated from PIV measurements (See \S \ref{PIVtext}), and injecting this typical value in the Doppler shifted dispersion relation (\ref{RDDOP}), a broadening of order $10^{-3}$\,mm$^{-1}$ is found, whereas the measured width $W$ is of order $10^{-2}$\,mm$^{-1}$.} \red{The width of the dispersion relation is therefore better interpreted as a correlation length due to the finite size of the wave packet at a given frequency, as discussed in \citet{Miquel2011}.} To facilitate the discussion, we convert this correlation length into a correlation time $\tau_c (f^*) = (v_g\,2\pi \, W(f^*)/2)^{-1}$, with $v_g=\partial \omega / \partial k$ the group velocity. This finite life time of the wave-packet is due to wave viscous dissipation and to nonlinear wave interactions which distributes the energy of the waves through the scales. $\tau_c$ is plotted in Fig.~\ref{RD} (e) for three wave amplitudes as a function of the frequency and is compared to the viscous dissipative time $\tau_{diss}=(\sqrt{2} \sqrt{\nu \omega} k /4)^{-1}$ (see \S \ref{Constraintsdiss}) and the linear time $1/\omega$. $\tau_c$ is found between $1/\omega$ and $\tau_{diss}$. Despite $\tau_{diss}$ is found around $40$ times larger than $1/\omega$ in the capillary range, $\tau_c$ values are about $5$ times $1/\omega$ and $\tau_c$ evolves nearly as $1/f$. \red{The correlation time is close to the smallest possible value. The wave} would not be defined if $\tau_c< 1/\omega$. Moreover, $\tau_c$ is usually interpreted as a typical time of nonlinear interaction \citep{Miquel2011}. From the dimensional analysis of the quadratic interaction term for pure capillary waves \citep{DeikeJFM2013}, this nonlinear time reads: $(\tau_{NL})^{-1} \sim \epsilon^{1/2}\,(\frac{\gamma}{\rho})^{-1/4}\,f^{1/2}$. Here, the evolution of $\tau_c$ as a function of $f$, follows relatively well the $f^{-1/2}$ power law in the range $25<f < 100\,$Hz for the two highest amplitudes, which is consistent with the interpretation of $\tau_c \sim \tau_{NL}$.\\

\red{Weakly nonlinear models of interacting waves suppose a time scale separation $1/f \ll \tau_{NL} \ll \tau_{diss}$. Nonlinear exchanges of energy must occur on a typical time much smaller than the viscous decay time and much larger than the wave period (otherwise the wave blows up, \textit{i.e.} disappears in a strongly nonlinear manner). } Here, $\tau_c \sim \tau_{NL}$ is approximately $8$ times smaller than $\tau_d$ and \red{$5$ times larger than $1/\omega$}. Therefore, the broad width of experimental dispersion relations show that the wave regimes are strongly nonlinear in terms of energy exchanges. A correlation time equal to the period corresponds typically to a number of visible crests $\sim5$, which is coherent with capillary parasitic waves generation with few visible ripples in the front of the long gravity wave.

Note, we do not observe supplementary branches related to the occurrence of bound waves, contrary to larger size experiments studying the gravity wave turbulence regime~\citep{Herbert2010}. This is consistent with the parasitic wave generation mechanism, for which in the range between $4$ and $6$ Hz, parasitic capillaries are found as free-waves \citep{Perlin1993,FedorovPOF1998}.

  \begin{figure}
 \begin{center}
 \includegraphics[width=10cm]{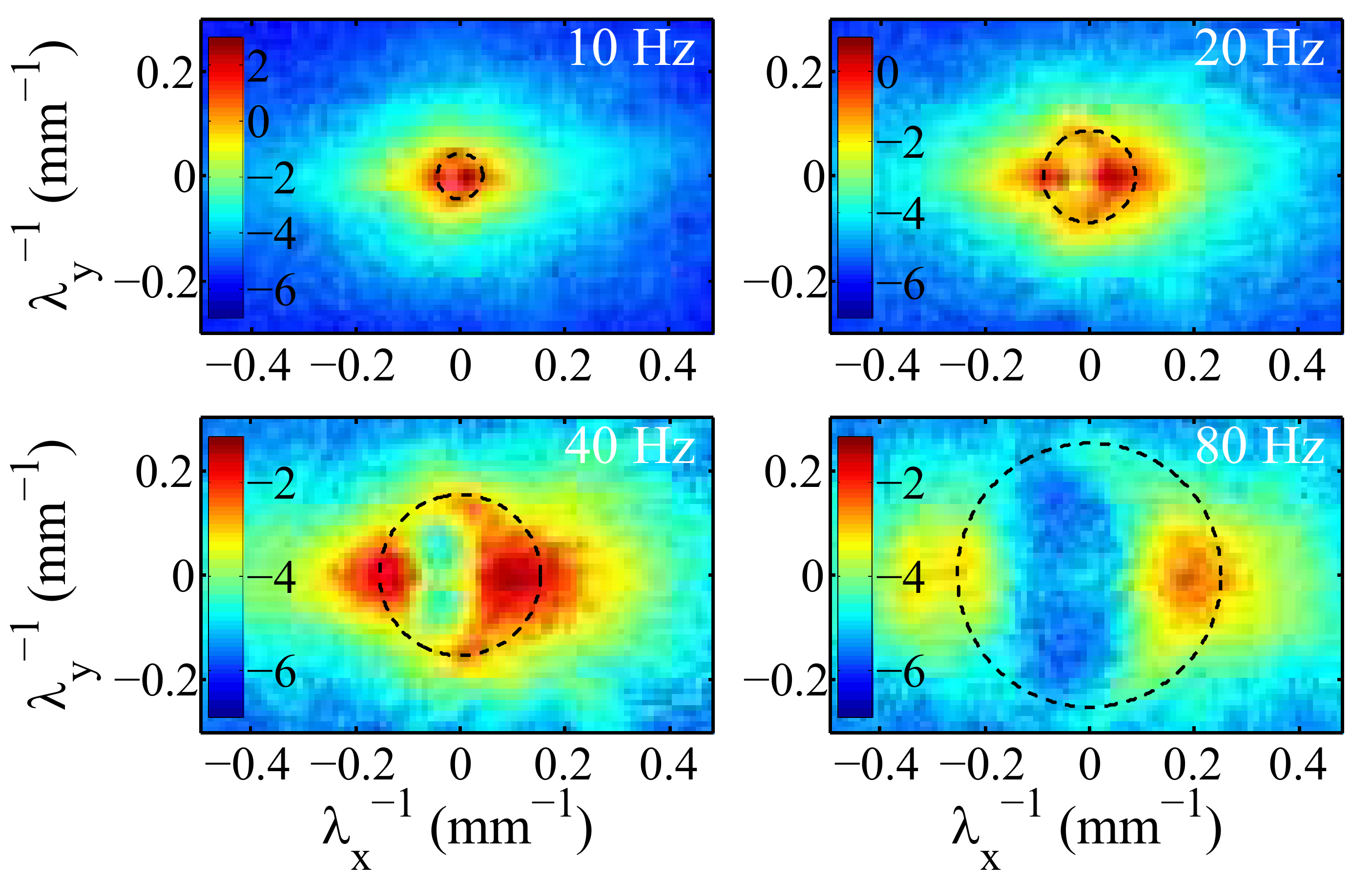} 
        \caption{(color online). Spectrum $S_\eta (\omega, {\lambda_x}^{-1}, {\lambda_y}^{-1})$ for fixed $f=10$, 20, 40 and 80 Hz ($log_{10}$ colorscale) showing the anisotropy of wave field at high forcing amplitude ($\sigma_h =3.6\,$mm and $\sigma_s=0.34$). The anisotropy due to the forcing is still present at high frequency. Dashed black circle: Linear dispersion relation of (\ref{RDL}). Wave amplitudes are larger in the direction of forcing ($x$-axis) revealing anisotropy in all frequency range.}
    \label{Aniso}
       \end{center}
 \end{figure} 
 
The directional properties of the wave field are investigated in Fig.~\ref{Aniso} by plotting $S_\eta (\omega,\mathbf{k})$ as a function of  ${\lambda_x}^{-1}$ and ${\lambda_y}^{-1}$ for fixed frequency $f$. A strong anisotropy of the wave field is observed along the $x$-axis, which is the forcing direction. This anisotropy is conserved regardless of the frequency scales inside the capillary waves range (see at $f=20$, $40$ or $80$ Hz). Capillary waves appear thus preferentially along the same direction as the long waves, \red{as it would be for parasitic generation. One may suppose that superposition of several Three-Wave interactions for a random excitation would restore the isotropy. However dominant one-dimensional interactions were also reported with waves close to the gravity-capillary crossover \citep{Aubourg2015,Aubourg2016}. This observation has been explained by quasi-resonant interactions, due to the significant width of the dispersion relation. Moreover viscous dissipation and nonlinear interactions reduce the life times of capillary waves. A short correlation time should favor interactions with forcing waves and thus should increase the anisotropy of the wave field.}

\subsection{Spatial and temporal spectra}  
 
\begin{figure}
 \begin{center}
 \includegraphics[width=13.5cm]{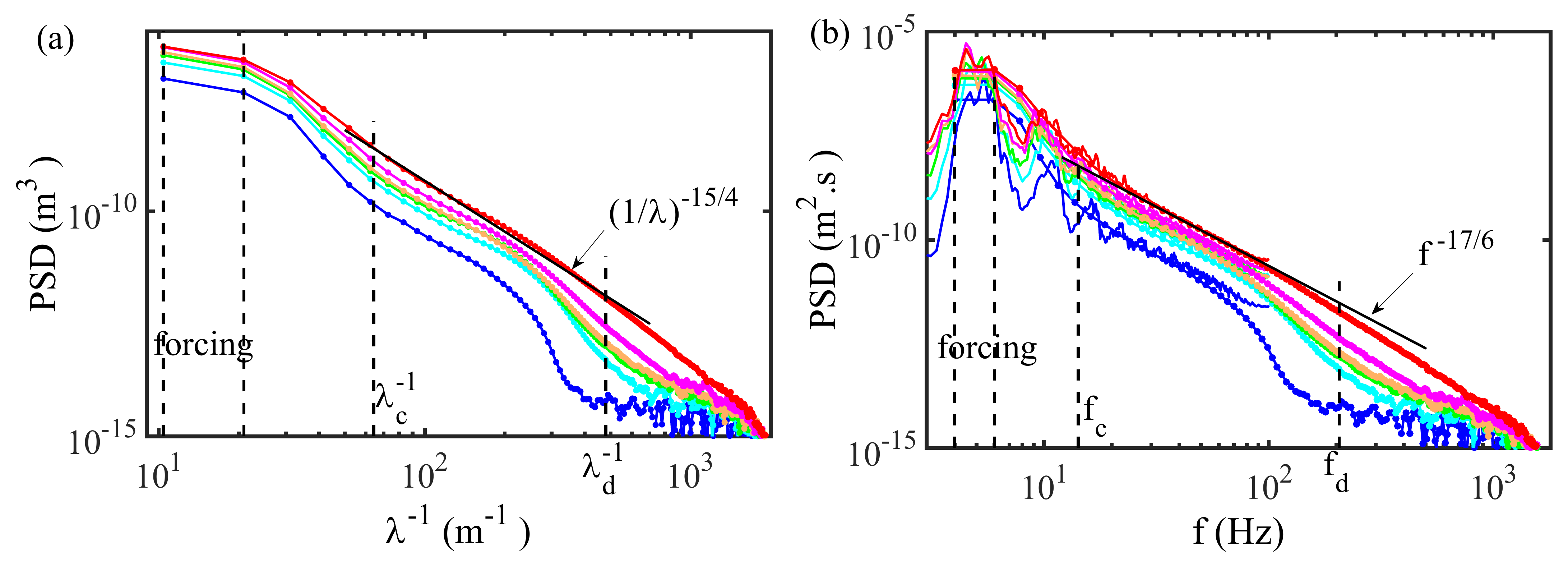} 
        \caption{(color online). (a) Spatial power spectra $S_\eta (k)$ (power spectrum density PSD) for different forcing amplitudes. From bottom to top: $\sigma_h = $1.3, 2.1, 2.7, 3.1, 3.4 and 3.6 mm, and $\sigma_s=0.15$, 0.19, 0.24, 0.27, 0.29 and 0.34. Solid black line is the capillary prediction $k^{-15/4}$. $\lambda_c=2\pi\sqrt{\gamma/(\rho\,g)} \approx 15.5$\,mm is the crossover scale between gravity and capillary waves. $\lambda_d \approx \, 2.1$\,mm is approximately for the highest amplitude, the dissipative scale, below it viscous dissipation dominates nonlinear interactions. (b) Temporal power spectra $S_\eta (\omega)$ for  the same measurements. Solid black line is the capillary prediction $f^{-17/6}$. $f_c = 14.2 $\,Hz and $f_d \approx 204\,$Hz, are the equivalent of $\lambda_c$ and  $\lambda_d$ in the frequency space. Two estimates of the frequency spectrum are shown, the first by integration over the wave-numbers of $S_\eta (\omega,k)$ (continuous curves) and the second by converting the spatial spectrum in the frequency space (dot curve) using the linear dispersion relation (\ref{RDL}).}
    \label{Spectra}
       \end{center}
 \end{figure}

By computing the spatial and temporal spectra of wave elevation,  $S_\eta (k)$ and  $S_\eta (\omega)$,  at a sufficiently high wave amplitude, we evidence power law spectra in the capillary wave range, \red{which would constitute} the capillary wave turbulence cascade as reported previously \citep{Berhanu2013}. \red{They are derived from $S_\eta (k,\omega)$ by the relations:
\begin{eqnarray*}
S_\eta (k)= \int_0^{\omega_{max}} \, S_\eta (k,\omega) \mathrm{d \omega} \\
S_\eta (\omega)= \int_0^{k_{max}} \, S_\eta (k,\omega) \mathrm{d}k
\end{eqnarray*} 
From the Parseval-Plancherel relation, the signal energy from the spectra is given by:
\begin{equation}\int_0^ {k_{max}} S_\eta (k) \mathrm{d}k =\int_0^{\omega_{max}} \, S_\eta (\omega) \mathrm{d \omega} \approx \sigma_h^2 \end{equation}
Due to the lack of the resolution at large scale, the last expression is only approximate. When expressed as a function of $\lambda^{-1}=k / (2 \pi)$ or $f=\omega /( 2 \pi)$, the correspondence between spectra writes: $S_\eta (\lambda^{-1}) =(2\,\pi) \, S_\eta (k) $ and  $S_\eta (f) =(2\,\pi)\,S_\eta (\omega) $}. \blue{The power spectra $S_\eta$ are also referred by the acronym P.S.D. (Power spectrum density).} The spatial spectrum $S_\eta (\lambda^{-1})$ is depicted in Fig.~\ref{Spectra} (a) for different forcing amplitudes. For high enough wave amplitude, a power-law spectrum is indeed observed in the capillary range \textit{i.e.} for $k > k_c=1/l_c$ or $\lambda < \lambda_c=2\pi\sqrt{\gamma/(\rho\,g)} \approx 15.5$\,mm, whose exponent is close to $-15/4$, the value predicted by the Wave Turbulence theory. Departure from the power law is observed at a scale $\lambda_d^{-1}$, equal to $0.48$ mm$^{-1}$ for the highest forcing and decreasing with wave amplitude. \red{We note that $\lambda_d$ is comparable but larger that the typical scale of the vorticity layer $l_\Omega$ (See \S \ref{PIVtext}).} For ${\lambda}^{-1}\geq {\lambda_d}^{-1}$, viscous dissipation balances nonlinear interactions \citep{Zakharov1967,Deike2012} and self similarity is broken. \red{Spatial Fourier spectrum is also useful to estimate the steepness due to capillary waves only: 
\begin{equation}\sigma_{s c}=\left(\int_{k_c}^{k_d} k^2 \, S_\eta (k) \,d k \right)^{1/2}\end{equation}
For $\sigma_h=3.6\,$mm, $\sigma_{s c} \approx 0.15 $, when $\sigma_s=0.34$. Therefore the contribution to $\sigma_s$ of large gravity waves is larger than the one of capillary waves, but the last one is not negligible, as found also in field observation~\citep{MelvilleJFM2015,Breon}. Moreover this order of magnitude confirms that capillary waves are also in a nonlinear regime.}

 \begin{figure}
 \begin{center}
 \includegraphics[width=6.3cm]{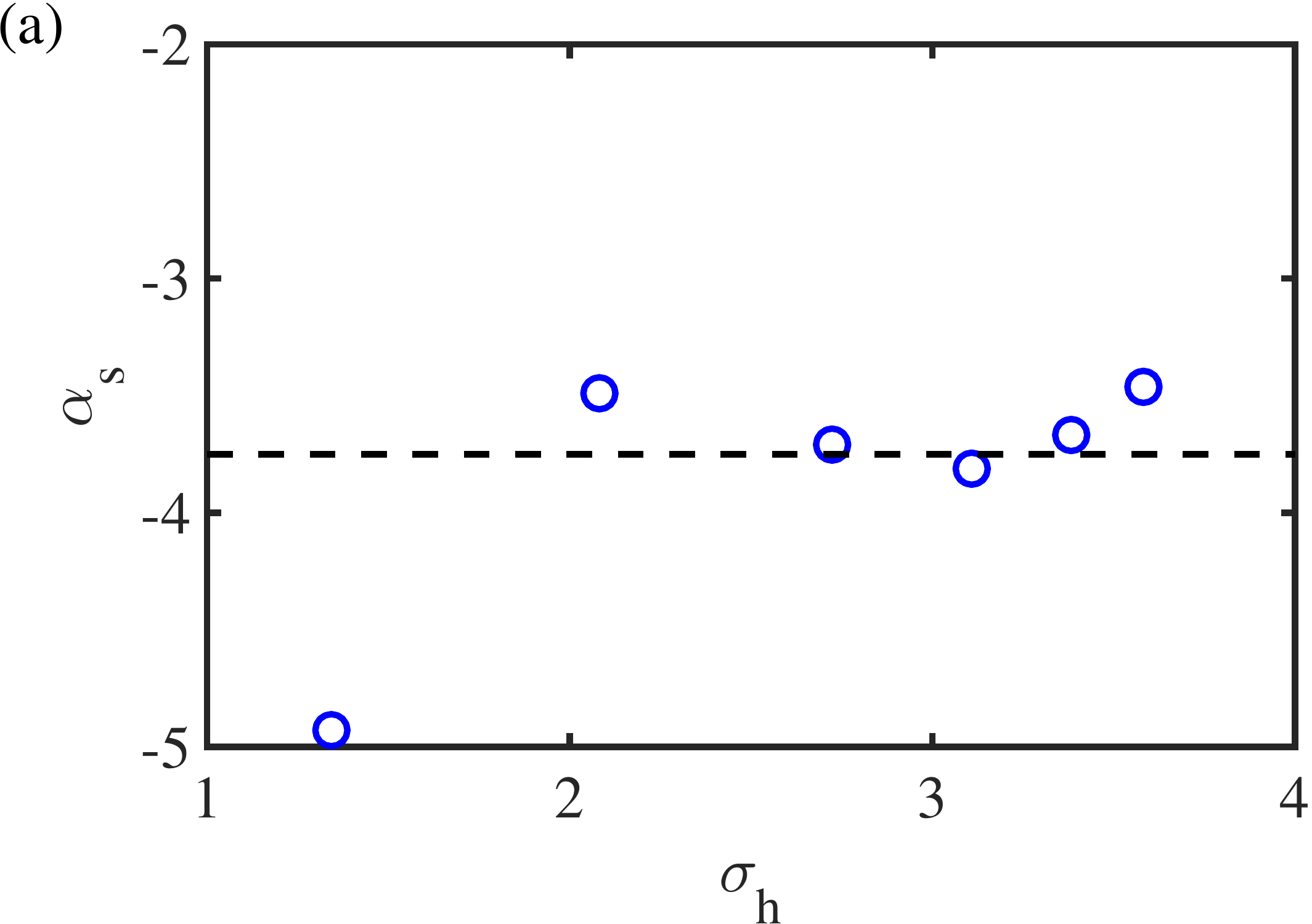} \hfill
  \includegraphics[width=6.3cm]{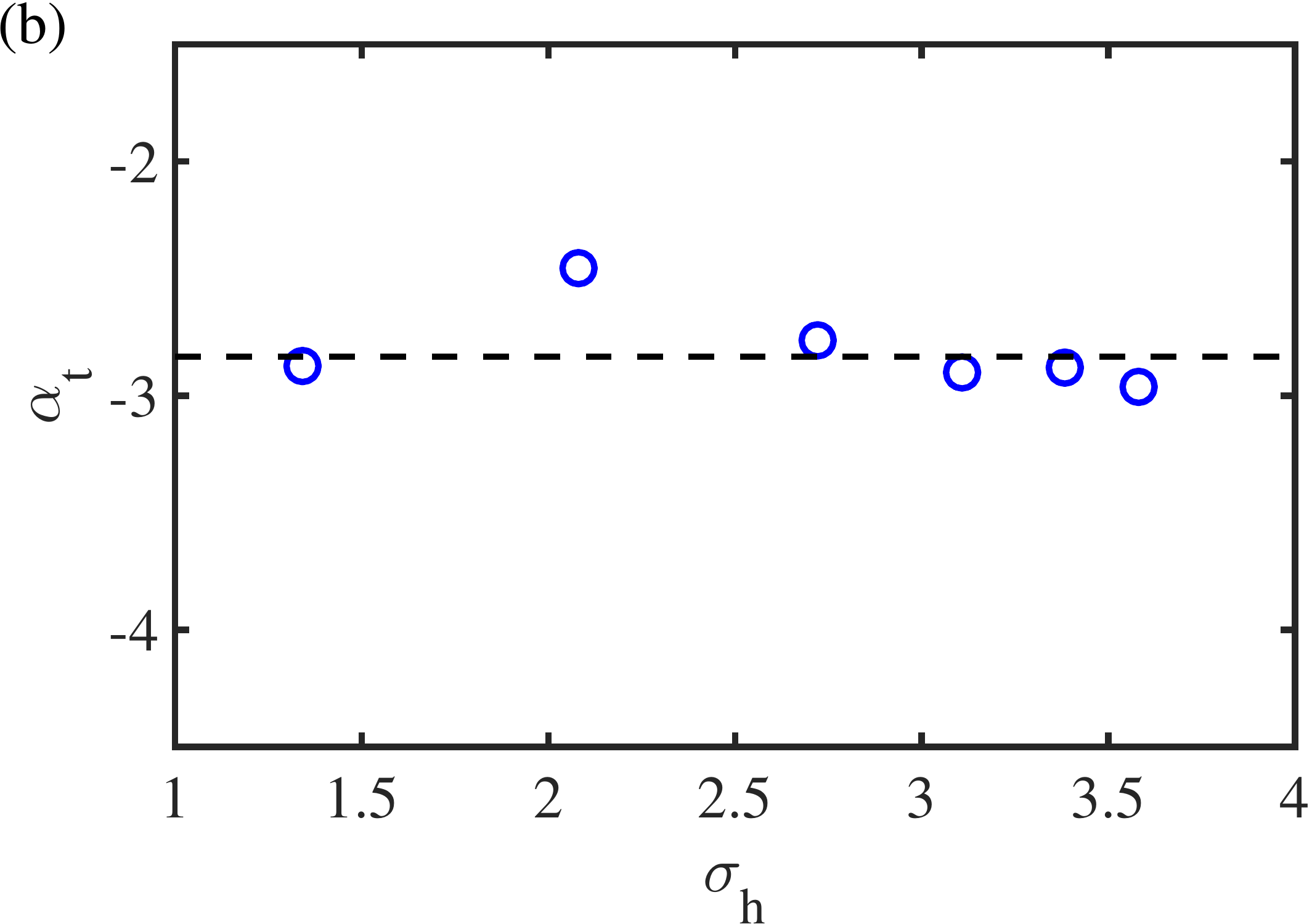} 
        \caption{(color online). \red{ (a) $S_\eta (k)$-exponents $\alpha_s$ vs. $\sigma_h$ (fits from $0.094 \leq \lambda^{-1} \leq 0.30$ mm$^{-1}$). Dashed line shows the theoretical value $-15/4$.  (b) $S_\eta (f)$-exponents $\alpha_t$ vs. $\sigma_h$ (fits from $20 \leq f \leq 100\,$Hz). Dashed line shows the theoretical value $-17/6$.}}
    \label{spectrumslope}
       \end{center}
 \end{figure}    

The temporal spectrum $S_\eta (\omega)$ is shown in Fig.~\ref{Spectra} (b) for the same set of data. Two estimations are given, the first by integration over the wave-numbers of $S_\eta (\omega,k)$ (continuous curves) and the second by converting the spatial spectrum in the frequency space (dot curve) using the linear dispersion relation (\ref{RDL}). Despite the shift of the experimental dispersion relation, both estimations are consistent, the first having a better resolution at low frequencies and the second reaching significantly larger frequencies. This matching of the spectra through the dispersion relation shows \red{ the consistency of our space-time measurements. As} with the spatial spectrum, for high enough wave amplitude a power law is observed  for the temporal spectrum between $f_c$ and $f_d$, with an exponent close to the value $-17/6$ predicted by capillary \red{Wave Turbulence Theory}. \red{We observe that outside the forcing range the frequencies of the eigen-modes (see \S \ref{randomforcing}) do not appear in the spectrum, due to wave viscous dissipation (See Appendix A) and to the random forcing. A secondary peak in the spectrum is located at twice the typical forcing frequency}. Notice that $f_d \approx 200\,$Hz for the highest forcing and \red{like $\lambda_d$ is interpreted as the balance between viscous dissipation and nonlinear interactions}.

\red{For scales between $\lambda_c$ and $\lambda_d$, power laws are fitted, $S_\eta (k) \sim k^{\alpha_s}$ and $S_\eta (f) \sim f^{\alpha_t}$. The corresponding values of $\alpha_s$ and $\alpha_t$ are given in Fig.~\ref{spectrumslope}. Except at the lowest excitation amplitude, $\alpha_s$ and $\alpha_t$ do not depend on $\sigma_h$ and are close of predictions of wave turbulence theory.} Thus as stated previously \citep{Berhanu2013}, the exponents of the spectra in the capillary wave range are in agreement with the predictions of Capillary Wave Turbulence \red{for $\sigma_h \geq 2$\,mm.}

\red{Power-law spectra of capillary waves in time and space indicate self-similar regimes of uncorrelated random waves, corresponding to an energy transfer from large scale to small scale, where energy is dissipated by dissipation. The turbulence of the capillary waves denote these regimes resulting from nonlinear wave interactions, which are not related to hydrodynamic turbulence. However the substantial viscous dissipation limits the inertial range ($[\lambda_c^{-1},\lambda_d^{-1}]$ or $[f_c,f_d]$) to only one decade in the spectrum, as observed in other experiments where capillary waves are forced by gravity waves \citep{Falcon2007,Deike2012,Deike2014}.}

\subsection{Wave correlations in time Fourier space}  
\label{bispectrum}

  \begin{figure}
 \begin{center}
\includegraphics[width=6.5cm]{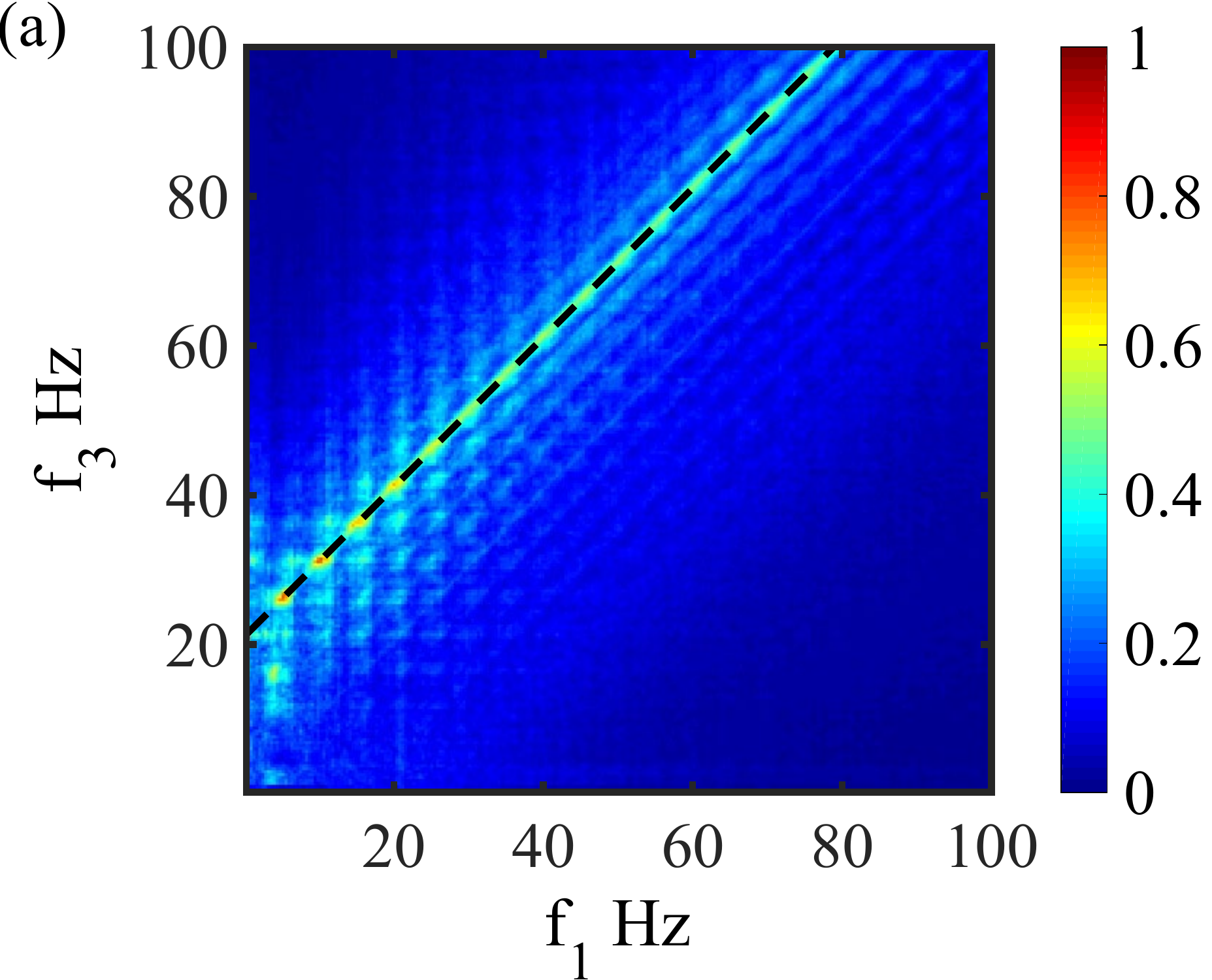} 
\hspace{.2cm}
 \includegraphics[width=6.5cm]{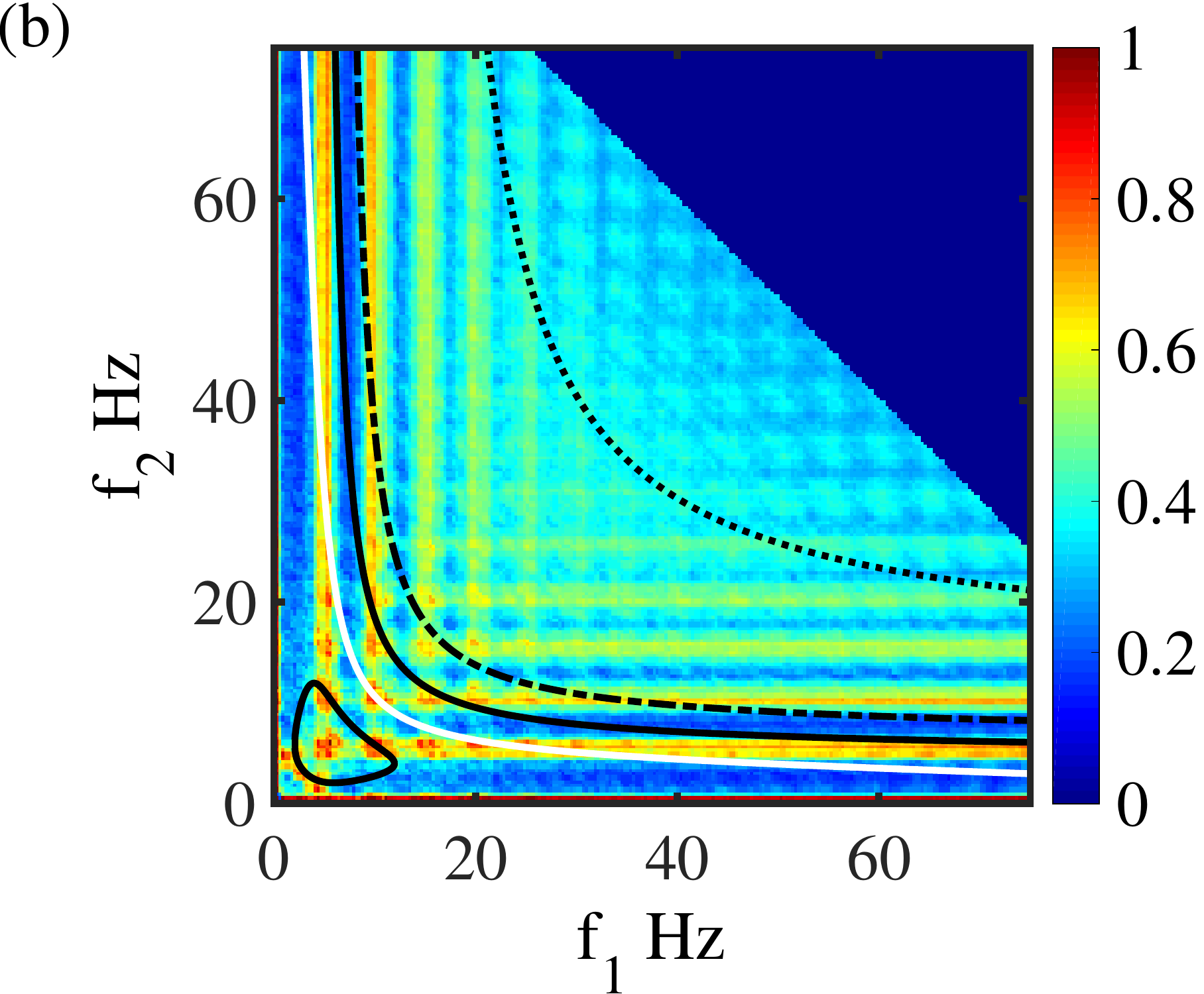} 
        \caption{(color online). (a) Three-Wave coherence or bispectrum  $B(f_1,f_2,{f_3})$ in colorscale, plotted for $f_2=21\,$Hz, for the highest forcing amplitude $\sigma_h =3.6\,$mm and $\sigma_s=0.34$. A significant correlation level is observed along the line $f_3=f_1+f_2$ (dash line), revealing presence of three-wave interactions. (b) Bicoherency level $B(f_1,f_2,f_3=f_1+f_2)$ for the same measurement. Strong level of three-wave correlation implies the forcing waves and their harmonics. \red{White curve displays the location of exactly resonant one-dimensional Three-Wave interactions. The black solid bounds non-resonant interactions with a mismatch $\Delta k =50\,$m$^{-1}$. Dashed line $\Delta k =100\,$m$^{-1}$. Dotted line  $\Delta k =300\,$m$^{-1}$.}}
    \label{correl}
       \end{center}
 \end{figure}    

\red{In a classic statistical description of a turbulent wave-field,} the spectra are built by coexistence of numerous and simultaneous Three-Wave resonant interactions, which in average transfer energy flux from the forcing scale to the dissipative scale.
It is assumed that after averaging on a long time, only the waves \red{verifying the resonant conditions (Eqs. ~\ref{Resonancek} and \ref{Resonance})} can exchange substantially energy~\citep{Janssen2004}. To detect occurrence of Three-Wave  interactions in the temporal domain, we track phase correlations in the Fourier space between the waves forming a triad, through the computation of the bispectrum $B(f_1,f_2,f_3)$ \citep{DudokdeWit2003,Punzmann2009,Aubourg2015,Aubourg2016}, which is defined by:
\begin{equation}
 B(f_1,f_2,f_3)  =\dfrac{\vert \langle \tilde{\eta} (\omega_1)\,\tilde{\eta} (\omega_2)\,\tilde{\eta}^* (\omega_3)  \rangle \vert }{ \langle \vert  \tilde{\eta} (\omega_1) \vert \rangle \,\, \langle\vert\tilde{\eta} (\omega_2) \vert \,\rangle \langle \vert \tilde{\eta} (\omega_3) \vert \rangle }
\end{equation}

In Fig.~\ref{correl} (a), for a given value $f_2=21\,$Hz and the highest wave amplitude (\textit{i.e.} the highest forcing), occurrence of Three-Waves interaction, appears as values of $B(f_1,f_2=21\,\mathrm{Hz},f_3)$ of order one, along the line $f_3=f_1+f_2$, \textit{i.e.} the frequency resonant condition. The same behaviour is also observed for other values of $f_2$ and lower forcing amplitudes. Then to compare the respective weight of the different triads, the bicoherency $B(f_1,f_2,f_3=f_1+f_2)$ \citep{Aubourg2015,Aubourg2016} is depicted in Fig.~\ref{correl} (b) by assuming the frequency resonant condition for the same measurement. \red{In the top-right corner, correlation is set to zero for  points where $f_1+f_2 > 100$\,Hz, as acquisition frequency is $200$\,Hz.} Knowing that the forcing is performed in the frequency range between $4$ and $6$ Hz, the bicoherency graph displays a periodicity linked with this range and the harmonics values. Due to the relatively high amplitude of the waves produced in the forcing range by the wavemaker, the triads involving these waves are privileged. \red{The pattern presents some resemblance with the one obtained at high steepness for waves near the gravity-capillary crossover as in \citet{Aubourg2016}. The analysis of the directional properties of the wave-field in Fig.~\ref{Aniso} shows that the main part of wave energy is due to waves propagating along the $x$-axis. Following the approach of \citet{Aubourg2015,Aubourg2016}, we investigate the role of three-wave interactions when $\mathbf{k_1}$, $\mathbf{k_2}$ and $\mathbf{k_3}$ are parallel to $O_x$. First one-dimensional exactly resonant interactions are considered. The corresponding curve is depicted in white in the bicoherency map in Fig.~ \ref{correl} (b) and appears to correspond to some local maxima of correlation. Then a mismatch $\Delta_k$ to the spatial resonant condition (\ref{Resonancek}) is enabled. The value of $\Delta_k$ is estimated equal to the width of the dispersion relation due to nonlinear broadening. In \S \ref{RDpart}, the experimental width $W$ of the dispersion relation was found to reach typically $0.04$\,mm$^{-1}$ at $f=30\,$Hz, which gives $\Delta k \approx\, 280$\,m$^{-1}$. The domains of allowed non-resonant interactions are delimited and plotted with black curves for $\Delta k =50\,$, $100\,$ and $300\,$m$^{-1}$. For the largest mismatch nearly all the couples $(f_1,f_2)$ can be populated with 1D three-wave interactions (\textit{i.e.} couples located below the dotted curve). Therefore the large width of the dispersion relation, shows that at high nonlinearity, three-wave interactions far from the resonance are possible. The distinction between quasi-resonant and non resonant interaction becomes here arbitrary, as both are transferring energy. A non-resonant or quasi-resonant interaction with a mismatch $\Delta_k$ produces an energy transfer oscillating in space on a length equal to $(2 \pi)/\Delta_k$. The wave packets have a typical length $1/W$ (auto-correlation length) fixed by the width of the dispersion relation. If $1/W < \Delta_k$ non-resonant interactions could lead thus to a net energy transfer. This hypothesis is similar to the results of Kenneth Watson \citep{WatsonJFM1996} describing a wave-field as interacting spatial modes. One-dimensional non-resonant three-wave interactions reproduce qualitatively parasitic capillary wave generation, with a  typical mismatch of $400\,$s$^{-1}$ from the temporal resonant conditions $|\omega_1 \pm \omega_2 -\omega_3 |=0$. }

\subsection{\green{Wave correlations in space Fourier space.}}

\begin{figure}
 \begin{center}
 \includegraphics[width=6.5cm]{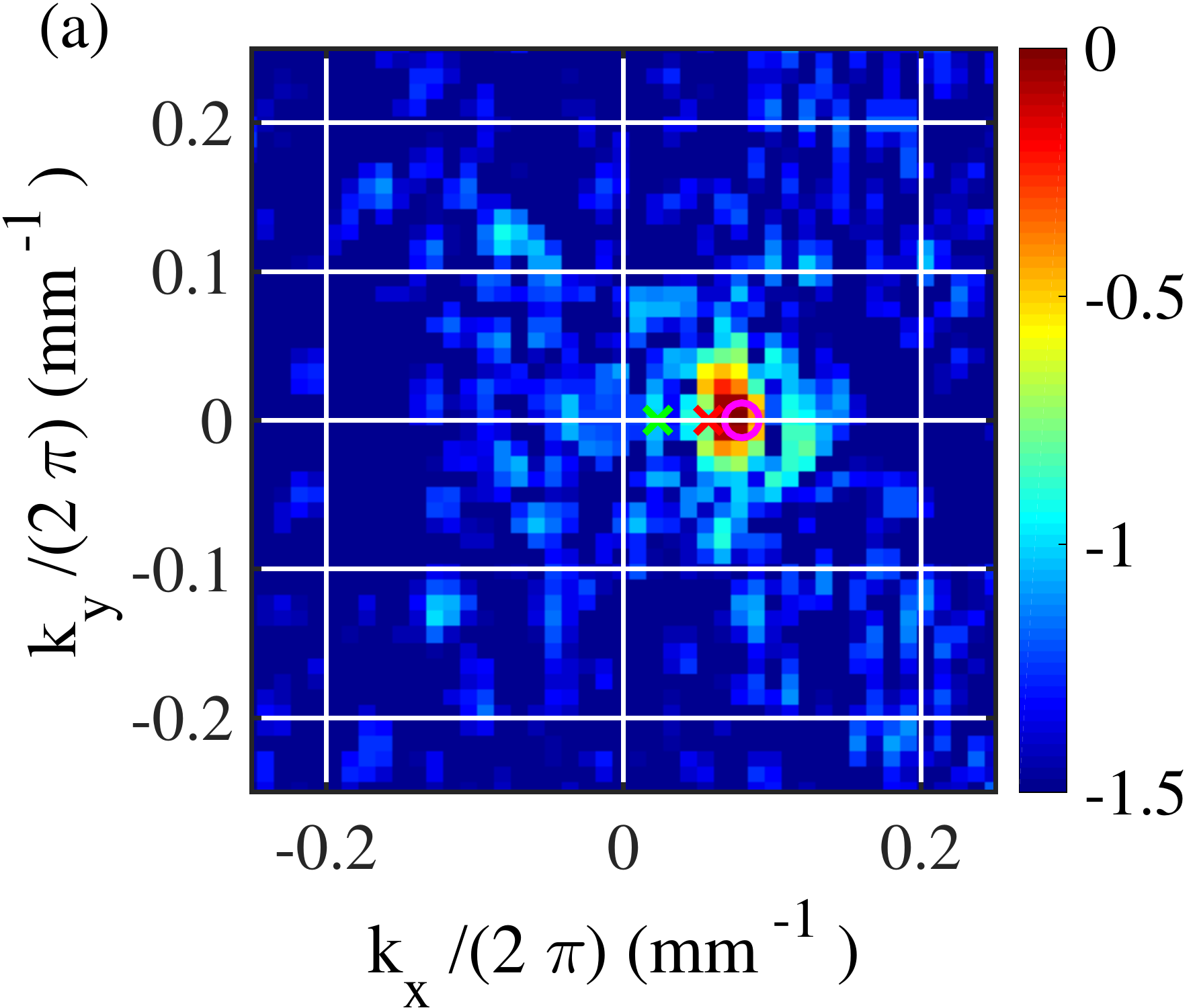} \hfill
     \includegraphics[width=6.5cm]{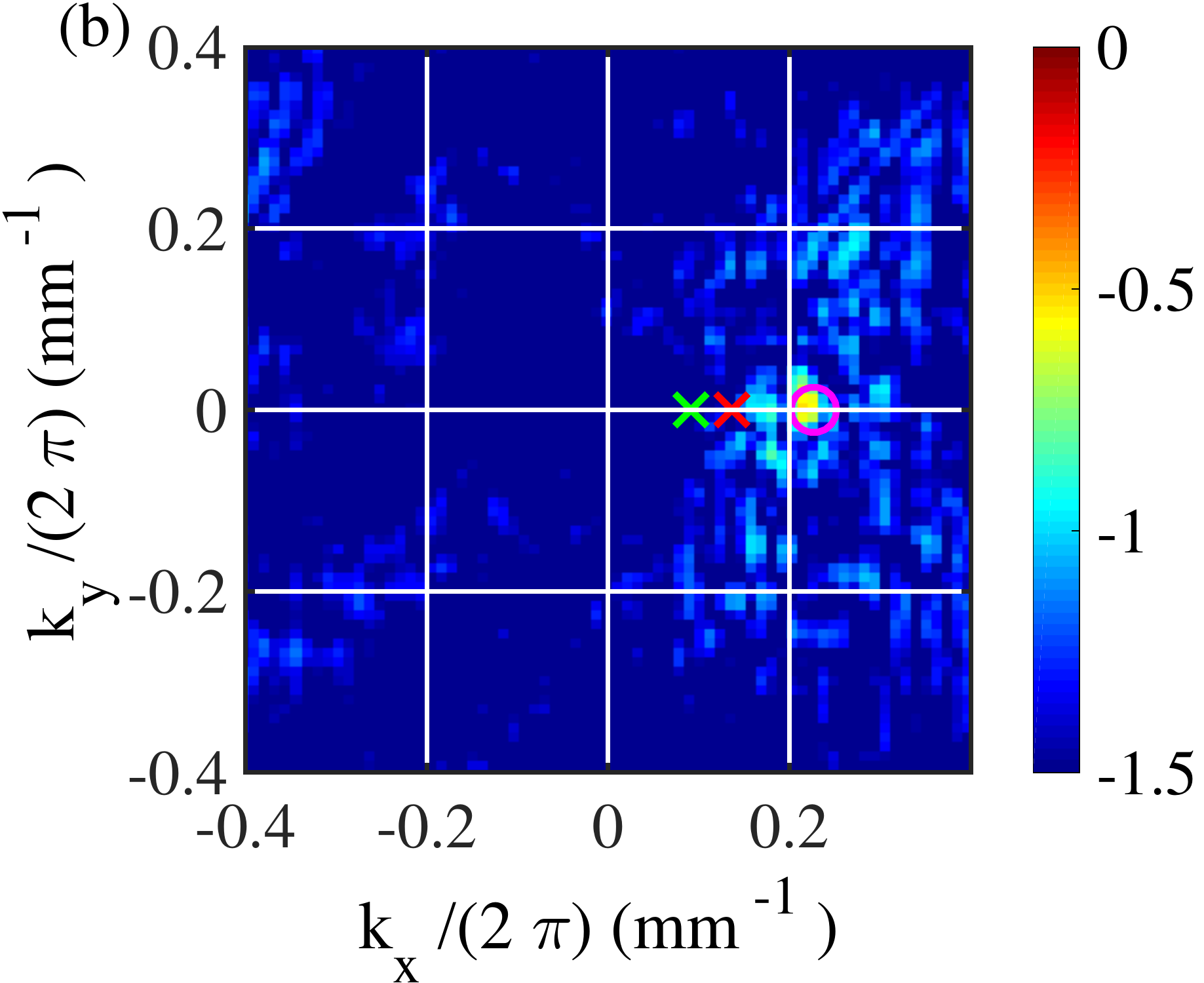} 
\includegraphics[width=6.5cm]{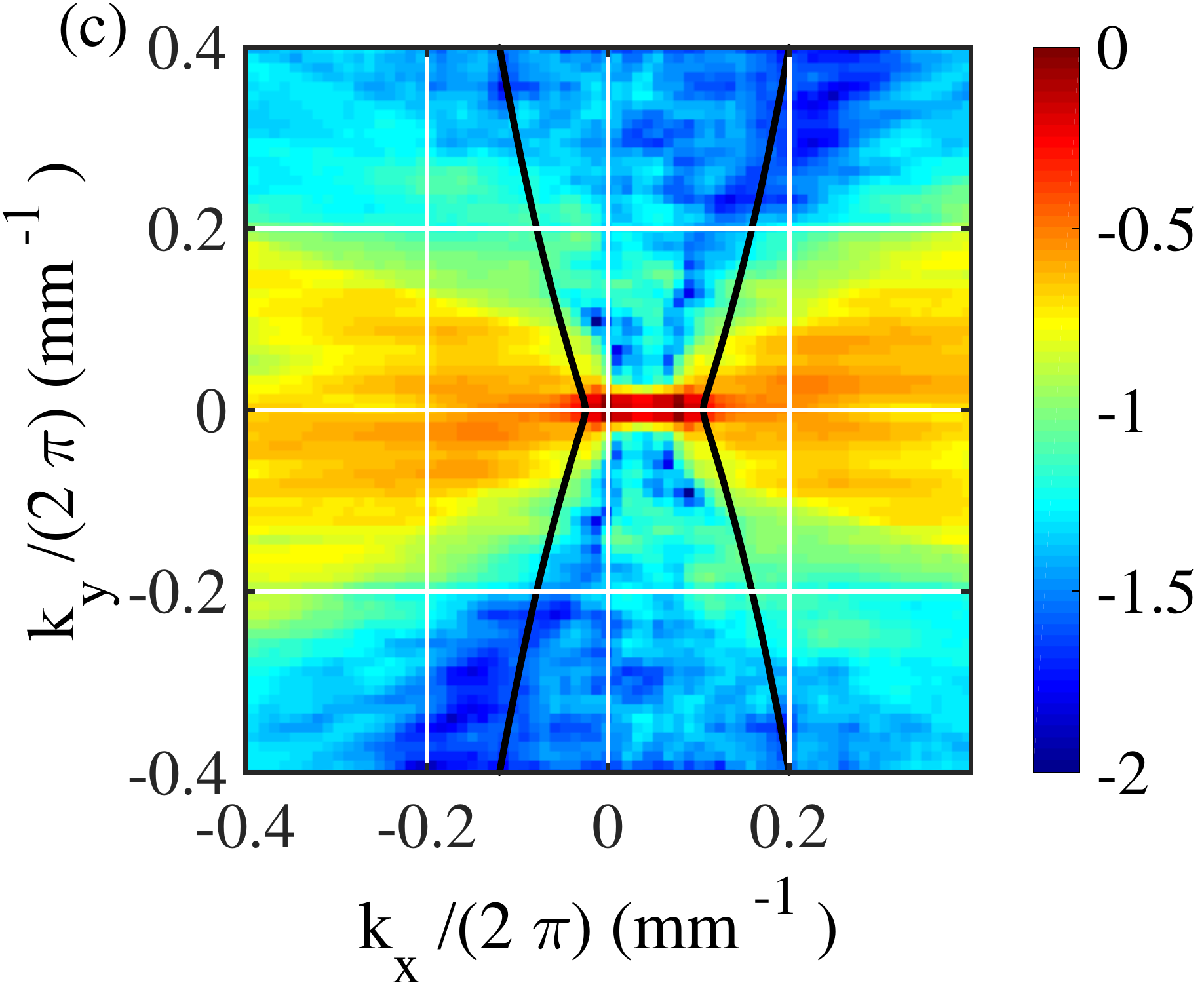} \hfill
  \includegraphics[width=6.5cm]{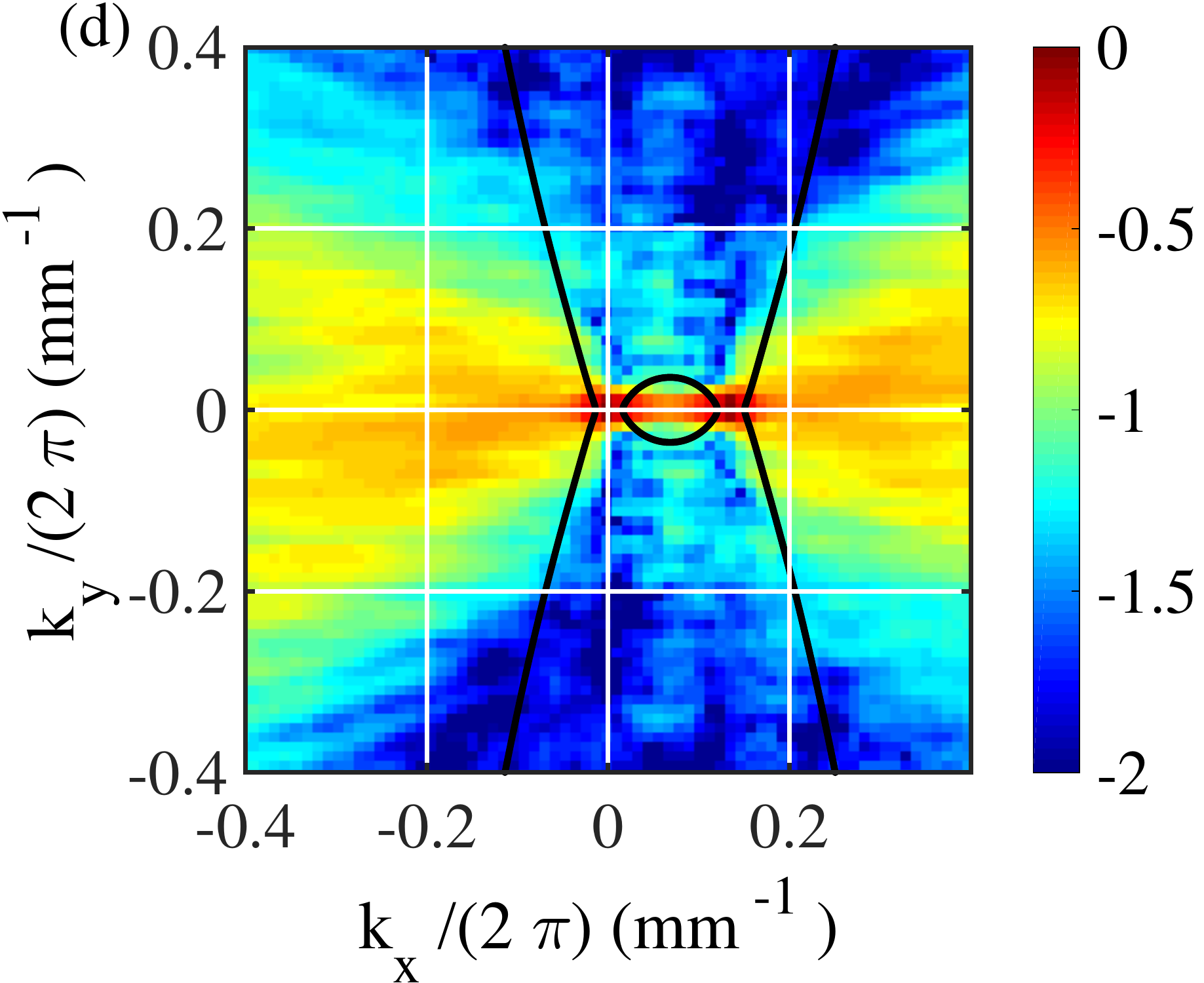} 
          \caption{\green{(color online). (a) Bispectrum $B_s(\mathbf{k}_1,\mathbf{k}_2,\mathbf{k}_3)$  plotted in $log_{10}$ colorscale as a function of $\mathbf{k}_1$ with $\mathbf{k}_2/(2\pi)=(0.0569,0) $\,mm$^{-1}$ (red $\times$) and $\mathbf{k}_3/(2\pi)=(0.0228,0)$\,mm$^{-1}$ (green $\times$). A stronger peak of correlation is observed around $\mathbf{k}_1/(2\pi)=\mathbf{k}_2/(2\pi)+\mathbf{k}_3/(2\pi)=(0.0797,0) $\,mm$^{-1}$ (pink $\circ$). The corresponding frequencies are $12.58$, $6.32$ and $17.74$\,Hz, thus this Three-Wave interaction is reasonably close to the resonance ($12.58+6.32 = 18.9 \approx 17.74$\,Hz). (b) Bispectrum $B_s(\mathbf{k}_1,\mathbf{k}_2,\mathbf{k}_3)$ as a function of $\mathbf{k}_1$ with $\mathbf{k}_2/(2\pi)=(0.1365,0) $mm$^{-1}$ (red $\times$)  and $\mathbf{k}_3/(2\pi)=(0.0910,0) $\,mm$^{-1}$ (green $\times$) . A peak of correlation is observed around $\mathbf{k}_3/(2\pi)=(0.2275,0) $mm$^{-1}$ (pink $\circ$). The corresponding frequencies are $34.25$, $20.65$ and $69.28$\,Hz, thus this Three-Wave interaction is non resonant ($34.25+20.65=54.9 \neq 69.28 $ Hz). (c) Bicoherence $B(\mathbf{k}_1,\mathbf{k},\mathbf{k}_1-\mathbf{k})$ plotted in $log_{10}$ colorscale as a function of $\mathbf{k}$ with $\mathbf{k}_1/(2\pi)=(0.0797,0)\, $mm$^{-1}$. The black curves are the loci of exact resonances according the linear dispersion relation. Non-resonant Three-Wave interactions are thus significant and mainly directed along the $x$ axis. (d) Bicoherence $B(\mathbf{k}_1,\mathbf{k},\mathbf{k}_1-\mathbf{k})$ as a function of $\mathbf{k}$ with $\mathbf{k}_1/(2\pi)=(0.2275,0) $\,mm$^{-1}$. The black curves are the loci of exact resonances according the linear dispersion relation. For a larger wavenumber $\mathbf{k}_1$, Three-Wave interactions are also found mainly non-resonant and directed along the $x$ axis.}}
    \label{bicoherencespatial}
       \end{center}
 \end{figure}   

\green{To complete our study of Three-Wave interactions, we perform a study of phase correlations at the third order in the spatial Fourier space. We compute first the space bispectrum to test the occurrence of Three-Wave interactions, then we use the spatial bicoherence to test the resonant or non-resonant character given the linear dispersion relation as performed by \citet{Aubourg2016} and by \citet{PanYueJFM2017}. To facilitate the comparison, we adopt their convention: the wave $1$ would be created by the interaction between the waves $2$ and $3$. The bispectrum writes thus:
\begin{equation}
B_s(\mathbf{k}_1,\mathbf{k}_2,\mathbf{k}) =\dfrac{ \vert \langle  \tilde{\eta} (\mathbf{k}_2)\,\tilde{\eta} (\mathbf{k}_3)\,\tilde{\eta}^* (\mathbf{k_1})   \rangle \vert}{ \langle \vert  \tilde{\eta} (\mathbf{k}_1) \vert \rangle \,\, \langle\vert\tilde{\eta} (\mathbf{k}_2) \vert \,\rangle \langle \vert \tilde{\eta} (\mathbf{k_3}) \vert \rangle }
\end{equation}
Two examples are given in Fig.~\ref{bicoherencespatial} (a) and (b) for the highest amplitude measurement by choosing the two vectors $\mathbf{k}_2$ and $\mathbf{k}_3$. If the spatial resonant condition is verified, (\ref{Resonancek}), a local maximum of correlation should be observed near $\mathbf{k}_1 \approx \mathbf{k}_2+ \mathbf{k}_3$. The lack of resolution at small wavenumbers due to the finite size of the images limit the accuracy of test. Moreover, a repetition of identical experiments could improve the convergence level. However, a clear correlation peak is observed at the sum wavenumber, showing a significant occurence of Three-Wave interactions. By computing the corresponding frequencies, given the linear dispersion relation (\ref{RDL}), we can also test the temporal resonant condition (\ref{Resonancek}). In the first case (Fig.~\ref{bicoherencespatial} (a)), the sum frequency is reasonably close to $f_1$, the interaction is quasi-resonant. But in the second case (Fig.~\ref{bicoherencespatial} (b)), the sum frequency differs notably from $f_1$, implying that the non-resonant interactions contribute also to the wave-field dynamics. To generalize this statement, knowing that the level of Three-Wave interaction is significant, we compute the bicoherence map defined by:
\begin{equation}
 B(\mathbf{k}_1,\mathbf{k},\mathbf{k}_1-\mathbf{k})  =\dfrac{ \vert \langle \tilde{\eta}^* (\mathbf{k}_1)\,\tilde{\eta} (\mathbf{k})\,\tilde{\eta} (\mathbf{k}_1-\mathbf{k})   \rangle \vert }{ \langle \vert  \tilde{\eta} (\mathbf{k}_1) \vert \rangle \,\, \langle\vert\tilde{\eta} (\mathbf{k}) \vert \,\rangle \langle \vert \tilde{\eta} (\mathbf{k}_1-\mathbf{k}) \vert \rangle }
\end{equation}
For a given $\mathbf{k}_1$, the condition $\mathbf{k}_1=\mathbf{k}_2+\mathbf{k}_3$ is automatically satisfied by taking $\mathbf{k}_2=\mathbf{k}$ and $\mathbf{k}_3=\mathbf{k}_1-\mathbf{k}$. Expressed as a function of $\mathbf{k}$, this statistical tool indicates the position of wavenumbers which are participating the more to the wavefield dynamics. The temporal resonance condition can be also tested under the form ($\omega_1=\omega(\mathbf{k})+\omega (\mathbf{k}_1-\mathbf{k})$) using the linear dispersion relation. Two cases are displayed in Fig.~\ref{bicoherencespatial} (c) and (d). The loci of exact temporal resonances are given by the blacks curves. We observe that the Three-Wave interactions in our systems are mainly one-dimensional, directed along the forcing axis ($O_x$) and non-resonant. A significant level of correlation is indeed observed far from the blacks curves. The typical width of the dispersion relation $\Delta k /(2\pi) \approx 0.05$\,mm$^{-1}$ is not sufficient to assume these interactions as quasi-resonant. The results are very similar for measurements at lower amplitude. We note also, that the lack of resolution in the $k$ space limits the accuracy of the method compared to the temporal bicoherence analysis which is better resolved. Assuming the linear dispersion relation, the non-resonant character of Three-Wave interactions is thus clearly demonstrated. The physical situation differs strongly from the work of \citep{PanYueJFM2017} simulating pure capillary waves, where the bicoherence level is significant only close to the exact resonance. In our experiments, as a deviation from the linear dispersion is reported in \S \ref{RDpart}, a bispectral analysis testing simultaneously both resonant conditions (Eqs \ref{Resonancek} and \ref{Resonance})  by computing a space-time bicoherence \citep{Aubourg2017} would be also useful to evaluate the role of non-resonant interactions and to detect a possible role of harmonics. However, a converged computation of space-time bicoherence require a large data set and a such study would deserve a systematic investigation.}

\section{{Fluctuations of Capillary Wave Turbulence}}
\subsection{Temporal evolution of spatial spectrum}
\label{Ekt}
\begin{figure}
 \begin{center}
 \includegraphics[width=13cm]{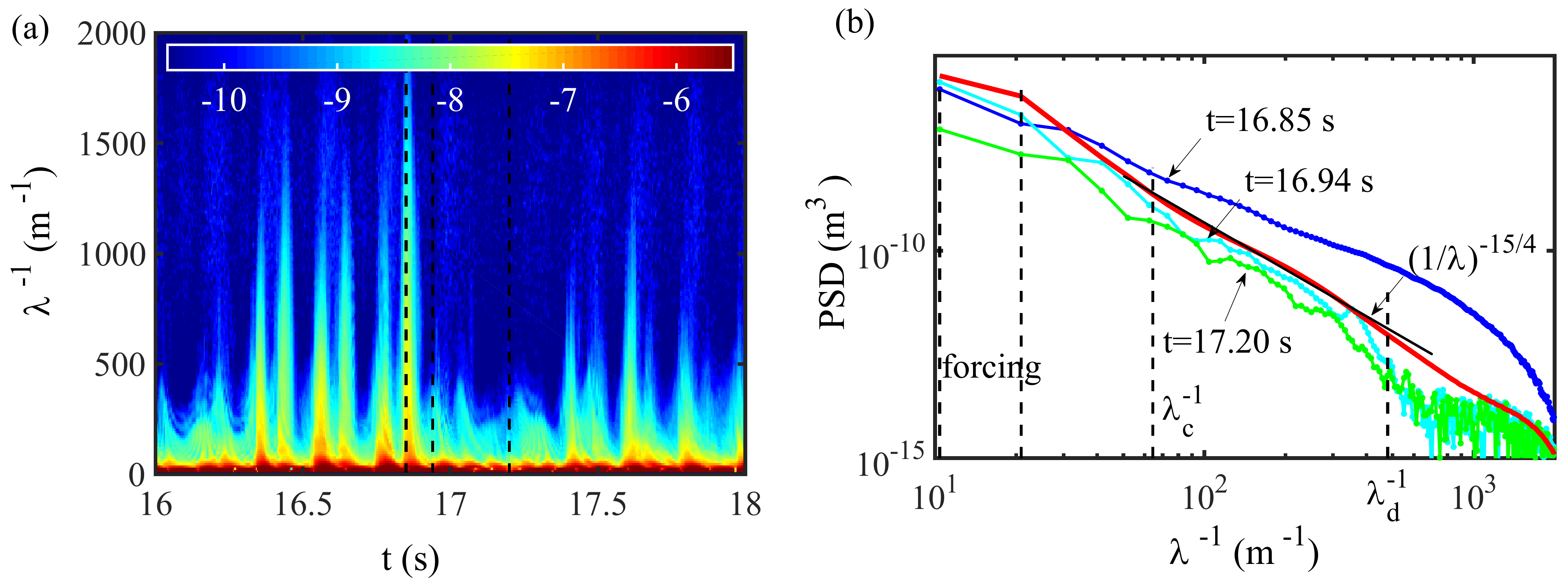} 
        \caption{(color online). (a) Temporal evolution of the spatial spectrum $S_\eta(k,t)$ in the ($t$,$\lambda^{-1}$) space for the highest wave amplitude measurement, $\sigma_h=3.6$\,mm. Color scale in $\log_{10}$ scale. Several bursts of energy are observed and affect a broad range of scales. Black dash vertical lines mark specific times $t^*$ used in (b). (b) Instantaneous spatial spectra $S_\eta(k,t^*)$ for $t^*=16.85$,$16.94$ and $17.20$\,s in dot-lines for the same measurement. Continuous thick red-line, spatial spectra averaged on the duration of the measurement $41.0$\,s. Black line indicates the power law $k^{-15/4}$. The instantaneous spectrum has a higher amplitude than the average spectrum during an energy burst and has a lower amplitude between the bursts.}
    \label{figspinsta}
       \end{center}
 \end{figure}

\red{In the previous section, we have characterized statistically stationary out-of-equilibrium turbulent states of interacting waves.} To better isolate the mechanisms at play, we now investigate the temporal behavior of the wave-field, by computing at each time the spatial power spectrum of wave elevation $S_\eta (k,t)$ (averaged over the directions). As previously discussed for the wave potential energy in \citet{Berhanu2013}, we observe in Fig.~\ref{figspinsta} (a) that $S_\eta (k,t)$ displays stochastic bursts transferring energy towards small spatial scales, as a consequence of the random forcing. \red{For the displayed measurement $196$ bursts can be counted on a duration of $41$\,s, which is coherent with a typical forcing frequency of $5$\,Hz.} Indeed, most of the time, the wave energy is confined near the forcing scales, but from time to time, wave energy is quickly transferred through all spatial scales. The autocorrelation time of $S_\eta (k,t)$ is found for $k$ inside of the capillary range, of order $0.05$\,s and the corresponding rise time is estimated about $0.02$\,s by computing the cross-correlation of $S_\eta (k,t)$ for $k$ within the forcing scale and $k$ in the capillary range. Such a short time shows that the typical nonlinear time characterizing wave interactions has to be smaller than the wave period for frequency below $50\,$Hz. The hypothesis of weak nonlinearity is thus experimentally violated, as it was shown by the analysis of the width of the dispersion relation in \S \ref{RDpart}. Moreover, we show three instantaneous spatial spectra $S_\eta (k,t^*)$ as a function of $\lambda^{-1}$ in Fig.~\ref{figspinsta} (b), for some typical values of $t^*$. These spectra are compared with the spectrum obtained after averaging on the duration of the measurement. This last spectrum is identical to the one obtained in Fig.~\ref{Spectra} (a) and is compatible with the power law $k^{-15/4}$ in the capillary range. When $t^*$ corresponds to a burst of energy, the spectrum is significantly higher in amplitude than the average spectrum, whereas it is slightly smaller in amplitude outside the bursts. The strong events seem to transfer directly energy from the forcing to the capillary scales, then the slower relaxation between bursts, appears to be closer of the weak wave turbulent regime \red{hypotheses}. \red{Similar observations have been reported for surface gravity wave turbulence \citep{Bedard2013}. Fluctuations of filtered spectral amplitude appear as a succession of random bursts, interpreted as wave breaking events, for which the instantaneous spatial spectrum is higher.}

\begin{figure}
 \begin{center}
 \includegraphics[width=6.5cm]{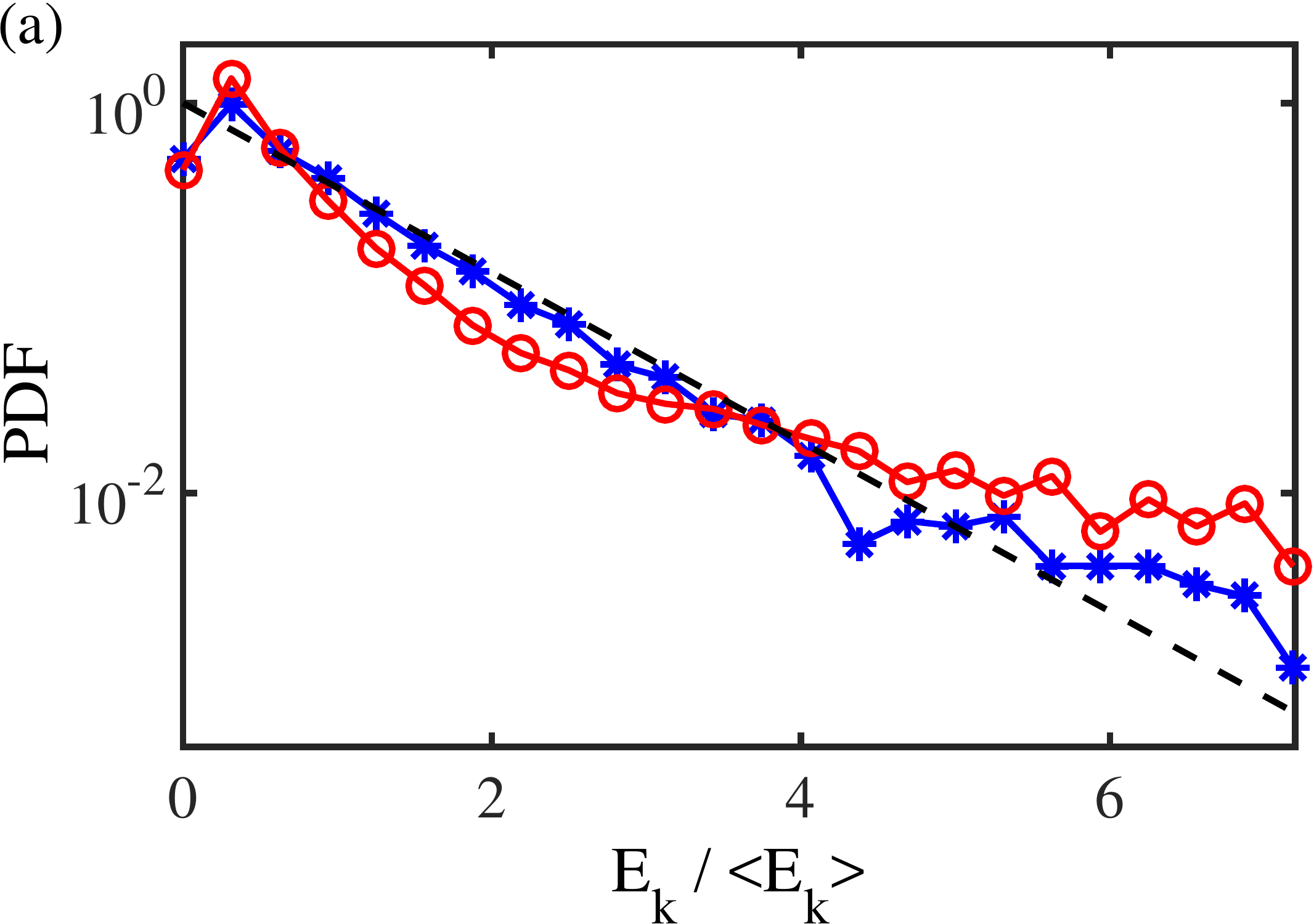} \hfill
  \includegraphics[width=6.5cm]{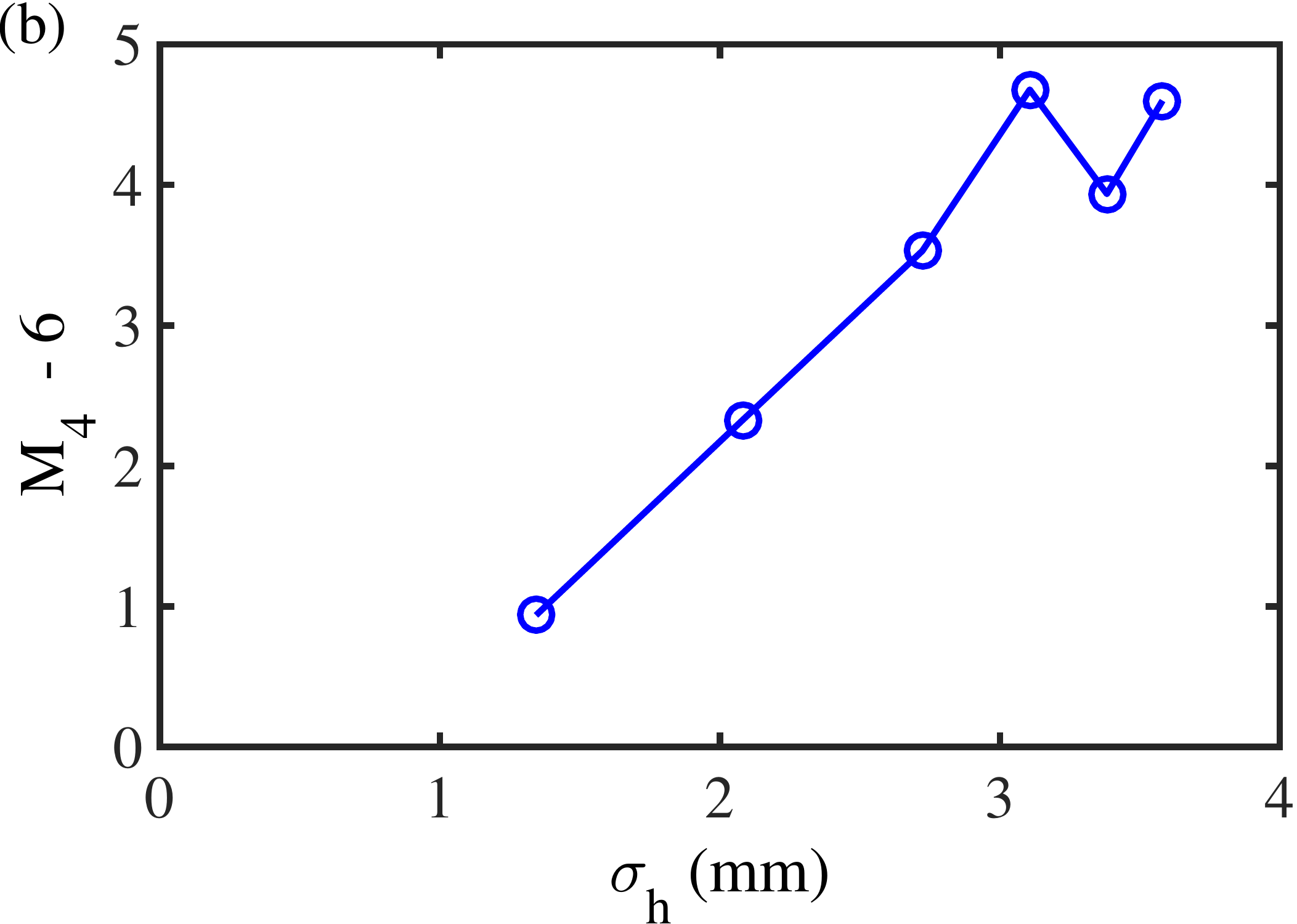} 
        \caption{(color online). \red{ (a) Probability density function (PDF) of wave energy $E(k_{\star},t)/\langle E(k_{\star},t) \rangle$ for $\lambda_{\star}^{-1}=0.094$\,mm$^{-1}$: ($\ast$) moderate forcing ($\sigma_h = 2.0\,$mm, $\sigma_s=0.19$), and ($\circ$) high forcing ($\sigma_h = 3.6\,$mm, $\sigma_s=0.34$). Dashed line is the exponential distribution: $\exp (-E_k/\langle E_k \rangle)$. (b) Deviation of the previous PDF from the exponential distribution, evaluated by computing the fourth order moment $M_4=\langle E_k^4 \rangle$ minus $6$, the exact result for the exponential distribution, as a function of the wave amplitude $\sigma_h$.}}
    \label{PDFEkt}
       \end{center}
 \end{figure}      
 
\red{In order to quantify the intermittency of energy fluctuations $E_k (t)$, the probability distribution function is plotted in Fig.~\ref{PDFEkt} (a) for a typical capillary wave mode, $\lambda_{\star}^{-1}=0.094$\,mm$^{-1}$.} \blue{The energy spectrum per density unit is first obtained from $S_\eta (k)$, assuming  that the wave energy is in average  two times the potential energy and neglecting higher order term in the surface extension:
\begin{equation}
  \int E_{k}(k,t) \mathrm{d}k =\frac{2}{\rho}\, \int \left[ \frac{1}{2} \rho g \, S_\eta (k) + \frac{1}{2} \gamma k^2 S_\eta (k) \right]\mathrm{d}k
\label{Ek}
 \end{equation}}
\red{In random regimes of capillary waves, a Gaussian distribution of wave amplitudes corresponds to a decreasing exponential distribution of $E_k (t)$, $\exp (-E_k/\langle E_k \rangle)$. At low forcing amplitude, the PDF of $E_k (t)$ follows the exponential distribution, whereas at high forcing amplitude, the PDF departs from the exponential distribution with an increased probability of strong events. This deviation from the Gaussian statistics observation defines the intermittency in the dynamics of random waves \citep{Choi2005}. It is attributed to the presence of large scale coherent structures caused by strongly nonlinear events \citep{NazarenkoJFM2010}. To quantify the intermittent behavior, we compute the fourth order moment $M_4=\langle E_k^4 \rangle$, a higher moment being more sensitive to the contribution of extreme events. For an exponential distribution, $M_4$ is exactly equal to $6$. The difference $M_4 - 6$ is plotted in Fig.~\ref{PDFEkt} (b)  as a function of wave amplitude $\sigma_h$. The departure from the exponential distribution increases with amplitude and saturates for the three last measurements. This criterion shows that capillary wave energy has an intermittent dynamics, due to occurrence of strong bursts transferring energy from large scales to small scales by a fast mechanism. This fast and non-local process corresponds likely to the generation of  parasitic capillary waves, as shown in the next part.} \\

\subsection{Evidence of parasitic capillary wave generation}

\begin{figure}
 \begin{center}
 \includegraphics[width=13cm]{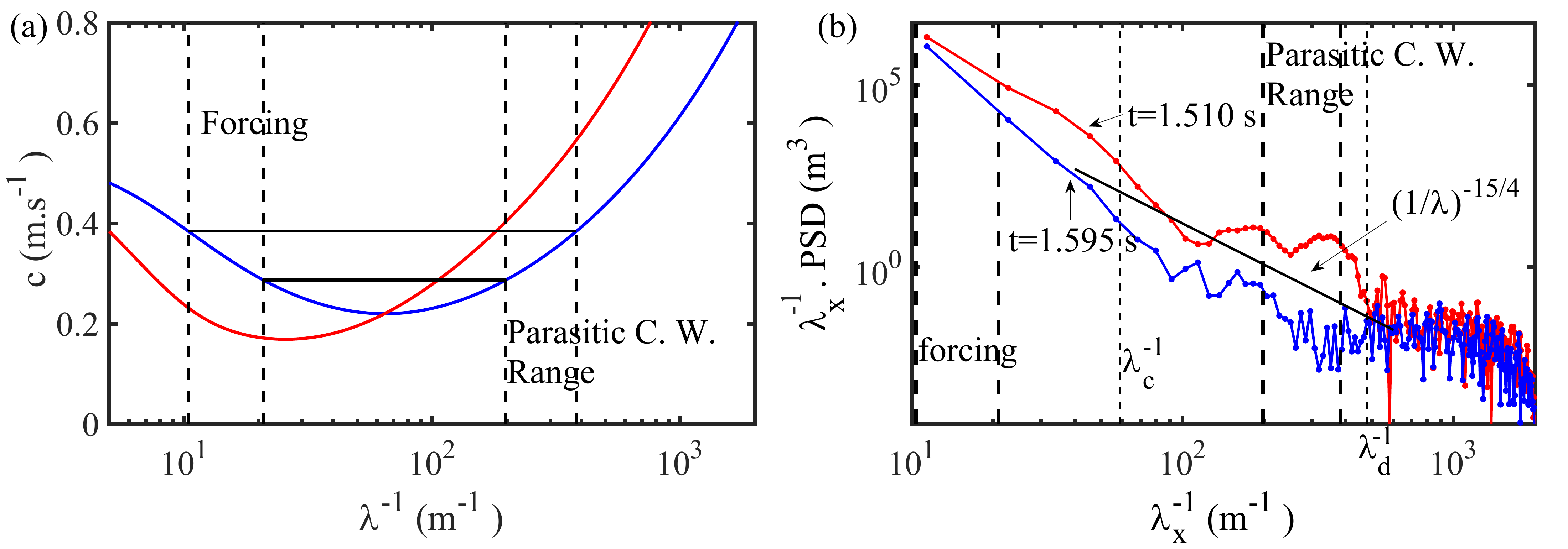} 
        \caption{(color online). (a) Schematic illustration of the condition of capillary wave \red{generation} by the parasitic capillary wave mechanism. Blue (dark gray) and red (light gray), respectively phase and group velocity of gravity-capillary waves as a function of $\lambda^{-1}$. The waves in the forcing range excite capillary waves having the same phase velocity. (b) Spatial \red{one-dimensional} spectrum $S_\eta(k_x,k_y=0)$ as a function of $\lambda_x^{-1}$ and multiplied by $\lambda_x^{-1}$ to facilitate the comparison with $S_ \eta (k)$ for two time values and $\sigma_h=3.6\,$mm. At $t=1.510\,$s, corresponding to the image in Fig.~\ref{Snapshot}, where a parasitic capillary wave appears, the spectrum is populated in the range given by the condition on the phase velocity. At $t=1.595\,$s, the spectrum is lower in the capillary wave range.}
    \label{Parasitic}
       \end{center}
 \end{figure}

In the physical space, we interpret the occurrence of strong bursting events of wave energy, as the generation of parasitic capillary waves. In order to reinforce this observation, a quantitative criterion on the wave-field analysis is needed. As the theoretical works are only performed for propagating and non decaying monochromatic
 gravity waves, it appears difficult to make a direct comparison with our experimental results. Nevertheless an approximate criterion of resonance is given by the equality of the phase velocity of the gravity wave and the one of the produced parasitic capillary wave. Knowing that waves are forced in the frequency range $[4,6]$\,Hz, a corresponding range of parasitic wave generation can be found from the expression of the phase velocity of gravity capillary wave $v_\phi=\dfrac{\omega}{k}$, using the linear dispersion relation (\ref{RDL}), as schematized in Fig.~\ref{Parasitic} (a). Expressed as the inverse of the wavelength, the forcing range reads $\lambda^{-1} \in [10.4,20.9]$\,m$^{-1}$ and the parasitic capillary wave generation range is found 
$\lambda^{-1} \in [198,382]$\,m$^{-1}$. \red{In their analysis of experimental parasitic wave generation, \citet{FedorovPOF1998}  claim that the interval of parasitic wave generation is extended due to nonlinear effects.} Taking into account the nonlinear correction of the phase velocity of the gravity waves treated as a Stokes wave, the last range \red{would be} multiplied by $(1+\sigma_s^2)$ , which leads to $\lambda^{-1} \in [221,426]$\,m$^{-1}$ as a range of maximal excitation of capillary waves. The Doppler shift due to the orbital current induced by the large gravity wave on the capillary wave \red{may} be also taken in account. In the laboratory frame of reference, the total range of observation of parasitic capillary waves, is obtained after multiplication by a factor $(1\pm 2\sigma_s)$ \citep{FedorovPOF1998}, giving $\lambda^{-1} \in [70.8,716]$\,m$^{-1}$. These scales appear clearly excited for strong bursts of the spatial spectrum displayed in Fig.~\ref{figspinsta} (a). As the instantaneous steepness can be higher than $\sigma_s$, a clear cut-off at a maximal wave number is not visible with a random excitation. Moreover a depletion of the spectrum in the intermediate range between $\lambda^{-1}=20.9$\,m$^{-1}$ and $\lambda^{-1}=70.8$\,m$^{-1}$ is not observed. To make the comparison with the work of \citet{FedorovPOF1998}, we plot in Fig.~\ref{Parasitic} (b) the 1D-spatial spectra $S_\eta(k_x,k_y=0)$  along the $x$ direction corresponding to the main direction of propagation of the gravity wave displayed in Fig.~\ref{Snapshot}, at the instant of parasitic wave generation ($t=1.510\,$s)  and after it. \red{The interval $[198,382]$\,m$^{-1}$ of parasitic wave generation from the criterion of equality of phase velocities is plotted by thick vertical dash lines.} Capillary waves are indeed excited in the parasitic capillary wave range at $t=1.510\,$s. Due to the finite size of the observation window, the exact wavelength of the forcing is not resolved in the spectrum. Then at $t=1.595\,$s the spectrum has a lower amplitude in the capillary wave range, but with a similar amplitude in the forcing range and for $\lambda^{-1}> \lambda_d^{-1}$ \textit{i.e.} the dissipative zone.

\red{In order to generalize this statement for more capillary wave generation events, a video animation has been made and can be found in supplementary materials ({Movie2}). On a duration of $2.5\,s$, the wave field gradient $||\mathbf{\nabla}h(x,y)||$ and the instantaneous spatial spectrum (one-dimensionnal $k_x\,S_\eta(k_x,k_y=0)$ in blue and directionally integrated $S_\eta (k)$ in green) are displayed in the same figure, showing there time evolution. At specific times, capillary wave trains are clearly visible on the crests of long steep gravity waves. Simultaneously in the spectrum, the range of parasitic wave generation experiences a sudden increase. The largest events correspond to frontal collisions of two counter-propagating gravity wave. This last parasitic capillary generation mechanism has been described experimentally for steep standing waves \citep{Schultz1998}, but has not been the object of a theoretical study.  It appears than the bursts of intermediate amplitude are more prone to follow the classical scenario of formation of parasitic capillary waves. Nevertheless,} parasitic wave generation mechanism appears to be a process relevant for the frequency range and steepness of forcing waves in our experiment. This mechanism transfers by a strongly nonlinear interaction wave energy from large to small scale. However, the available theoretical work does not seem applicable directly to a random field of waves in a closed tank. Moreover after averaging on a sufficient duration, the spatial spectrum can be described as a power-law in capillary wave range. The periods between the bursts with parasitic wave generation, have a significantly lower wave amplitude and steepness. A relaxation process on a longer time scale seems to populate continuously the spectrum with direct and inverse transfer, to approach the wave turbulent spectrum in average. This description is similar to the turbulent cycle of gravity wave turbulence \citep{NazarenkoJFM2010}, where strongly nonlinear coherent structures distributing energy on a large range of scales coexist with a randomly distributed background of waves, producing in average the self-similar spectra. 
\blue{Thus in this part, we show experimentally that the capillary spectrum is built by short time events corresponding to parasitic capillary wave train generation and corresponds to the wave-turbulence spectrum. Between the bursts the capillary spectrum is below the wave turbulent spectrum. This last spectrum is only found after temporal average on a duration incorporating parasitic events and wave relaxation between the bursts. Therefore, experimentally, the power-law spectrum of capillary waves is primarily formed by parasitic capillary waves and not by local interactions of capillary waves.}

\section{\red{Comparison of experimental results with the Wave Turbulence Theory }}
\subsection{\red{Context}}
\label{Context}
\red{For strong enough forcing, spectra of wave elevation have been found in \S \ref{Cascade} to follow power-law whose exponents are in agreement with those given by the Wave-Turbulence Theory. However these spectra appear to be built in average by a random succession of strong nonlinear events, for which parasitic capillary trains are emitted. This statement differs from the classic picture of Wave Turbulence, where randomly distributed weakly nonlinear Three-Wave interactions construct the spectra. We propose now a more thorough comparison of our experimental results with the predictions of the Wave Turbulence Theory~\citep{Zakharovbook,Nazarenkobook,Newell2011}, which provides a statistical description of weakly nonlinear dispersive waves. We provide briefly the theoretical results, justifying the later analysis of our experimental data.}

\red{In a statistically stationary regime, power-law spectra are predicted in wave turbulence theory. They correspond to a transfer of a conserved energy flux $\epsilon$, from an injection scale to a dissipative smaller scale.} \blue{For pure capillary waves, by computing the contribution of resonant Three-Wave interactions, Wave Turbulence Theory predicts the wave action spectrum $n_\textbf{k}$ \citep{Zakharov1967,Pushkarev2000}. $n_\textbf{k}$ expresses qualitatively the number of quasi-particles populating the scale defined by the wave vector $\textbf{k}$.}
\blue{ \begin{equation}
{n_\textbf{k}}=C_{KZ}\,(\gamma / \rho)^{-1/4}\,\epsilon^{1/2}\,k^{-17/4}
\label{Spn}
\end{equation}}
\Green{Using the assumption of an isotropic system, the exact value of the dimensionless Kolmogorov-Zakharov constant $C_{KZ}$ can be found \citep{Pushkarev2000,PanYueJFM2017}. To relate the wave action spectrum to the measured wave elevation spectrum, few steps are needed. First the wave action spectrum is integrated over the direction in the hypothesis of an isotropic system, which leads to: ${n_k}=2 \pi \, C_{KZ}\,(\gamma / \rho)^{-1/4}\,\epsilon^{1/2}\,k^{-13/4}$. The wave energy spectrum per density unit is defined by $E_k=\omega(k)\,n_k$, where $\omega(k)$ is given by the linear dispersion relation of pure capillary waves $\omega^2=(\gamma/\rho)\,k^3$. Averaging over several periods, the wave energy  writes $\langle E \rangle =\langle E_c + E_{pot} \rangle= 2\, \langle E_{pot}\rangle$. For pure capillary waves, in the weakly non-linear limit $E_{pot}=(1/2)\,(\gamma /\rho)\,k^2\,S_\eta (k)$. The spatial spectrum of wave elevation can be thus computed from the wave action spectrum. The temporal spectrum is obtained from the spatial spectrum by using the relation dispersion for pure capillary waves and by performing a variable change in the relation:}
\blue{\begin{equation} \langle \eta^2 \rangle=\int\, S_\eta (k) \mathrm{d} k= \int\, S_\eta (\omega) \mathrm{d} \omega \end{equation}}
\blue{Finally, we find:
 \begin{equation}
S_\eta (k)=2\pi\,C_{KZ}\,\epsilon^{1/2}\,(\gamma/\rho)^{-3/4}\,k^{-15/4} 
 \quad \quad 
S_\eta (\omega)=4\pi/3\,C_{KZ}\,\epsilon^{1/2}\,(\gamma/\rho)^{1/6}\,\omega^{-17/6}
\label{Spc}
\end{equation} }
\Green{We note that the values of dimensionless pre-factors differ in a recent theoretical work \citep{PanThesis}.}
\red{The spectra (\ref{Spc}) can also be deduced from dimensional analysis, knowing that the scaling on $\epsilon$ is set by the order of the resonant interaction (a $N$ waves process implies an energy flux scaling under the form: $\epsilon^{1/(N-1)}$) \citep{Connaughton2003}, but without the values of the dimensionless constants $C_{KZ}$.}
\red{Simulations of the Zakharov hamiltonian formulation of water waves have been first presented \citep{Pushkarev1996,Pushkarev2000} and recent direct numerical simulations have confirmed the validity and the accuracy of the predictions given by the Wave Turbulence Theory \citep{DeikeDNS,PanYuePRL2014} for pure capillary waves and with a low dissipation level. By adding a weak level of broad-scale dissipation, a numerical study of the decay of pure capillary wave turbulence \citep{PanYueJFM2015} displays an exponential decay of amplitude and a decrease of the slope of the capillary cascade. The role of nonlinear broadening has been also investigated \citep{PanYueJFM2017} for numerical simulations performed in finite domains.}

\red{To approach the natural situation, a direct numerical simulation of turbulence of surface waves, when capillary waves are excited by gravity waves, is of interest. Recently in a numerical study of air-water sheared flow, \citet{ZontaJFM2015,Zonta2016} observed that the spatial spectrum of the surface deformation presents a power law whose exponent is compatible with the value given by Wave Turbulence Theory for capillary waves. In contrast with the previous numerical studies, the complete dispersion relation of gravity capillary waves is implemented, but due to the shear forcing, the resulting wave field is anisotropic. For the case with the strongest capillary effects, the conversion of gravity waves into capillary ripples is supposed to occur through the parasitic capillary waves mechanism \citep{Zonta2016}. These last numerical results present some similarities with our measurements. However to determine quantitatively if Wave Turbulence Theory is relevant in these situations, a measurement of energy flux $\epsilon$ is necessary.}

\subsection{Dissipation spectrum and estimation of energy flux}
\label{Disspart}
\begin{figure}
 \begin{center}
 \includegraphics[width=13cm]{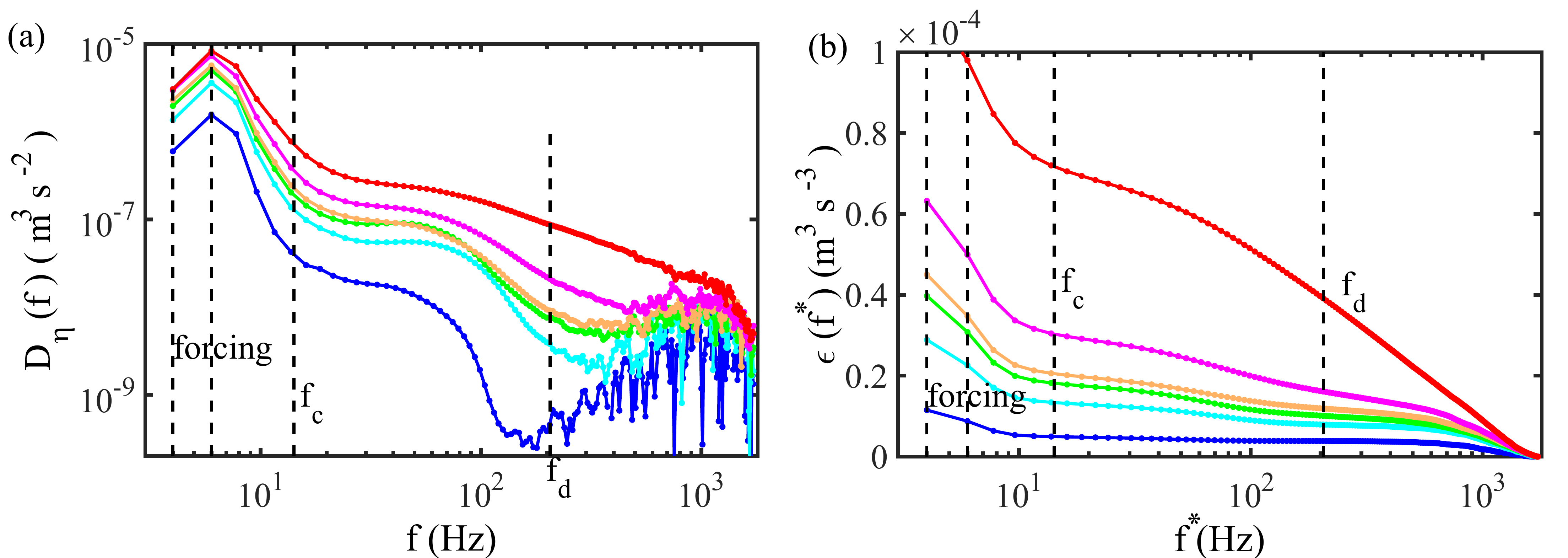} 
        \caption{(color online). (a) Wave dissipation spectrum $D_\eta(f)$ (see Text for definition) for different forcing amplitudes. From bottom to top: $\sigma_h = $1.3, 2.1, 2.7, 3.1, 3.4 and 3.6 mm, and $\sigma_s=0.15$, 0.19, 0.24, 0.27, 0.29 and 0.34. (b) Estimation of energy flux $\epsilon (f^*)$ as a function of the frequency $f^*$  by integration of dissipation spectrum for the same measurements. The energy flux is not constant, showing that dissipation occurs at all scales, even over the range of frequencies where a power law spectrum is observed.}
    \label{FigDiss}
       \end{center}
 \end{figure}     
In order to estimate the energy flux, we follow the approach proposed by \cite{Deike2014}, in a previous study of capillary wave turbulence using a local capacitive probe, and compute the dissipation spectrum within the capillary wave spectrum. \blue{A similar approach has been performed in the context of elastic wave turbulence experimentally \citep{Humbert2013} and numerically \citep{Miquel2014}.} \red{First, on average over a large number of wave periods the kinetic energy $E_c$ is equal to the potential energy $E_{pot}$. Assuming  $\langle E \rangle =\langle E_c + E_{pot} \rangle= 2\, \langle E_{pot}\rangle$, the spectrum of energy per density unit $E_{k}(k)$ writes:
\begin{equation}
  \int E_{k}(k) \mathrm{d}k =\frac{2}{\rho}\, \int \left[ \frac{1}{2} \rho g \, S_\eta (k) + \frac{1}{2} \gamma k^2 S_\eta (k) \right]\mathrm{d}k
\label{Ek2}
 \end{equation}
}
Here the analysis is carried out by using spatial spectrum $S_\eta (k)$ averaged over the time, but to facilitate the comparison with previous works using local probes, the results are then expressed as a function of the frequency using the linear dispersion relation, as it was shown in Fig.~\ref{Spectra} (b). The dissipation spectrum is obtained by computing the total dissipated power $D$, from the dissipation rate of energy $\Gamma$ at the frequency $f$:
\begin{equation} D=\int D_\eta(f)\,df=\int E_{f}(f)\,\Gamma (f)\,df \end{equation}
For gravity-capillary waves at a frequency above $4$\,Hz, we assume that most of the dissipation can be described, by the linear inextensible film model \citep{Lamb1932,VanDorn,Deike2012,HendersonSegur2013}. \red{In the capillary range the experimental dissipation rate $\delta$ is indeed well described by this model \citep{Deike2012,Haudin2016}. The energy dissipation} writes $\Gamma (f)=\sqrt{2} \sqrt{\nu \omega}\, k(f) /2$. It can be shown \citep{Deike2012,Deike2014}, than the contributions of the tank walls and bottom are negligible in the total dissipation for gravity waves of frequency above $4$\,Hz and even more for capillary waves. The dissipation spectrum is plotted in Fig.~\ref{FigDiss} (a). As stated previously \citep{Deike2014}, most of the dissipation occurs in the gravity wave range. Then the energy budget in the frequency space can be written as:
\begin{equation}
\frac{\partial E_f}{\partial t}+\frac{\partial \epsilon}{\partial f}= - D_\eta(f),
\label{Ebilan}
\end{equation} 
This relation is adapted from a similar equation in the $k$ space \citep{Nazarenkobook}. Consequently, \red{after a temporal averaging suppressing the temporal dependency,} the energy flux at a given frequency $\epsilon(f)$ reads:
\begin{equation}
\epsilon(f^*)=\int_{f^*}^{f_m}D_{\eta}(f)df.
\end{equation}
Here $f_m \approx 1800\,$Hz is the maximal frequency corresponding to the highest value of $k$ scanned in the spatial Fourier analysis. By convention, we impose $\epsilon(f^*=f_m)$ to be zero. As the analysis is performed \red{after temporal averaging for a statistically} stationary regime, it is not necessary to consider unsteady energy transfer as in \cite{PanYueJFM2015}. This estimation of energy flux $\epsilon(f^*)$ obtained by integration of dissipation spectrum is shown in Fig.~\ref{FigDiss} (b). As stated in previous studies of wave turbulence, in presence of dissipation, the energy flux  $\epsilon(f^*)$ decreases with frequency and thus is not conserved inside the power-law capillary spectrum. \green{In numerical simulations of capillary wave turbulence with broad-scale dissipation, the varying non-linearity level leads to smaller spectral slopes, than the theoretical values \citep{PanYueJFM2015}. For elastic plate turbulence, experimentally \citep{Humbert2013} and numerically \citep{Miquel2014}, non conservation of energy flux due to dissipation leads to a steepening of the spectra slope. A similar observation observation is reported experimentally for gravity wave turbulence~\citep{Campagne2018}. In contrast here and in a similar experiment \citep{Deike2014} for low enough viscous fluids}, the spectral slope remains close to the Wave Turbulence Theory value ($-17/6$ for the temporal spectrum). Due to the smaller container size and the higher excitation level, the values of $\epsilon(f^*)$ are found here two orders of magnitude larger than in \cite{Deike2014,DeikeDNS}. We note that this estimation of the energy flux is valid only for \red{a statistically stationary} measurement, as the formula providing the dissipation $\Gamma(f)$ are obtained after averaging over several wave periods~\citep{HendersonSegur2013}. 

 \begin{figure}
 \begin{center}
 \includegraphics[width=6.5cm]{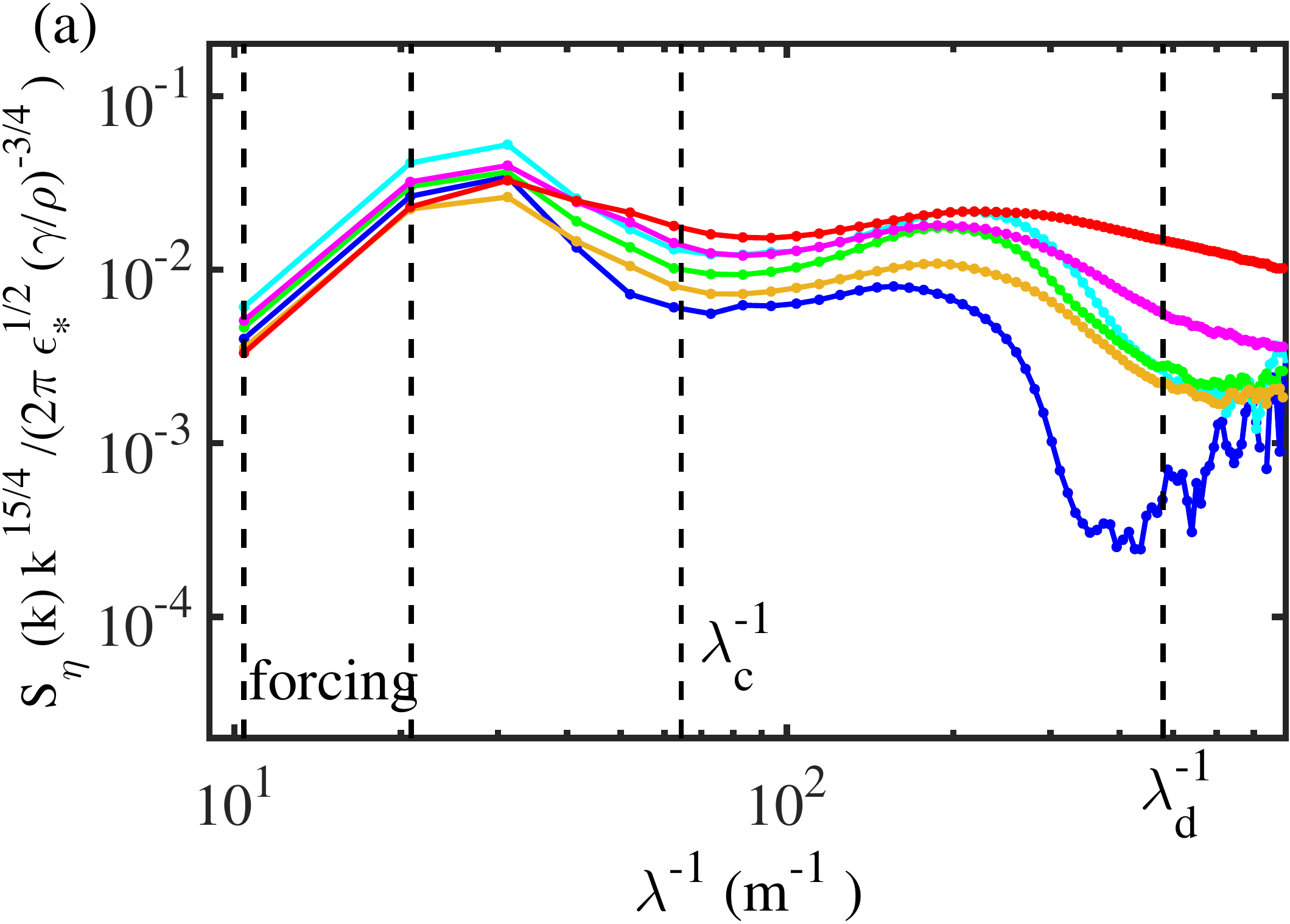}\hfill 
 \includegraphics[width=6.5cm]{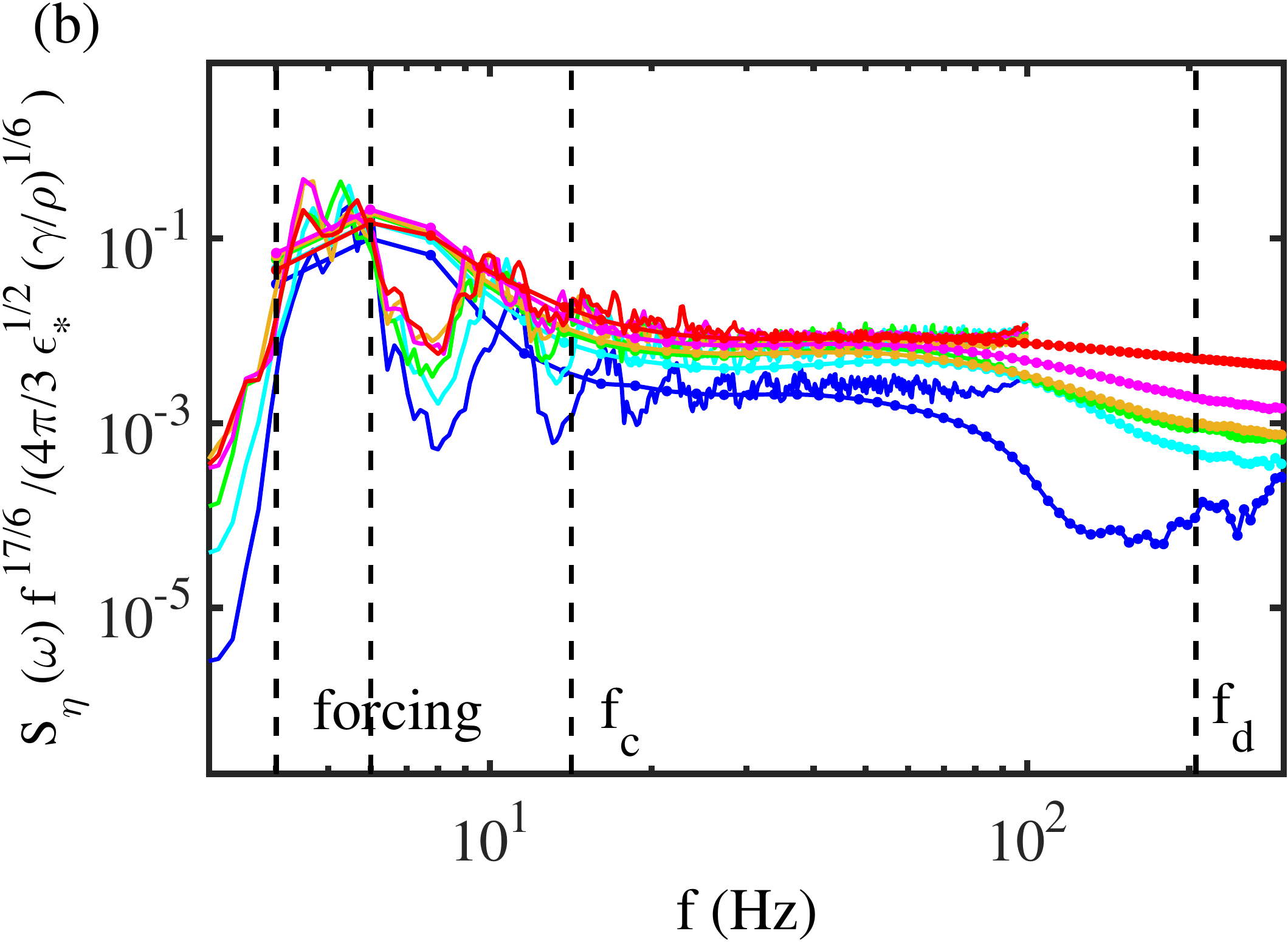} 
  \includegraphics[width=6.5cm]{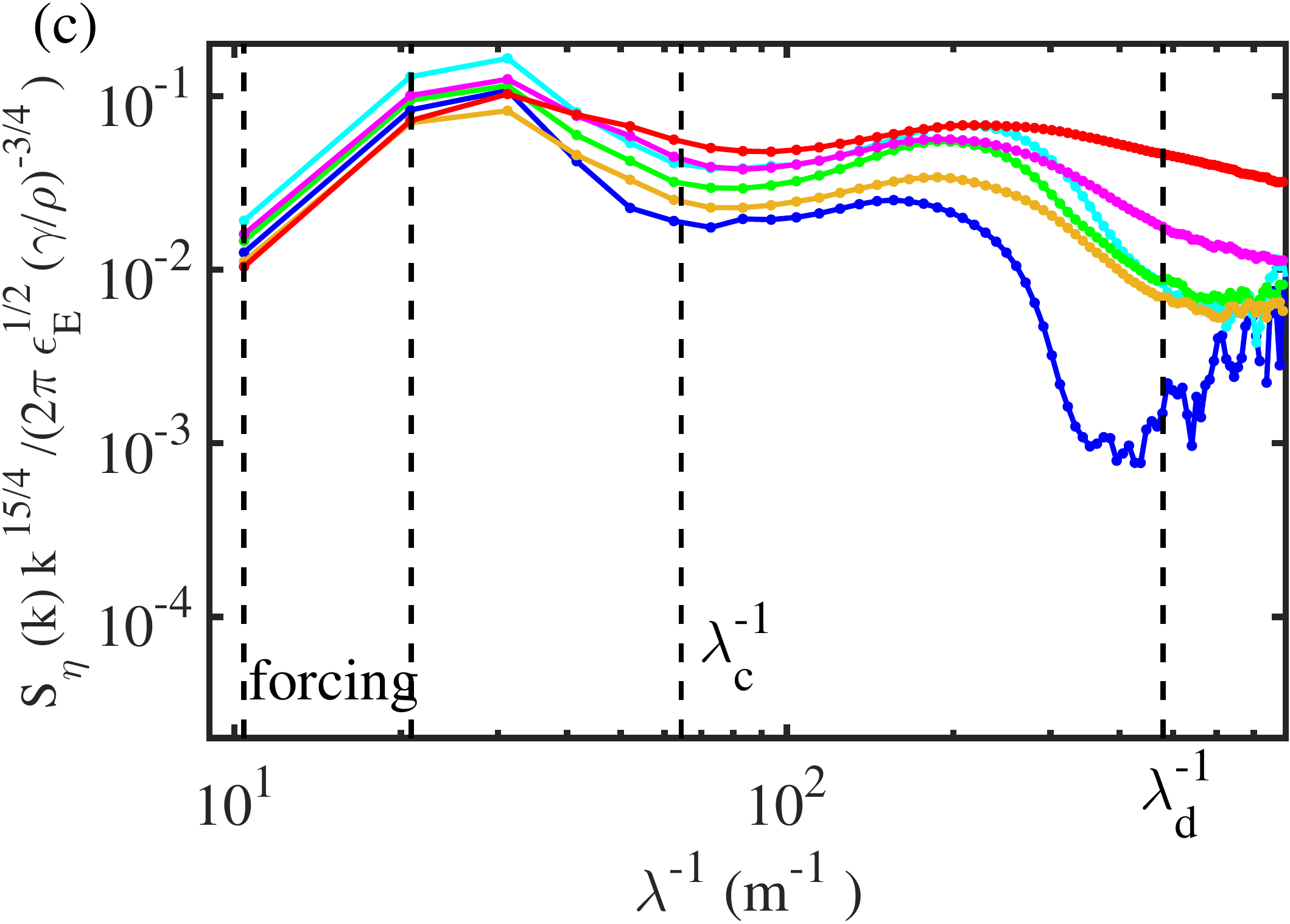}\hfill
 \includegraphics[width=6.5cm]{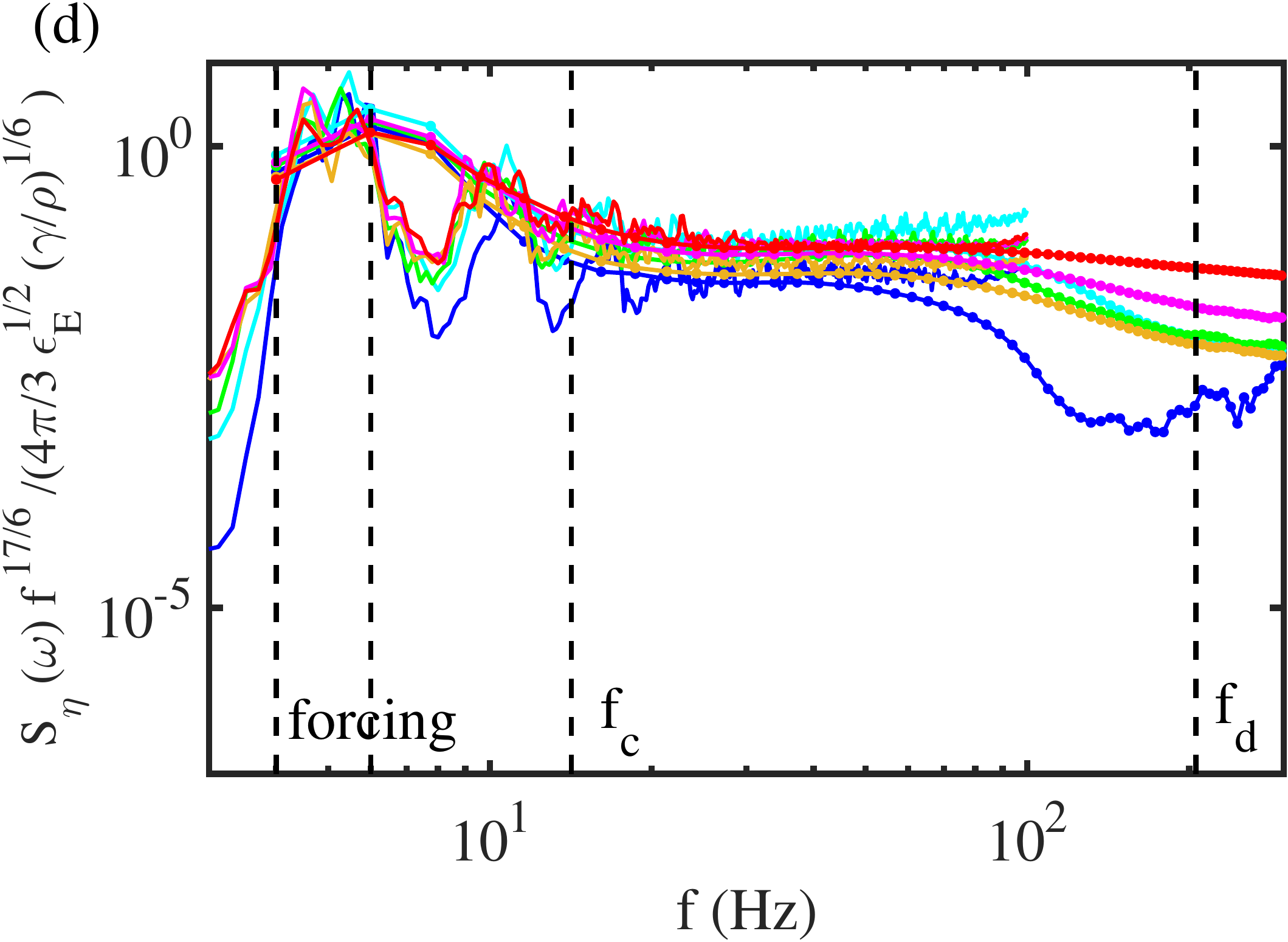} 
  \includegraphics[width=12cm]{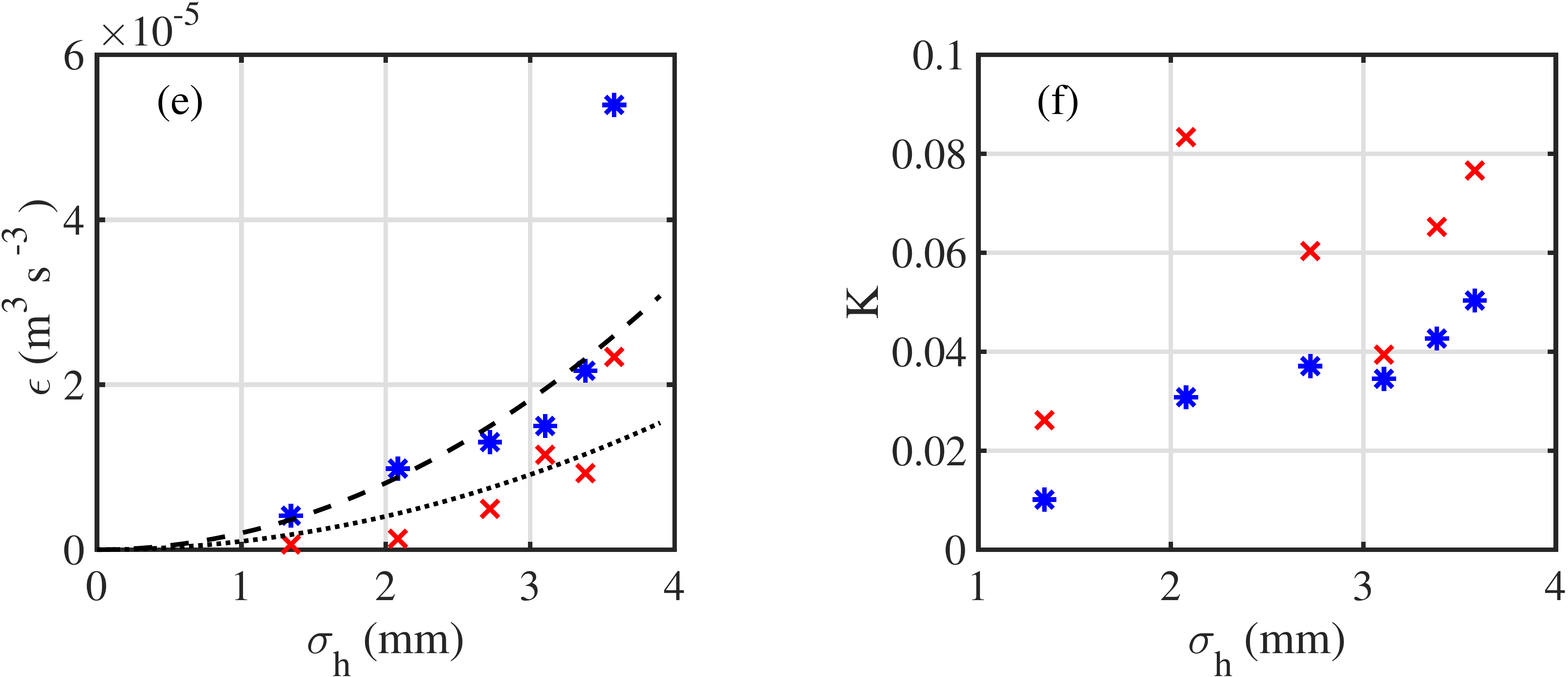} 
         \caption{(color online). (a) Spatial spectra $S_\eta (k)$, compensated by the theoretical expression (\ref{Spc}) for varying wave amplitude (see Fig.~\ref{Spectra}), using $\epsilon=\epsilon^*$. (b) Temporal spectra $S_\eta (\omega)$, compensated by the theoretical expression (\ref{Spc}) for varying wave amplitude, using $\epsilon=\epsilon^*$. \Green{The non-dimensional pre-factor, $K$, of the rescaled spectra, can be evaluated between the crossover scale ($\lambda_c^{-1}$ or $f_c$) and the dissipative scale ($\lambda_c^{-1}$ or $f_c$), see (Eq.~\ref{SpK}))} \blue{(c) same plot as in (a) with $\epsilon=\epsilon_E$. (d) same plot as in (b) with $\epsilon=\epsilon_E$.}\blue{ (e) Effective energy fluxes in the capillary range $\epsilon^*$ (defined in \S \ref{Disspart}, blue  $\ast$) and $\epsilon_E$ (defined in \S \ref{epsilonE}, red $\times$) for varying wave amplitudes $\sigma_h$. Dash and dot lines expected scaling law $\epsilon \propto \sigma_h^{2}$. } \Green{(f) Non-dimensional pre-factor $K$ (see (Eq.~\ref{SpK})) obtained experimentally from compensated spectra as a function of $\sigma_h$. $\ast$, $K$ using $\epsilon=\epsilon^*$ (energy flux from dissipation spectrum) and $\times$, $K$ using $\epsilon=\epsilon^*$ (energy flux from energy fluctuations). The non-dimensional pre-factor $K$ varies with the forcing strength, in disagreement with the Capillary Wave Turbulence Theory.}}
    \label{Figspcomp}
       \end{center}
 \end{figure}      
  
\blue{To compare in more details the experiments with the theoretical framework, we estimate for each measurement, an effective energy flux $\epsilon^*$ for turbulence of capillary waves by taking the average of $\epsilon(f^*)$ between $f_c$ and $f_d$. $\epsilon(f^*)$ is nearly linear with the total dissipated power $D$ in the capillary range. In Fig.~\ref{Figspcomp} (e), $\epsilon^*$ (blue stars) is found to grow with the wave amplitude $\sigma_h$, as this quantity is an indirect measurement of forcing amplitude. $\epsilon^*$ is defined from the energy balance (\ref{Ebilan}) and $E_f$ is a quadratic quantity of wave amplitude, thus $\epsilon^*$ should behave as $\sigma_h^2$. For the range of measurements $\epsilon^*$ follows experimentally a scaling in $\sigma_h^{2.1}$, except for the last point corresponding to the highest forcing amplitude. }

\Green{Following the expressions of capillary wave turbulence spectra (\ref{Spc}), compensated spatial and temporal spectra can be computed. For both the spatial and temporal compensated spectra in Fig.~\ref{Figspcomp} (a,b), we do not observe a satisfying collapse of the different measurements. Even if we exclude the lowest amplitude measurement, there is roughly a factor two between the spectrum for $\sigma_h=3.6$\,mm and the one for $\sigma_h=2.1$\,mm. We note also that the estimation of energy flux $\epsilon^*$ is indirect and relies on strong hypotheses on the dissipative processes, and remains thus approximate in experiments.}

\subsection{\blue{Estimation of energy flux from energy fluctuations}}
 \label{epsilonE}
\blue{To consolidate the estimation of energy flux, we propose here an other method to evaluate experimentally the energy flux, relying on the temporal fluctuations of the energy spatial spectrum $E_k (t)$ deduced from $S_\eta (k,t)$ (see \S \ref{Ekt}). The energy budget of the capillary wave cascade written in the wavenumber space provides theoretically also an alternative method to compute the energy flux~\citep{Nazarenkobook,Miquel2014,PanYueJFM2015}:
\begin{equation}
 \dfrac{\partial E_k}{ \partial t} + \dfrac{\partial \epsilon }{ \partial k} = - D_\eta (k),
\end{equation}
 with  $E_k$ the spatial spectrum of energy per density unit (\ref{Ek2}), $\epsilon$ the energy flux and $D_\eta (k)$ the spatial dissipation spectrum. $\frac{\partial E_{k} (k,t)}{\partial t}$ can be obtained from the instantaneous spatial spectrum $S_\eta(k,t)$ (plotted in Fig.~\ref{figspinsta}) and it would be thus interesting to access to the fluctuations of the energy flux $\epsilon$. However experimentally the dissipation rate $\Gamma (k)$ at the scale $k$ is estimated through the inextensible film model valid only after averaging over several periods. The dissipation spectrum $D_\eta (k)=E_k(k)\,\Gamma(k)$ is thus not resolved in time (with $\Gamma(k)$ the dissipation rate at the wavenumber $k$). However, the energy fluctuations could provide a complementary estimation of the energy flux. We find experimentally that the temporal r.m.s value of energy fluctuations integrated over the wavenumbers has the same order of magnitude than $\epsilon^*$. This quantity writes: 
\begin{equation}
\epsilon_E= \sqrt{ \langle \left( \int_{k c}^{k_d} \dfrac{\partial E_k}{ \partial t} \, \mathrm{d} k \right)^2 \rangle },
 \end{equation}
with $k_c$ the wavenumber corresponding to the gravity-capillary crossover, $k_d$ the dissipative scale and $\langle \, \, \rangle$ denotes a temporal average. $\epsilon_E$ is plotted as a function of $\sigma_h$ in Fig.~\ref{Figspcomp} (e) (red crosses) and compared with $\epsilon^*$. We observe that $\epsilon_E$ has the same order of magnitude than $\epsilon^*$, but is systematically smaller. $\epsilon_E$  follows also roughly the scaling in $\sigma_h^2$.}\\

\blue{Then replacing $\epsilon^*$ by $\epsilon_E$, we plot the corresponding compensated spectra in Fig.~\ref{Figspcomp} (c) and (d). The compensated spectra are slightly better gathered than in Fig.~\ref{Figspcomp}, but the ratio between the top and the bottom curve is also of order two. This method of estimation of $\epsilon$ from the energy fluctuations is not completely theoretically justified, because a rigorous estimation of the effective energy flux $\epsilon$ should balance dissipation and energy fluctuations. However, these results suggest than in average these two quantities have the same order of magnitude.}\\
%
 
\blue{We note also that $\epsilon$ could be directly computed from the non-linear interaction kernel \citep{Nazarenkobook}. This can be achieved in numerical simulations when having access to both the time dependent Fourier modes of the wave amplitude but also of the wave potential. This is not available experimentally so far. In numerical simulation, a such budget has been computed by \cite{Yokoyama2014} in the context of elastic plates.}

\subsection{\Green{Test of the scaling of spectra according Wave Turbulence Theory.}}

\Green{Using the two estimations of effective energy flux $\epsilon^*$ and $\epsilon_E$ for turbulent capillary waves, we have plotted the compensated spectra in Fig.~\ref{Figspcomp} according the theoretical expressions given by the wave turbulence theory. From these energy flux estimations, a good rescaling is not observed. Moreover, experimentally the hypotheses of the Wave Turbulence theory (isotropy, weak-non linearity, conserved energy flux ...) are not met. Therefore, we emphasize that we cannot evaluate the numerical values of the Kolmogorov-Zakharov constant $C_{KZ}$ in our experiment, from the compensated spectra. Nevertheless to summarize our results, we can estimate the non-dimensional pre-factor of the wave elevation spectra using the wave action spectrum (\ref{Spn}):} 
\Green{ \begin{equation}
S_\eta (k)=2\pi\,K\,\epsilon^{1/2}\,(\gamma/\rho)^{-3/4}\,k^{-15/4} 
 \quad \quad 
S_\eta (\omega)=4\pi/3\,K\,\epsilon^{1/2}\,(\gamma/\rho)^{1/6}\,\omega^{-17/6}
\label{SpK}
\end{equation}}

 \Green{We obtain $K$ by taking the average of the spatial compensated spectrum (with $\lambda^{-1} \in [94,312]\,$m$^{-1}$ to take in account the increase of $\lambda_d^{-1}$ with $\sigma_h$), of the temporal spectrum ({average between $f \in [15,100]$\,Hz}) and of the temporal spectrum derived from spatial measurements. Due to the approximate rescaling in frequency or wave-numbers, the value of $K$ for each spectrum is evaluated up to a factor two. We plot in Fig.~\ref{Figspcomp} (f) the average value between these three rescaled spectra as a function of $\sigma_h$ , in the case where the flux is given by the dissipation spectrum $\epsilon=\epsilon^*$ (blue stars) and in the case where the flux is given by the energy fluctuations $\epsilon=\epsilon_E$ (red crosses). In the case $\epsilon=\epsilon^*$, $K$ is found to increase with $\sigma_h$. In the case $\epsilon=\epsilon_E$ the values of $K$ are systematically larger and more scattered. According to the Wave Turbulence Theory, the pre-factor $K$ should not depend on the wave amplitude (which is contained in $\epsilon$) and the values of $K$ should all collapse towards a constant. Our experimental results demonstrate that at high level of nonlinearity, with parasitic capillary wave formation and with an important broad-scale dissipation, the spectra are not quantitatively in agreement with the Wave Turbulence Theory predictions, although the experimental spectral slopes have the values given by the theory. }
  

\subsection{\red{Applicability of Wave Turbulence Theory in capillary wave experiments}} 
\red{Given the presented results, we discuss now the applicability of the Zakharov's theory in capillary waves experiments. }
\red{Several independent experiments reported observation of the spectra exponents given by the Wave Turbulence Theory \citep{Putterman1996,Henry2000,Brazhnikov2002,Falcon2007,FalconFalcon2009,Xia2010}. Nevertheless test of the energy flux scaling $\epsilon^{1/2}$ have been performed in only one experimental study \citep{Deike2014} for a very low viscosity liquid (mercury). Most of previous laboratory works have tried to isolate capillary wave turbulence from the gravity wave regime by using a parametric forcing \citep{Putterman1996,Putterman1997,Henry2000,Brazhnikov2002,Snouck2009,Xia2010}, by operating under microgravity \citep{FalconFalcon2009} or by studying waves at the interface between two fluids \citep{During2009,IssenmannEPL2016}.
In experiments where capillary waves are forced by random gravity waves, the situation is puzzling. Spectra of capillary waves measured with \red{local} capacitive probes verify the power law $f^{-17/6}$ \citep{Falcon2007,Deike2012,Deike2014}.
However the study of the decay of turbulence has shown that the decline of the capillaries follows the viscous damping of the largest container eigen-mode \citep{Deike2012}. A similar observation was also reported for a parametric excitation with cryogenic liquid of low viscosity \citep{Brazhnikov2002}. Consequently dissipation occurs at all scales of the turbulent cascade leading to a non conserved energy flux \citep{Deike2014}. This last point constitutes a severe drawback of the applicability of Wave Turbulence Theory for capillary wave experiments, as the formalism of the theory relies on a Hamiltonian description of the wave-field dynamics.}
     
\red{Here, our spatio-temporal measurements of free-surface deformation complete these previous observations. Capillary waves are generated by nonlinear wave interaction from the randomly excited gravity waves produced by the wave-maker. At high enough forcing amplitude, for the scales belonging to the capillary wave range, spectra of wave elevation behave as power-laws, defining the turbulence of capillary waves. Exponents of spectra are in agreement with the Wave Turbulence Theory ($-15/4$ in $k$-space and $-17/6$ in $\omega$-space). However the level of nonlinearity given by the wave steepness and the significant broad-scale dissipation suggests that the experimental capillary wave turbulence does not likely follow the weakly nonlinear mechanisms described in the theory. Due to the large viscous dissipation in capillary scales, the level of forcing corresponding to a randomization of the initial conditions, corresponds to a high level of nonlinearity. Using two different methods, we show that the time-scale separation is broken. First the nonlinear times have similar values than the linear time \textit{i.e.} the wave period. Then by analyzing the temporal fluctuations of the spectra, we demonstrate that the wave energy transfers occur through successive bursts distributing quasi-instantaneously energy from forcing scales to a large range of smaller scales. Between bursts, a relaxation process takes place which could be closer to the weakly nonlinear theory. We interpret those strong nonlinear events in the physical space as a manifestation of parasitic capillary wave generation. This non-local wave interaction mechanism produces capillary waves from gravity waves commonly observed at the surface of the oceans. Evidence of parasitic capillary waves trains are found in the experimental reconstructions of the wave field and due to the random forcing, capillary waves are excited on a broad range of scales. The process generating the capillaries is strongly nonlinear and explains the short correlation time scale observed in the capillary turbulent spectra.}

\Green{Then by estimating the dissipated power through the viscous wave damping, we estimate the energy flux $\epsilon(f^*)$ in the capillary wave range. This energy flux is not conserved through the scales, displaying again the importance of viscous dissipation and a strong disagreement with the hypothesis of Wave Turbulence Theory which supposes that dissipation is negligible in the inertial range, where power-law spectra are expected (and observed). The rescaling of the wave spectra by the mean energy flux for different excitation amplitudes is approximate both when considering the spatial spectra (in Fig.~\ref{Figspcomp} (a) ) and the temporal one (in Fig.~\ref{Figspcomp} (b) ). These observations confirm that the capillary wave turbulence experimentally obtained in strongly nonlinear regime is not described by the weakly nonlinear theory.}

\red{However we evidence presence of Three-Wave interactions in temporal and spatial Fourier spaces and the exponents of power-law spectra are close from those given by the Wave Turbulence Theory. How can we explain these statements ? We assumed that wave energy transfer occurs mainly through strong and fast nonlinear events, which are interpreted as parasitic capillary wave generation, when gravity waves of forcing are sufficiently steep. In fact, parasitic wave generation can be described at leading order as a non-resonant Three-Wave process \citep{WatsonJFM1996,WatsonJFM1999}. Moreover it was found that the energy decay of parasitic waves follows a law  $E(t)/E(t=0)=[1+t/\tau]^{-1}$ \citep{DeikeJFM2015cap}, with $\tau$ a characteristic nonlinear time. This temporal evolution in decaying regime is a direct consequence of the quadratic nonlinearity of equations governing free-surface waves. Then, the bispectral analysis shows a clear evidence of Three-Wave interactions in frequency space. Therefore at the leading order, the wave energy transfer occurs through Three-Wave interactions, which may be resonant or not. This statement implies a energy flux scaling in $\epsilon^{1/2}$, \blue{which is only approximately verified by plotting the compensated spectra (Fig.~\ref{Figspcomp} (a), (b), (c) and (d)) with $\epsilon$ estimated from the dissipation spectrum or from the fluctuations of the energy spectrum}. Consequently from dimensional analysis and the linear dispersion relation, the exponents of power-law spectra for capillary waves in turbulent regime are $-15/4$ in $k$-space and $-17/6$ in $\omega$-space, like in Wave Turbulence Theory \citep{Connaughton2003}, under the condition that nonlinearity is high enough to develop a self-similar regime on a sufficiently large range, defining clear power-law spectra. Incorporation of higher order nonlinear interactions (Four, Five ... Wave processes), could improve the description especially in the rescaling of spectra by the estimated average energy flux $\epsilon^*$. This experiment with nonlinear random regimes of capillary waves demonstrates that the verification of spectral slopes is not sufficient to test accurately Wave Turbulence Theory in experiments. Although difficult and approximate in experiments, estimation of energy flux appears necessary.}     
    
\section{\red{Experimental constraints on capillary wave turbulence observation}}
\label{Constraints}
\red{Wave Turbulence Theory fails to describe quantitatively random regimes of capillary waves in our experiments. We argue that this statement can be generalized to other experimental or natural situations, where random capillary waves on Earth are generated in water from gravity waves by nonlinear interactions, due to two main reasons, the significant viscous dissipation and the gravity-capillary waves crossover.}
\subsection{\red{Effect of viscous dissipation}} 
\label{Constraintsdiss}
\red{First the viscous dissipation is important for capillary waves and cannot be neglected. The inextensible film model \citep{VanDorn,HendersonSegur2013}, provides the wave decay rate $\delta=\sqrt{2} \sqrt{\nu \omega} k /4$ in deep water condition and is verified for capillary waves. The transition between gravity and capillary waves occurs for a critical wave number $k_c=\sqrt{\rho\,g/\gamma}$ which corresponds to a critical frequency $f_c \approx 14$\,Hz. For this frequency $f_c$ from which starts the capillary wave range, the dissipative time $1/\delta_c \approx 0.69$\,s. This time can be interpreted spatially by computing the corresponding dissipative length $l_{att}=v_g / \delta$, with $v_g=\frac{\partial \omega}{\partial k}$, which gives for $f_c$, $l_{att\, c} \approx 0.15$\,m. This length constitutes the typical distance travelled by a wave produced by the wavemaker or a nonlinear interaction. If we consider the highest frequency of the capillary power-law spectrum, $f_d=200\,$Hz, $1/\delta_m \approx 0.024$\,s and  $l_{att\, d} \approx 0.015$\,m. The size of the tank being $0.165$\,m, it appears consequently impossible to obtain a homogeneous wave-field by exciting the waves in the capillary range, except when the forcing is homogeneous like when capillary waves are forced through the Faraday instability. Finite size effects are also often discussed as a quantization of the wavenumbers in the tank, but we emphasize that this quantization occur by multiple reflections on the boundaries of the tank, like a $N$ waves interference process (See Appendix A). Moreover the reflection coefficient in the presence of a contact line in capillary regime is not well documented, but a recent study \citep{Michel2016} shows that is lower than one. The damping by the meniscus and the strong surface dissipation by viscosity, explain thus that no quantization in wavenumbers is observed in capillary wave experiments as it can been seen in Fig.~\ref{RD} (a), because a capillary wave disappears after travelling a distance of order $2\,l_{att\, c} \approx 0.3$\,m. A small sized experiment helps to obtain a sufficient energy flux by surface unit $\epsilon$ to observe power law spectra, but for capillary waves, the boundaries are nearly not felt. However difficult to quantify, the strong dissipation, implies likely a strong forcing amplitude to fill spectrum in the capillary wave range, which explains that to obtain the self-similar wave-turbulent regimes, a steepness of order $0.2$ is needed. Moreover it is not efficient to produce directly capillary waves through the wavemaker, because they would be observable only in the vicinity of the wavemaker due to the viscous damping. The capillary waves are then produced from gravity waves by nonlinear wave interactions.}\\
\Green{The substantial viscous dissipation occurs thus at all scales and implies that energy flux is not conserved through the scales. Consequently the constant flux solutions of Wave Turbulence Theory cannot be applied directly to experimental field of random capillary waves. To take in account the decay of energy flux through the capillary scales, an average energy flux $\epsilon^*$ ({or $\epsilon_E$}) can be used to rescale the spectra. With a low viscosity liquid like mercury is used, the rescaling is acceptable \citep{Deike2014}, but here with water, using the same method {the rescaling is less convincing}.}

\subsection{\red{Effects of gravity-capillary wave crossover}}
\red{The nonlinear conversion of gravity waves into capillary waves is also problematic to apply Wave Turbulence Theory to experimental capillary waves. The power-law spectra are indeed obtained for self-similar dispersion relations expressed also as power-laws. The analytically predicted spectra  are thus expected either for pure gravity waves or for pure capillary wave turbulence, thus far from the crossover wavenumber $k_c$. Several experiments~\citep{Falcon2007,Deike2012,DeikeJFM2015} in water or mercury have shown an abrupt transition around $f_c$ between a gravity wave power-law spectrum and a capillary wave power law spectrum. However, for a given energy flux $\epsilon$, both spectra are incompatible as the values of the Kolmogorov-Zakharov constants are fixed by the theory.  For gravity wave turbulence, according Wave Turbulence Theory,  the power spectrum of the wave elevation $\eta (\textbf{x},t)$ writes in time \citep{Zakharov1967grav}: $S_\eta (\omega)= C_{KZ}^{(g \omega)}\,\epsilon^{1/3}\,\omega^{-4}$. The value of $C_{KZ}^{(g\,\omega)}$ was analytically found equal approximately to $2.75$ \citep{Zakharov2010}. If we assume that at the gravity-capillary crossover $f_c$, the gravity wave spectrum  equals the capillary wave spectrum, $S_\eta (\omega)=4\pi/3\,C_{KZ}\,\epsilon^{1/2}\,(\gamma/\rho)^{1/6}\,\omega^{-17/6} $, with $4\pi/3\,C_{KZ} \approx 41$ \citep{Pushkarev2000}, then the same energy flux cannot be transferred from gravity scales to capillary scales. A simple order of magnitude illustrates this statement. For example, in gravity wave turbulence experiments, energy flux was found to be equal to $\epsilon=1\,10^{-4}\,$(m/s)$^3$ for the gravity wave cascade \citep{DeikeJFM2015}. A continuity of temporal spectrum $ S_\eta (\omega)$ for $\omega_c=2\pi\,f_c$ implies that $\epsilon$ becomes equal to $6.9\,10^{-7}$\,(m/s)$^{3}$ in the capillary wave regime, a value significantly lower. This results holds for other values of the energy flux. }\green{  If we assume that $f_c$ is a free parameter, continuity of gravity and capillary wave turbulent spectra gives a gravity capillary crossover at $f=1.65\,$Hz for $\epsilon=1\,10^{-4}\,$(m/s)$^3$, which appears too small. The order of magnitude of $f_c$ would be found only for energy flux of order $\epsilon=1\,10^{-10}\,$(m/s)$^3$.} \red{Therefore only a small amount of the energy in the wave system could be transferred to the capillary waves by a Wave Turbulence mechanism assuming local interactions in $k$ space. An estimation of the injected power at large scale will thus fail to estimate the energy flux of the capillary wave turbulence. Whereas it was proposed that for small enough energy flux ($\epsilon < (\gamma g / \rho)^{3/4} \approx 4.3\,10^{-3}$\,(m/s)$^{3}$), gravity and capillary spectra could be connected \citep{Connaughton2003}, it appears thus that for given values of the Kolmogorov-Zakharov constants, either energy accumulation is expected around $f_c$ either energy transfer to the small scales occur by non-local interactions, \textit{i.e.} the involved scales are significantly separated in frequency or in wavenumber space. Our experiments illustrate the last case. Steep gravity waves produce short capillary wave trains by a fast and direct mechanism at small scale, like the parasitic capillary wave generation mechanism. As illustrated in Fig.~\ref{figspinsta} (a), energy bursts populate quasi-instantaneously a large range of capillary scales. For a sufficient steepness we expect thus that gravity waves generate nonlinearly capillary waves by direct non-local interactions, whereas Wave Turbulence Theory assume local interactions in wavenumber space~\cite{Nazarenkobook}.  Few studies have investigated the statistical study of gravity-capillary waves, using a kinetic equation including Three-Wave interactions numerically \citep{WatsonJFM1993,WatsonJFM1996,WatsonJFM1999,Dulov2009,Kosnik2010} and analytically \citep{Stiassnie1996}. However, the resulting spectra are consequently not expressed as power-laws and so are more difficult to test experimentally.}

\section{Conclusion}
\red{An experimental study of turbulent regimes of capillary wave forced by steep gravity waves is presented here, using a spatio-temporal measurement of free-surface deformation. Gravity waves are generated by a wavemaker with a random forcing in a small tank. Capillary waves are generated by nonlinear wave interactions, principally by transient generation of capillary trains in a way similar to the parasitic capillary wave generation mechanism. In spatial Fourier space, capillary wave generation is associated with intermittent bursts of energy transferring quasi-instantaneously energy from large to small scales. After a temporal average, capillary waves appear in average uncorrelated, justifying a statistical analysis based on the spatial and time spectra of wave elevation. At high enough amplitude of excitation, for the scales belonging to the capillary wave range, spectra of wave elevation behave as power-laws, whose exponents are in agreement with the Wave Turbulence Theory ($-15/4$ in $k$-space and $-17/6$ in $\omega$-space). However due principally to the substantial viscous dissipation and the significant level of nonlinearity, the Wave Turbulence Theory fails to describe these experimental measurements. The observed power-laws may thus be explained by dimensional analysis in presence of a quadratic nonlinearity.}

\red{In this work, we experimentally characterize an example of strong wave turbulence, which differs in the mechanisms at play with the weakly nonlinear Wave Turbulence Theory. In particular capillary waves are mainly produced by the parasitic capillary wave generation mechanism, which can be interpreted in first approximation as a strongly non local and non-resonant Three-Wave interaction~\citep{WatsonJFM1996}. \blue{Taking in account higher order nonlinear interactions could be necessary to describe quantitatively parasitic wave generation. Moreover,} non-resonant interactions are often not considered in the study of random waves interacting nonlinearly, because contributions of non-resonant interactions in wave energy vanish for sufficiently large system or large observation time. Nevertheless, for capillary waves, due to viscous dissipation, life time of wave packets is short and tank size is limited, therefore contribution of non-resonant interactions has no reason to be negligible. The determination of the relative contribution between resonant and non-resonant interactions could constitute a new direction in wave turbulence study, beyond the weakly-nonlinear limit. A such study is indeed relevant in most experimental systems with interacting waves. A similar statement could be drawn in the nature, for the small scale spectrum of a random sea, when capillary waves are generated by the parasitic capillary wave generation mechanism.} 

\begin{acknowledgments} We thank N. Mordant for discussions and help in data processing. \green{We acknowledge also S. Nazarenko, F. P\'etr\'elis and S. Fauve for discussions}, L. Gordillo for help in PIV measurements and A. Lantheaume for technical assistance. \blue{We thank the referees for insightful comments which greatly improve the quality of this article.} This work was funded by ANR-12-BS04-0005 Turbulon. 
\end{acknowledgments}
\appendix
\section{Effect of wave dissipation on finite size effects}

Using a simplified model we study in this appendix the effect of wave viscous dissipation in the wave quantization phenomenon, which is expected when the wave-field is confined in a finite-sized container. We consider a one-dimensional domain along $O_x$ between $x=0$ and $x=L$, limited by rigid walls. An initial monochromatic surface wave of wavenumber $k$  and angular-frequency $\omega$  is continuously injected in $x=0$ with an amplitude $A_0$. The motion of the wall creating the wave is supposed sufficiently small to be neglected. Due to linear viscous dissipation the wave decays spatially with a rate $\beta$. The wavelength is supposed small in front of the typical dissipation length $1/\beta$ and the system length. The initial free-surface deformation writes using complex formalism $\underline{\eta_0} (x,t)=A_0\,\e^{-\beta\,x} \e^{\mathrm{i} (\omega t -k x)}$. When the wave reaches the position $x=L$, to vanish the horizontal velocity at any time, it can be shown that a reflected wave labelled $1$  is created, propagating backward with the same amplitude than the incident wave. The wave $1$ is then reflected in $x=0$ to creates the forward wave $2$ and so on and so forth. The total free-surface deformation can be expressed in stationary regime by the sum:
\begin{equation}
\underline{\eta} (x,t)=A_0 \e^{-\beta\,x} \e^{\mathrm{i} (\omega t -k x)}+A_1 \e^{\beta\,(L-x)} \e^{\mathrm{i} (\omega t +k x)}+A_2 \e^{-\beta\,x} \e^{\mathrm{i} (\omega t -k x)}+A_3 \e^{\beta\,(L-x)} \e^{\mathrm{i} (\omega t +k x)} \ldots
\end{equation}
with the following relations between the wave amplitudes,
\begin{eqnarray*}
A_0 \e^{-\beta\,L}\e^{-\mathrm{i} k L} & = & A_1 \e^{\mathrm{i} k L} \\
A_1 \e^{-\beta\,L}& = & A_2 \\
A_2 \e^{-\beta\,L}\e^{-\mathrm{i} k L} & = & A_3 \e^{\mathrm{i} k L} \\
&\ldots & 
\end{eqnarray*}
Then amplitude of forward and backward waves write respectively :
\begin{equation}A_{2p}=A_0 (\e^{-2\beta L } \e^{-2 \mathrm{i} k L})^p \quad \quad A_{2p+1}=A_1 (\e^{-2\beta L } \e^{-2 \mathrm{i} k L})^p \end{equation}
Consequently $\underline{\eta}$ can be seen as the sum of two geometric sequences with the same ratio $ \e^{-2\beta L } \e^{-2 \mathrm{i} k L}$. By taking the infinite limit in the sum, like in a $N$-wave interference problem, we obtain:
\begin{equation}\underline{\eta} (x,t) =\dfrac{A_0}{1-\e^{-2\beta L } \e^{-2 \mathrm{i} k L}}\,\left(\e^{\mathrm{i} (\omega t -k x)}+ \e^{-\beta L} \e^{-2 \mathrm{i} k L}\e^{\mathrm{i} (\omega t +k x)}\right)  \end{equation}
The space and time average amplitude is obtained by taking the square-root of the product of $\underline{\eta}$ with its complex conjugate. After some algebra, we obtain:
\begin{equation}
\left\langle \eta \right\rangle =\sqrt{\underline{\eta}\, \underline{\eta}^*}=\left( \dfrac{A_0^2\,\,\dfrac{1+\e^{-\beta L}}{1-\e^{-\beta L}}}{1+4 \dfrac{\e^{-\beta L}}{(1-\e^{-\beta L})^2}\,\sin^2 (k L) }\right) ^{1/2} 
\label{EqQuantdiss}
\end{equation}
 \begin{figure}
 \begin{center}
\includegraphics[width=0.32\textwidth]{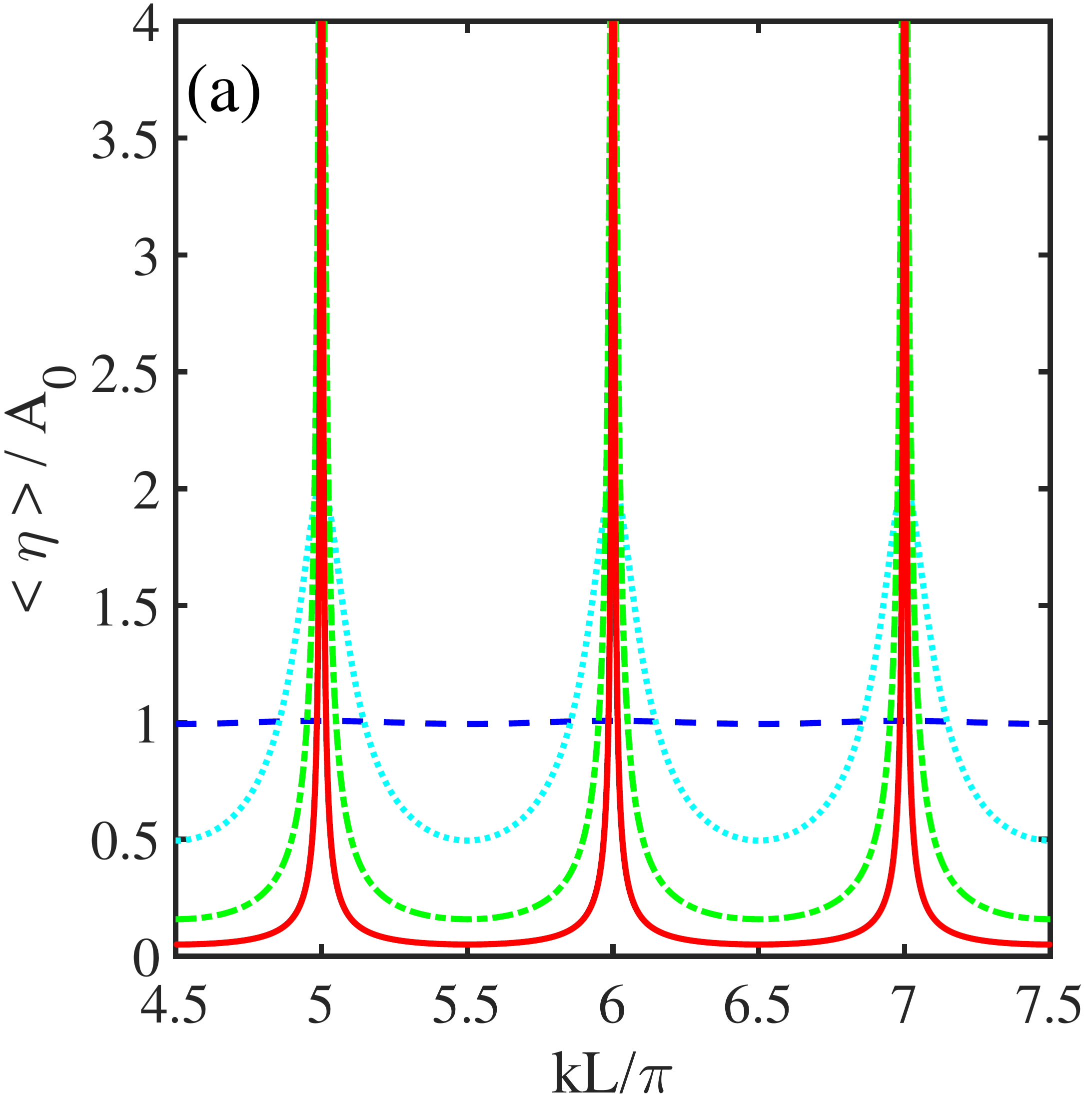} \hfill
\includegraphics[width=0.32\textwidth]{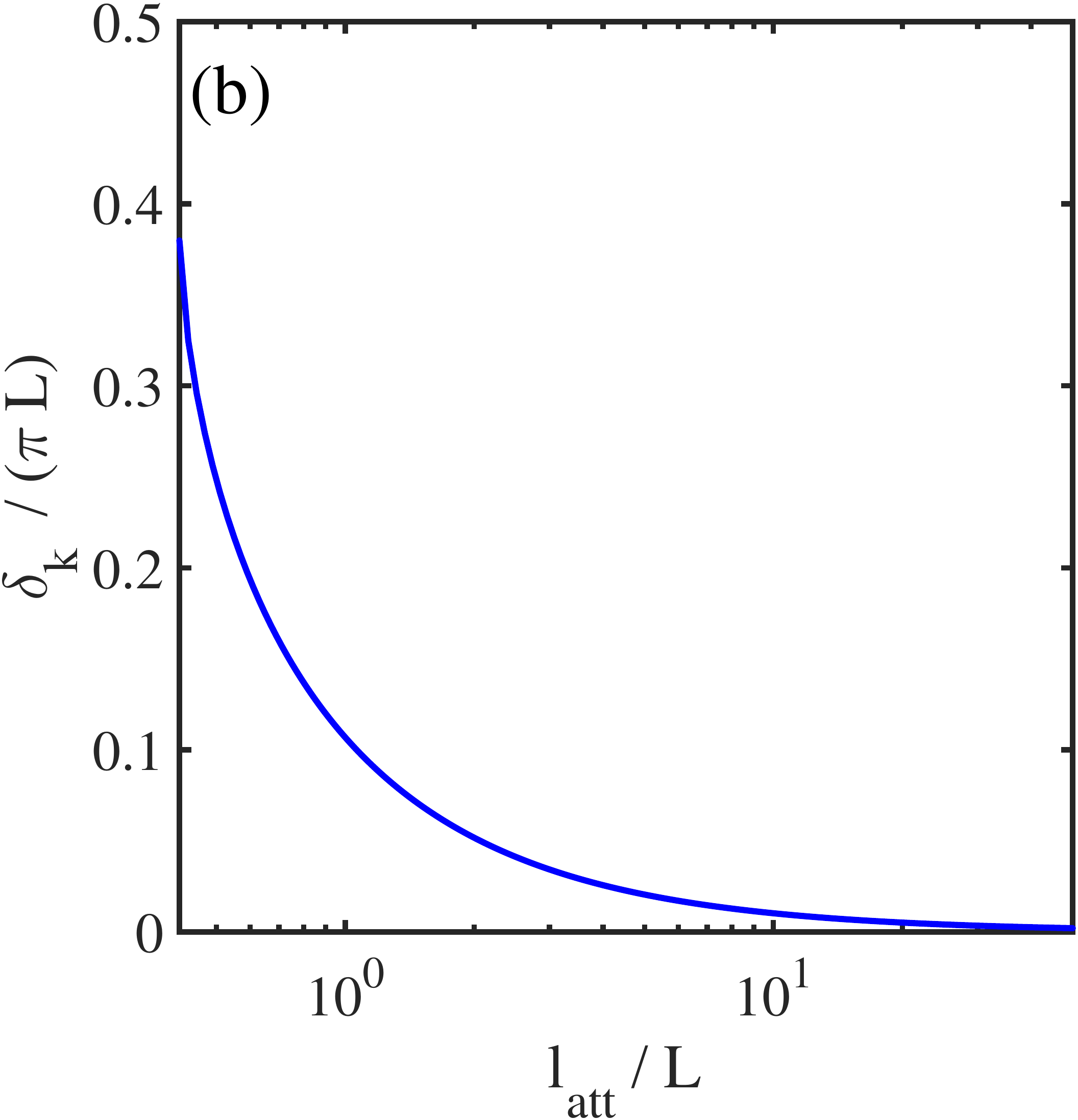} \hfill
\includegraphics[width=0.32\textwidth]{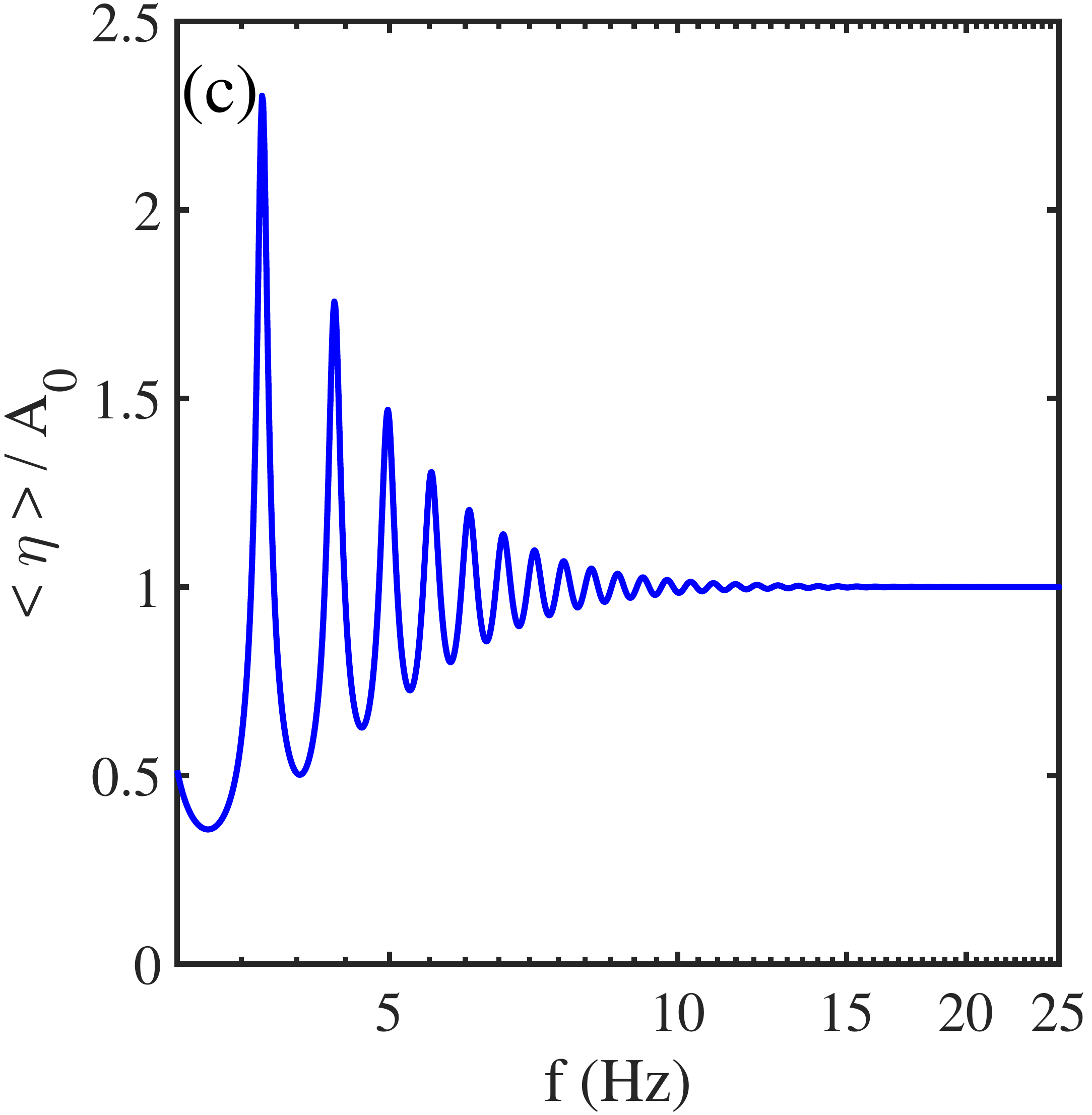} 
        \caption{(color online). (a) $\left\langle \eta \right\rangle /A_0$ as a function of $k L / \pi $ for different level of wave dissipation quantified by the attenuation length $l_{att}=0.2\,L$ (blue dash line), $l_{att}=2\,L$ (cyan dot line), $l_{att}=20\,L$ (green dash-dot line) and $l_{att}=200\,L$ (red plain line). For small enough dissipation wavenumber quantization for $k$ multiple of $\pi/L$ is found. (b) Half width at half maximum of the resonant peak as a function of $l_{att}$. (c) $\left\langle \eta \right\rangle /A_0$ as function of the wave frequency using the experimental parameters: container size, dispersion relation and increasing dissipation rate with frequency. Wave-mode quantization disappears above $f>10\,$ Hz.}  
    \label{quantdiss}
       \end{center}
 \end{figure}    
The dependency of $\left\langle \eta \right\rangle$ with the wavenumber $k$ is plotted in Fig.~\ref{quantdiss} (a) for varying dissipation level. The attenuation length $l_{att}=1/\beta$ is used to facilitate the comparison with the system size. The solution is analogue to the resonance of a cavity in which a wave is injected. When dissipation is small or $l_{att}$ large, resonance occurs for the eigen-mode of the system given by the condition $k=\dfrac{p\,\pi}{L}$ with $p$ a positive integer. The peak amplitudes saturate due to the non-zero dissipation. If dissipation is increased, the attenuation length decreases and the width of the peaks increases as it can be seen in Fig.~\ref{quantdiss} (b). For $l_{att} \lesssim 0.4 \, L $ the width become comparable with the distance between peaks and they become indistinguishable. $\left\langle \eta \right\rangle$ is thus nearly flat for $l_{att}=0.2\,L$ in Fig.~\ref{quantdiss} (a).
These results can be applied to the experimental situation, by taking $L=0.165\,$m, using the dispersion relation Eq~\ref{RDL} and expressing $l_{att}=v_g / \delta$ with $\delta=\sqrt{2} \sqrt{\nu \omega} k /4$ the decay rate in deep water in the limit of fully contaminated surface (see \S \ref{Constraintsdiss}). The average wave amplitude  $\left\langle \eta \right\rangle$ is displayed as a function of the frequency of the injected wave in Fig.~\ref{quantdiss} (b). Resonance due to the finite size of the tank, becomes insignificant for $f>10$\,Hz, thus in the capillary regime. This simple model shows that the quantization of the wavenumbers in finite-sized domain is a limit result for vanishing viscous dissipation, in presence of forcing. The eigen-modes are indeed stationary wave solutions in free regime (without forcing by a wavemaker) and obtained by applying a Helmholtz equation on the domain. With forcing and small dissipation, these modes are created physically by an interference process due to the multiple reflections of the waves on the domain boundaries. In presence of significant dissipation, the container modes are less defined or even disappear completely when the waves are too much damped during their propagation to feel the boundaries. Experimentally energy dissipation at the reflection in capillary wave regime, due to the motion of the contact line, increases even more the total amount of dissipation and the wave-mode quantization becomes even less observable.

\bibliographystyle{jfm}
%

\end{document}